\documentclass[twocolumn]{aastex62}

\pdfoutput=1
\usepackage{multirow}
\usepackage{graphicx}
\usepackage{grffile}
\usepackage{epsfig}
\usepackage{epstopdf}
\usepackage{amsmath}
\usepackage{psfrag}
\usepackage{units}
\usepackage[]{natbib}
\usepackage{url}
\DeclareMathAlphabet{\mathpzc}{OT1}{pzc}{m}{it}

\def\sun{\hbox{$\odot$}}
\def\nh{{n_{\rm H}}}

\def\nh2{{n(\rm H_2)}}
\def\h2{${\rm H_2}$}

\def\3cm{\rm {cm^{-3}}}
\def\2cm{\rm {cm^{-2}}}
\def\s-1{\rm {s^{-1}}}
\def\etal{et al.}

\def\Msun{$M_{\sun}$}
\def\Lsun{$L_{\sun}$}
\def\mum{{$\mu\rm m$}}
\def\kms{\hbox{${\rm km\,s}^{-1}$}}
\def\Kkms{\hbox{${\rm K}\,{\rm km}\,{\rm s}^{-1}$}}
\def\ndv{\hbox{${\rm cm}^{-2}\,{\rm km}^{-1}\,{\rm s} $}}

\def\Wm{${\rm {W}~{{m^{-2}}}}$}

\def\hcop{{{HCO$^+$}}}

\def\hcn{{{HCN}}}

\def\hh{{{H$_2$}}}
\def\h2o{{{H$_2$O}}}
\def\hho{{{H$_2$O}}}
\def\oh2o{{{o-H$_2$O}}}
\def\ph2o{{{p-H$_2$O}}}
\def\ohhop{{{o-H$_2$O$^+$}}}
\def\h2s1{{\rm {H$_2$}S(1)}}
\def\h2s2{{\rm {H$_2$}S(2)}}
\def\twco{{$^{12}$CO}}
\def\thco{{$^{13}$CO}}

\def\oi{{[O~{\scriptsize I}]}}

\def\oiii{{[O~{\scriptsize III}]}}

\def\ci{{[C~{\scriptsize I}]}}
\def\PzeroOi{{[O~{\scriptsize I}]~${}^3P_0-{}^3P_1$}}
\def\PoneOi{{[O~{\scriptsize I}]~${}^3P_1-{}^3P_2$}}
\def\CiPone{{[C~{\scriptsize I}]~${}^3P_1-{}^3P_0$}}
\def\CiPtwo{{[C~{\scriptsize I}]~${}^3P_2-{}^3P_1$}}
\def\cii{{[C~{\scriptsize II}]}}

\def\nii{{[N~{\scriptsize II}]}}
\def\niii{{[N~{\scriptsize II}]}}
\def\NiiPone{{[N~{\scriptsize II}]~${}^3P_1-{}^3P_0$}}

\def\nev{{[Ne~{\scriptsize V}]}}

\def\siii{{[S~{\scriptsize III}]}}
\def\siv{{[S~{\scriptsize IV}]}}

\def\FeII{{[Fe~{\scriptsize II}]}}

\def\c18o{{C$^{18}$O}}

\defcitealias{wada02}{WN02}
\defcitealias{wada05}{WT05}
\defcitealias{yamada07}{YWT07}
\defcitealias{wada09}{WPS09}

\newcommand{\pb}{P\'erez-Beaupuits}
\newcommand{\myemail}{jperezbe@eso.org}

\received{}
\revised{}
\accepted{}
\submitjournal{ApJ}

\shorttitle{The nuclear region of NGC253}
\shortauthors{\pb\ \etal}

\begin{document}


\title{A thorough view of the nuclear region of NGC~253 - \\
Combined Herschel, SOFIA and APEX dataset}


\author{J.P. \pb }
\affil{European Southern Observatory, Av. Alonso de Cordova, 3107, Santiago, Chile}
\affil{Max-Planck-Institut f\"ur Radioastronomie, Auf dem H\"ugel 69, 53121 Bonn, Germany}
\email{\myemail}
\author{R. G\"usten}
\affil{Max-Planck-Institut f\"ur Radioastronomie, Auf dem H\"ugel 69, 53121 Bonn, Germany}
\author{A. Harris}
\affil{Department of Astronomy, University of Maryland, College Park, MD 20742, USA}
\author{M. A. Requena-Torres}
\affil{Department of Astronomy, University of Maryland, College Park, MD 20742, USA}
\author{K.M. Menten}
\affil{Max-Planck-Institut f\"ur Radioastronomie, Auf dem H\"ugel 69, 53121 Bonn, Germany}
\author{A. Wei{\ss}}
\affil{Max-Planck-Institut f\"ur Radioastronomie, Auf dem H\"ugel 69, 53121 Bonn, Germany}
\author{E. Polehampton}
\affil{RAL Space, Rutherford Appleton Laboratory, Chilton, Didcot, Oxfordshire OX11 0QX, UK}
\affil{Institute for Space Imaging Science, University of Lethbridge, 4401 University Drive, Lethbridge, Alberta T1K 3M4, Canada}
\author{M.H.D. van der Wiel}
\affil{Institute for Space Imaging Science, University of Lethbridge, 4401 University Drive, Lethbridge, Alberta T1K 3M4, Canada}
\affil{ASTRON, the Netherlands Institute for Radio Astronomy, 7990 AA Dwingeloo, The Netherlands}






\begin{abstract}
{We present a large set of spectral lines detected in the $40''$ central region of the 
starburst galaxy NGC~253. Observations were obtained with the three instruments SPIRE, PACS and HIFI on 
board the Herschel Space Observatory, upGREAT on board of the SOFIA airborne observatory, 
and the ground based APEX telescope. Combining the spectral and photometry products of 
SPIRE and PACS we model the dust continuum Spectral Energy Distribution (SED) 
and the most complete \twco\ Line SED reported so far toward the nuclear region of NGC~253. 
Properties and excitation of the molecular gas were derived from a three-component non-LTE 
radiative transfer model, using the SPIRE \thco\ lines and ground based 
observations of the lower-$J$ \thco\ and \hcn\ lines, to constrain the model 
parameters.
Three dust temperatures were identified from the continuum emission, and three 
components are needed to fit the full CO LSED. Only the third CO component 
(fitting mostly the \hcn\ and PACS \twco\ lines) is consistent 
with a shock/mechanical heating scenario. A hot core chemistry is also argued as 
a plausible scenario to explain the high-$J$ \twco\ lines detected with PACS. The 
effect of enhanced cosmic ray ionization rates, however, cannot be ruled out, and is expected to play a significant role 
in the diffuse and dense gas chemistry. This is supported by the detection of ionic species 
like OH$^+$ and \hho$^+$, as well as the enhanced fluxes of the OH lines with respect to 
those of \hho\ lines detected in both PACS and SPIRE spectrum.}

\end{abstract}





\section{Introduction}

Several results have been published in the last few years, reporting observations with the SPIRE Fourier Transform Spectrometer (FTS) \citep{griffin10} on board the Herschel Space Observatory\footnote{Herschel is an ESA space observatory with science instruments provided by European-led Principal Investigator consortia and with important participation from NASA.}, towards various galaxies, e.g. M~82, 
NGC~1068, Mrk~231, Arp~220, NGC~6240, NGC~253 \citep{panuzzo10, kamenetzky12, spinoglio12, vdwerf10, rangwala11, meijerink13, rosenberg14}, and towards the Galactic Center 
\citep{goicoechea13}. The 
availability of a comprehensive number of transitions in the \twco\ ladder (up to the 
$J=13\to12$ transition) provided by the SPIRE-FTS, have made possible the analysis of the \twco\ line spectral 
energy distribution (LSED), using a variety of models of photon dominated regions (PDRs), hard X-ray dominated 
regions (XDRs), cosmic-ray dominated regions (CRDRs), and shocks. These models were used to estimate the source 
of excitation, and the ambient conditions of the molecular gas in all these galaxies, while constraining their 
parameters using not only the \twco\ spectral lines, but also the \thco\ lines \citep[e.g.][]{kamenetzky12}, as 
well as complementary ground-based observations of other molecules, like HCN \citep[e.g.][]{rosenberg14}. 

From these galaxies, so far only NGC~1068 has been studied combining the \twco\ (and other molecules) lines from SPIRE-FTS and PACS spectrometry data \citep{spinoglio12, hailey-dunsheath12}. The PACS \twco\ lines were shown to trace 
very different ambient conditions driven by X-rays in the nuclear region of NGC~1068 \citep{spinoglio12}.
Since the PACS data for NGC~253 are also available, we can now re-visit and extend the analysis and 
interpretation of the \twco\ LSED done by \citet{rosenberg14} based on SPIRE-FTS data only.

The Sculptor galaxy NGC~253 is a nearby (\textit{D}$\approx$3.5 Mpc; e.g., \citealt{mouhcine05, rekola05}), nearly edge-on (i$\sim$72$^o$--78$^o$; \citealt{pence81, puche91}), isolated spiral galaxy of type 
SAB(s)c, and is one of the closest galaxies outside the Local Group. 
Its angular size in the visible range is 27$'$.5$\times$6$'$.8. Together with M~82, it is the best 
example of a nuclear starburst \citep{rieke88}. Although, an active galactic nucleus (AGN) has also been 
suggested to coexist with the nuclear starburst \citep[e.g.,][]{weaver02, muller-sanchez10}, the corresponding 
low-luminosity AGN is not energetically dominant \citep{forbes00, weaver02}, 
and its IR/radio luminosity ratio indicates that the nature of its AGN candidate is more similar to the low accretion rate super-massive black hole Sgr~A$^∗$ at the center of the Milky Way galaxy that to an
an actual AGN driven by a more luminous central super-massive black hole \citep{fernandez-ontiveros09}.

The bulk of the NGC~253 starburst is confined to a $\sim$60~pc region centered southwest of the dynamical nucleus, 
according to the distribution of the 10--30 \mum\ continuum \citep{telesco93}. 
The estimated star formation rate in the nuclear region is $\sim$2--3~\Msun~yr$^{-1}$ \citep{radovich01, 
ott05}.
The 1-300~\mum\ luminosity of NGC~253 is, within $\sim$30$''$, 1.6$\times$10$^{10}$ \Lsun\ \citep{telesco80}. 
The total IR luminosity detected by IRAS is $L_{\rm IR}\approx 2 \times 10^{10}$ \Lsun\ \citep{rice88}. Four 
luminous super star clusters were discovered with the \textit{Hubble Space Telescope}, and a bolometric 
luminosity of $1.3\times10^9$~\Lsun\ was estimated for the brightest cluster \citep{watson96}. 
A bar can be seen in the Two Micron All Sky Survey (2MASS) image of NGC~253, and the kinematics show 
evidence for orbits in a bar potential \citep[][and references therein]{scoville85, das01}. 
Thus, the starburst activity is thought to be supported by the material brought to the nucleus by 
this bar \citep[e.g.,][and references therein]{engelbracht98, jarrett03}. 

Current estimates of the neutral gas mass in the central 20$''$-50$''$ range from 2.5$\times$10$^7$ \Msun\ 
\citep{harrison99} to 4.8$\times$10$^8$ \Msun\ \citep{houghton97}.
Studies of near-infrared and mid-infrared lines showed that the properties of the 
initial mass function (IMF) in the starburst region are similar to those of a 
Miller-Scalo IMF that has a deficiency in low-mass stars \citep{engelbracht98}.

A supernova rate of $\leq$0.3~year$^{-1}$ has been inferred for the entire galaxy from radio \citep{antonucci88, ulvestad97} and infrared \citep{vanBuren94} observations. 
The rate is 
most pronounced in the central starburst region, where a conservative estimate yields a rate of supernovae of 
$\sim$0.03~year$^{-1}$, which is comparable to that in our Galaxy \citep{engelbracht98}. This suggests a 
very high local cosmic-ray energy density. The mean density of the interstellar gas in the central starburst 
region is \textit{n}$\approx$600 protons~$\3cm$ \citep{sorai00}, which is about three orders of magnitude 
higher than the average density of the gas in the Milky Way. This extraordinary combination of high density gas 
and the enhanced local cosmic-ray energy density, was predicted to produce gamma rays at a detectable level 
\citep{paglione96, domingo-santamaria05, rephaeli10}. In fact, very high energy 
(VHE) ($>$100 GeV) gamma rays were effectively detected later in the nuclear region of NGC~253 with the High 
Energy Stereoscopic System (H.E.S.S.) array of imaging atmospheric Cherenkov telescopes \citep{acero09} and 
with the Large Area Telescope on board the \textit{Fermi Gamma-ray Space Telescope} \citep{abdo10}. The 
integral gamma-ray flux of the source above 220~GeV is $\sim$5.5$\times$10$^{-13}~\2cm$ s$^{-1}$, which 
corresponds to $\sim$0.3\% of the VHE gamma-ray flux observed in the Crab Nebula \citep{aharonian06}, and is 
consistent with the original prediction \citep{paglione96}. The detection of VHE gamma rays in NGC~253 
implies a high energy density of cosmic rays in this galaxy. Based on the observed gamma-ray flux, the cosmic
ray density was estimated to be 4.9$\times$10$^{-12}~\3cm$, with a corresponding energy density of cosmic rays
of $\sim$6.4 eV~$\3cm$ \citep{acero09}. Thus, the cosmic ray density in NGC~253 is about 1400 times the 
value at the center of the Milky Way \citep{aharonian06a}.

Observations of H$\alpha$ emission, and earlier Einstein and ROSAT X-ray data \citep[and references therein]{ptak97}, revealed a starburst-driven wind emanating from the nucleus along the minor axis of the 
galaxy. This wind was also detected later by Chandra \citep{strickland00}. 
Extraplanar outflowing molecular gas was also mapped in the \twco~$J=1\to0$ line with the high 
spatial resolution provided by the Atacama Large Millimeter/submillimeter Array (ALMA), and it was found to follow closely the 
H$\alpha$ filaments \citep{bolatto13}. The estimated molecular outflow rate is 3--9~\Msun~yr$^{-1}$, 
implying a ratio of mass-outflow rate to star-formation rate of about 1--3. This ratio is indicative of 
suppression of the star-formation activity in NGC~253 by the starburst-driven wind \citep{bolatto13}.



NGC~253 is also the brightest extragalactic source in the submm range, so its nucleus has been observed in 
various lines of CO and C \citep{gusten06a, bayet04, bradford03, israel02, sorai00, harrison99, israel95, gusten93, wall91, harris91}, and it was the selected target for the first unbiased molecular line survey of an extragalactic source \citep{martin06}. This survey showed that NGC~253 has a very rich molecular chemistry, with 
strong similarities to that of the Galactic central molecular zone, and even molecules like H$_3$O$^+$ have been 
detected in emission in the nuclear region of NGC~253 \citep{aalto11}.
High resolution SiO observations show bright emission resulting from large scale shocks, as well as gas 
entrained in the nuclear outflow \citep{garcia-burillo00}.  High resolution observations of \hcn\ and 
\hcop\ $J=1\to0$ show strongly centrally concentrated emission \citep{knudsen07}.

After the \cii~58~\mum\ and \oi~63~\mum\ lines, the \twco\ transitions are the most important cooling lines of 
the molecular gas in the interstellar medium (ISM). Therefore, several studies of the line spectral energy 
distribution (LSED) of \twco\ have been done to estimate the ambient conditions and excitation mechanisms of the 
molecular gas in the nuclear region of NGC~253 \citep[e.g.,][]{bradford03, bayet04}. Gas 
temperatures $T_{kin}\approx60$~K and \hh\ density of $\sim$10$^4~\3cm$  were found to be sufficient to explain 
the velocity-integrated intensities of the lower-$J$ (up to $J=7\to6$) \twco\ transitions, using a single 
component (temperature/density) Large Velocity Gradient (LVG) model \citep{gusten06a}. More recent 
observations using SPIRE on Herschel \citep{pilbratt10} provided a more extended \twco\ LSED, including 
transitions up to $J=13\to12$, which was reproduced using three gas components \citep{rosenberg14}. Three 
possible combinations of excitation mechanisms were explored, and all were found to be plausible explanations of 
the observed molecular emission. However, mechanical heating was found to be more dominant than cosmic ray 
heating in some of the models by \citet{rosenberg14}.

We present a combined set of updated and extended Herschel SPIRE-FTS spectrometer, PACS \citep{poglitsch10} and HIFI \citep{deGraauw10} observations of the nuclear region of NGC~253. 
These observations are part of the Herschel EXtra GALactic (HEXGAL, PI: R. G\"usten) Guaranteed Key Program. 
We also present data obtained with the Stratospheric Observatory For Infrared Astronomy (SOFIA, \citealt{Young12}), as well as from the ground based Atacama Pathfinder EXperiment 
(APEX\footnote{This publication is based on data acquired with the Atacama Path\-finder EXperiment. APEX
is a collaboration between the Max-\-Planck-Institut f\"ur Radioastronomie, the European Southern Observatory, 
and the Onsala Space Observatory.}; \citealt{gusten06}) telescope. Together, this set of data advances previous results on NGC~253 in the (sub-)mm, far- and mid-IR ranges, providing much more information about the excitation processes at play, and their effects on the molecular and atomic line emissions, than previously reported from ground based observations and SPIRE spectra alone. 


\begin{table*}[htp]
   \centering  
  \caption{NGC253 observations from HEXGAL KP GT}
   \label{tab:obsid} 
  \begin{tabular}{l r r l c} 
            \hline\hline

OBSID		& Duration(s)	& Instrument	 & Obs. mode 	&  SPG v \\

\hline

1342210652	&11311  	 	&PACS    	&Pacs Range Spectroscopy B2B mode	& 14.2.0 \\
1342212531	&5643		&PACS  	&Pacs Range Spectroscopy B2A mode & 14.2.0	\\
1342210671	&5578		&HIFI  	&Hifi	Mapping CO(9-8)	DBS	Raster	& 14.1.0 \\
1342210766	&510		&HIFI  	&Hifi CI(1-0) Point Mode Position Switching & 14.1.0 \\
1342210788	&5435		&HIFI  	&Hifi Mapping [CII] 		DBS Raster  & 14.1.0 \\
1342212140	&3833		&HIFI  	&Hifi Mapping CI(2-1) 	DBS	Raster  & 14.1.0 \\
1342210846	&11115  	 	&SPIRE 	&Spire Spectrum Point, intermediate imaging & 14.1.0  \\
1342210847	&13752  		&SPIRE 	&Spire Spectrum Point, sparse imaging  & 14.1.0  \\
            \noalign{\smallskip}	        
            \hline	

	 \end{tabular}
\end{table*}

The organization of this article is as follows. In Sec.~\ref{sec:obs} we 
describe the observations and the procedure followed to reduce and extract the spectral and photometry data. 
The results (maps, spectra and fluxes of detected and identified lines) are presented in Sec.~\ref{sec:results}. 
Analysis and modelling of the combined SPIRE and PACS continuum emission is discussed in Sect.~\ref{sec:continuum}.
In Sec.~\ref{sec:model} we propose a new non-LTE radiative transfer model for the \twco\ SLED, and present the analysis and discussions of the model results, and the interpretation of line ratios with other molecular lines.
A summary and final remarks are presented in Sec.~\ref{sec:remarks}.

\section{Observations and data reduction}\label{sec:obs}

A summary of all the observations that provided data used in this work, that are available in the Herschel Science Archive (HSA\footnote{\url{http://archives.esac.esa.int/hsa/whsa/}}), is shown in Table~\ref{tab:obsid}. All the Herschel data presented here were obtained as part of the key program guarantee time HEXGAL: KPGT\_rguesten\_1 (P.I. Rolf G\"usten). 
Herschel data was processed and reduced using the Herschel Interactive Processing Environment (HIPE\footnote{HIPE is a joint development by the Herschel Science Ground Segment Consortium, consisting of ESA, the NASA Herschel Science Center, and the HIFI, PACS and SPIRE consortia. \url{https://www.cosmos.esa.int/web/herschel/hipe-download}}) v.14 and v.15 \citep{ott10}. The difference in calibration obtained for the SPIRE line fluxes is 
less than 1\% between these two versions, but it can be as large as 15\% to 20\% with respect to older versions (e.g., HIPE v.11).

\subsection{Correcting the SPIRE intensity levels}\label{sec:spire-corrected}

The SPIRE spectral cubes calibrated with the point-like source pipeline were used because they lead to lower noise levels and the match between the two spectral bands is better. This is consistent with the large SPIRE beam compared with the estimated size of the emitting region, as discussed below.

Most works found in the literature present SPIRE spectra corrected using a 
photometric flux at one or more wavelengths, integrated in an aperture equivalent to the largest beam size (at the lowest frequency) of the SPIRE FTS.
Instead we used the {\it semiExtendedCorrector} task (available in HIPE since v.11), which stitch together the long- and short-wavelength FTS bands (SLW and SSW, respectively). 
This tool corrects the SPIRE spectra by simulating a source size convolved to a $\sim$40$''$ beam.
Details of this task and a description of the method used were reported by \citet{wu13}, and our application of it is described in Appendix~\ref{sec:appendix-SPIRE-reduction}. 
There are two main reasons to prefer this method: first, we can obtain an estimate for the size of the emitting region, and second, we identify frequency ranges in the SPIRE spectra that may still be affected by some calibration inaccuracies \citep{swinyard14}. 

\begin{figure*}[htp]
\centering
\hspace{-0.4cm}\includegraphics[angle=0,width=0.51\textwidth]{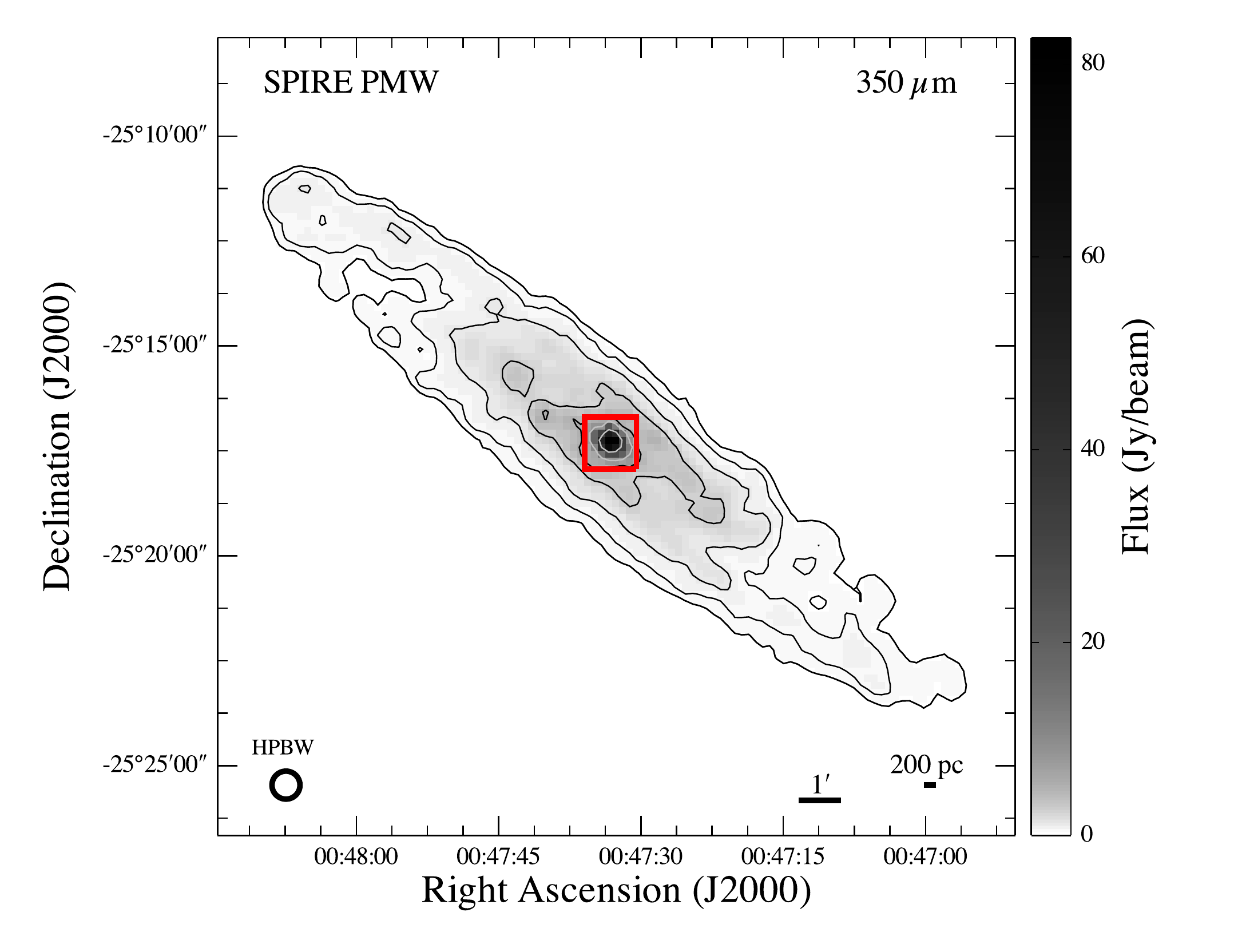}%
\hspace{+0.0cm}\includegraphics[angle=0,width=0.51\textwidth]{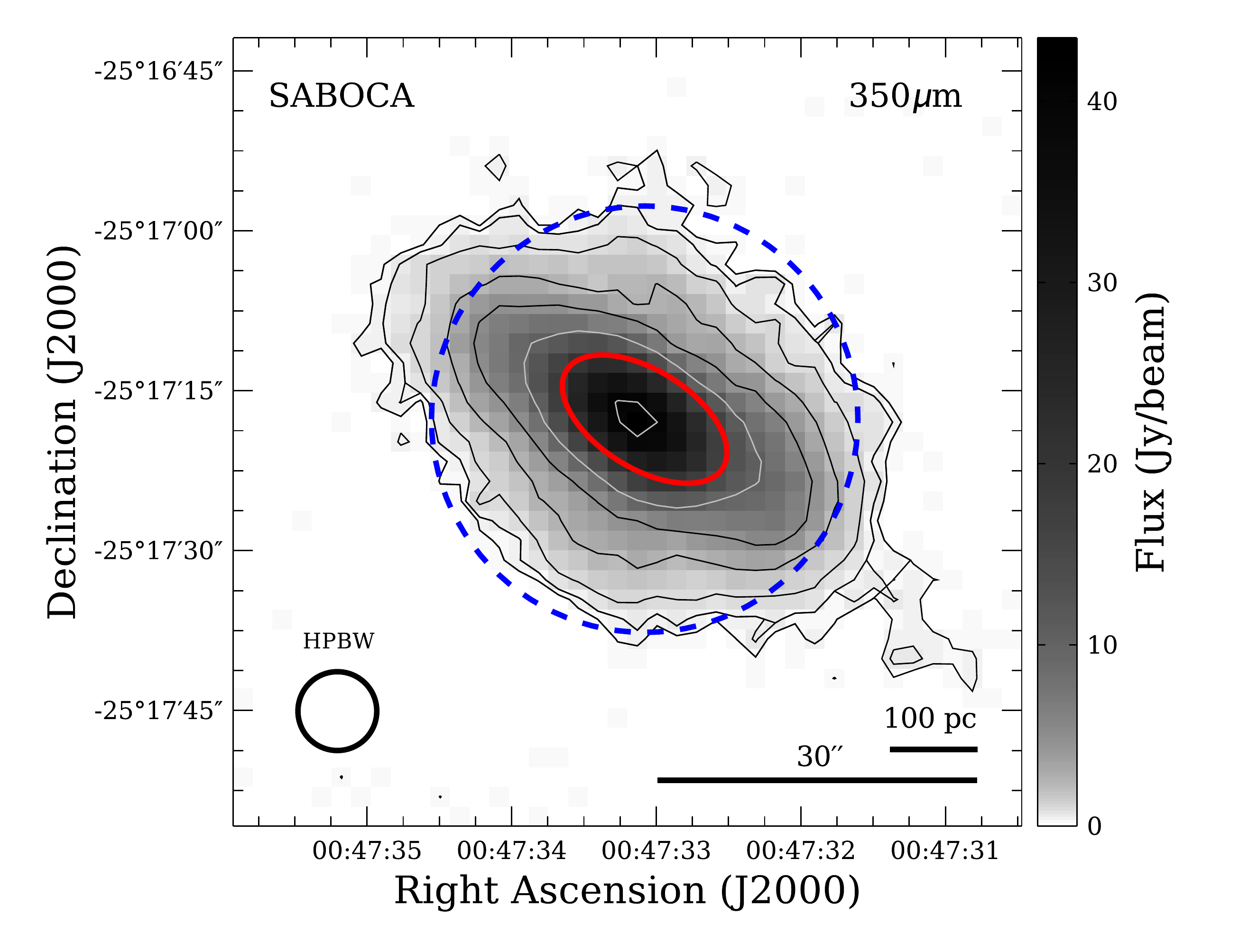}%

\vspace{-0.4cm}

\caption{{\footnotesize Dust continuum emission of NGC253 observed with the SPIRE photometric medium wavelength 
(PMW) ({\it left}) and with SABOCA on APEX ({\it right}) at 350~\mum, at their respective original resolution 
(HPBW indicated). The field of the SABOCA image is shown with a square on the SPIRE map. The contours are 0.30, 0.50, 1.0, 2.5, 5.0 (black) and 10, 40 Jy/beam (grey). A 40$''$ aperture 
is depicted with a (dashed) circle and the (thick) ellipse represents the source size (FWHM) of 
$17.3''\times9.2''$ obtained from a two-dimensional Gaussian fit.}}
\label{fig:continuum-maps}
\end{figure*}

To cross-check this method, choose a source size and, hence, the final 
spectra to use, we compare the continuum level of the corrected spectra with actual dust continuum emission at given wavelengths, as observed with an equivalent beam size (or aperture).
We first extracted the fluxes of the SPIRE photometric maps of NGC~253 (obs. ID 1342199387), 
obtained from the HSA, 
as well as the integrated flux at 350~\mum\ obtained with the Submillimeter APEX Bolometer Camera (SABOCA) on the APEX telescope \citep{siringo10}. The SPIRE photometric maps were re-processed with the pipeline for large photometer maps provided in HIPE v.15. Details of the re-processing can be found in Apendix~\ref{sec:appendix-SPIRE-reduction}.
Fig.~\ref{fig:continuum-maps} shows the SPIRE ({\it left}) and SABOCA ({\it right}) 350~\mum\ flux density maps. 
The SABOCA bolometer, with higher spatial resolution (HPBW$\sim$8$''$), was used to map only the nuclear region of the galaxy. 

The 40$\arcsec$ annular sky aperture integrated fluxes are 
278.7$\pm$19.0 Jy, 93.6$\pm$10.8 Jy, and 20.1$\pm$5.1 Jy
for the 250~\mum, 350~\mum, and 500~\mum\ images, respectively. 
The integrated APEX/SABOCA flux for an aperture of 40$''$ is 131.4$\pm$11.5 Jy, i.e., $\sim$30\% higher than the corresponding SPIRE flux. Note, however, that using the \textit{DAOphot} algorithm with automatic aperture correction the SPIRE flux at 350~\mum\ would be 117.0$\pm$13.5 Jy instead; more consistent, to within their (1~$\sigma$) uncertainties, with the SABOCA flux.
The difference in absolute values may be due to the different calibration schemes, the fact that an atmospheric contribution affects the SABOCA map, or that the SPIRE photometric map at 350~\mum\ is also affected by small calibration uncertainties.

\begin{figure*}[htp]
\centering
\hspace{-0.0cm}\includegraphics[angle=0,width=0.47\textwidth]{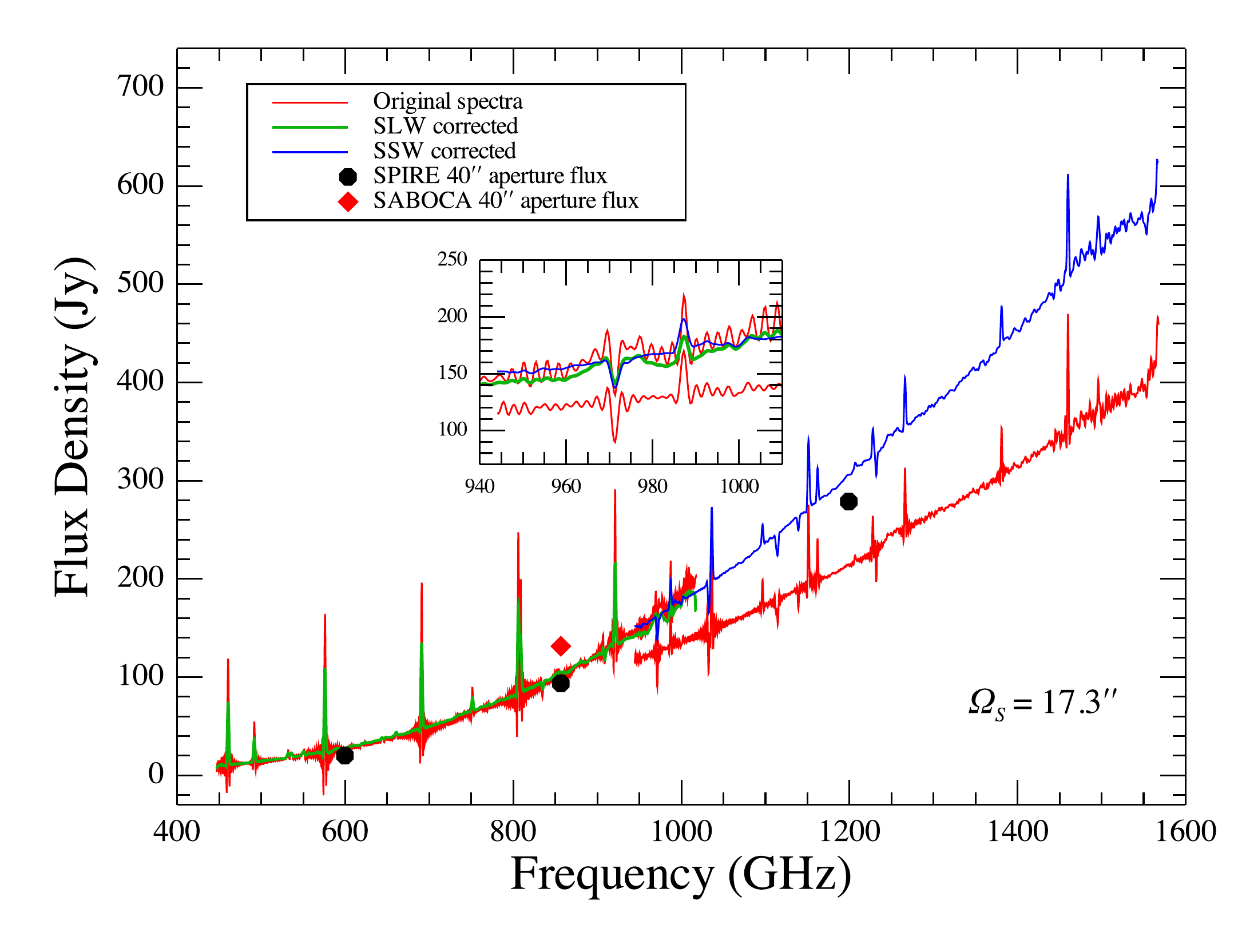}%
\hspace{-0.0cm}\includegraphics[angle=0,width=0.47\textwidth]{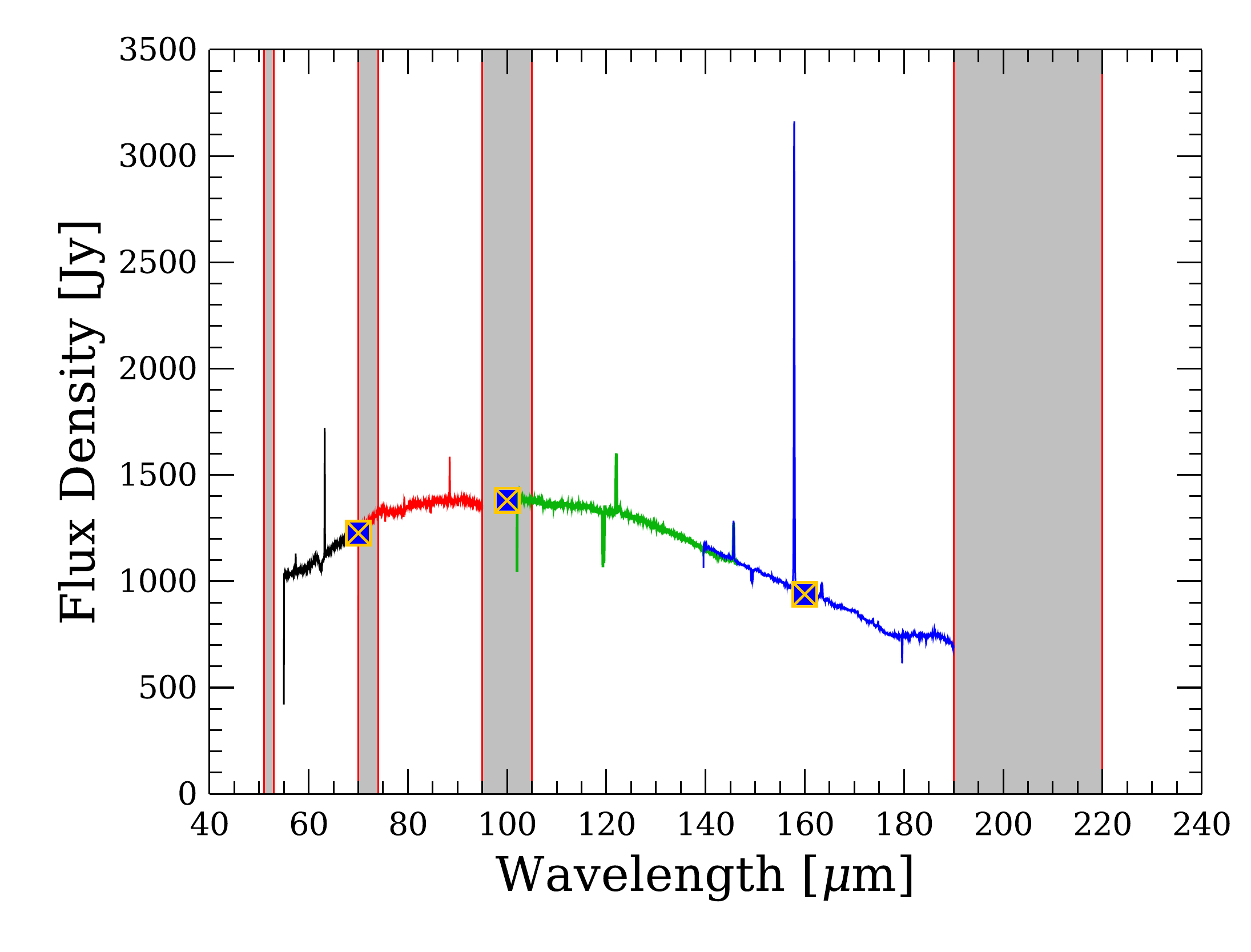}%

\vspace{-0.4cm}

\caption{{\footnotesize \textit{Left} - SPIRE FTS spectra of the nuclear region of NGC253, corrected to a $\sim$40$''$ beam assuming a source size ($\Theta_S$) with a semi-major axis of FWHM=$17.3''$ and eccentricity 0.85, as obtained from a 2-D Gaussian fit of the APEX/SABOCA continuum emission at 350~\mum. The green and blue lines are the corrected {\it apodized} data from the long- and short-wavelength FTS bands, respectively, while the original {\it unapodized} spectra is shown in red. The dots show the $40''$ aperture integrated fluxes from the SPIRE photometric maps (black) and APEX/SABOCA (red). The inset shows a zoom into the overlap region between the SLW and SSW bands.
 \textit{Right} - PACS spectra of the nuclear region of NGC253 extracted from the 5$\times$5 spaxels and corrected for point source losses. 
Using the telescope background normalization, and excluding the spectral leakage regions (vertical filled areas), results in a good match with the PACS 40$''$ aperture photometry fluxes at 70~\mum, 100~\mum, and 160~\mum\ (squares-plus-cross), and between the four wavelength ranges.
}}
\label{fig:spire-pacs-corrected-spectrum}
\end{figure*}

Fitting a 2-D Gaussian distribution to the SABOCA map yields a 
beam-deconvolved source size (FWHM) of $17.3'' \times 9.2''$ 
(about $210 \times 112$ pc at an assumed distance of 2.5 Mpc, \citealt{houghton97}), 
which corresponds to an elliptical shape (shown in Fig.~\ref{fig:continuum-maps}, 
{\it right}) with eccentricity 0.85. 
This source size is about 24\% smaller than the size ($30''\times16''$) found 
by \citet{weiss08} from the APEX/LABOCA 870~\mum\ map, with a larger beam size ($19''.2$). 
However, the eccentricity of the later case is the same (0.85), 
which indicates that the 2-D Gaussian intensity distribution of the continuum 
emission is consistent at this two wavelengths with the two different beam sizes. 
The SPIRE FTS spectra corrected assuming the source size estimated from the SABOCA map is shown in Fig.~\ref{fig:spire-pacs-corrected-spectrum}.

\subsection{Extracting the PACS spectra}




Data were obtained between on 2010 December 01 and 2011 January 11. The observations were made with a small chopping angle (1.5 arcmin). The calibrated PACS Level-2 data products (processed with latest SPG v14.2.0) were retrieved from the HSA.

PACS includes an integral field unit spectrograph observing in the $\sim$50--200~\mum\ range, with a spectral resolving power in the range of R = 1000--4000 ($\Delta v$ = 75--300~\kms), depending on wavelength. PACS comprises 5$\times$5 squared spaxel elements with a native individual size of 9\farcs4$\times$9\farcs4 each, and an instantaneous field of view (FoV) of 47\arcsec$\times$47\arcsec.
A correction for extended sources was introduced in the standard pipeline from HIPE v.13. Details of the corrections can be found in the PACS calibration history and the corresponding Wiki\footnote{\url{http://herschel.esac.esa.int/twiki/bin/view/Public/PacsCalTreeHistory}}. The corrections affects the continuum level by about 30\% in the blue band and about 5\% in the red band (Elena Puga, PACS Calibration Team,  \textit{private communication}). 
Before any extraction of the spectra it is recommended to undo the extended source correction factor applied in the PACS Level-2 products. This can easily be done in HIPE v.14 and v.15 by using the task $undoExtendedSourceCorrection$ included in the \textsf{herschel.pacs.spg.spec} module.

For consistency with the 40$''$ \textit{beam corrected} SPIRE spectra (Sect.~\ref{sec:spire-corrected}), 
the PACS spectral ranges were obtained as the total cumulative spectra from the 
5$\times$5 spaxels, corrected by the 3$\times$3 \textit{point-source losses} included in the  SPG 
v14.2.0 calibration tree. Note that a 5$\times$5 correction for \textit{point-source losses} leads to an over-estimate of the spectral continuum level because the nuclear region of NGC~253 is a semi-extended source in the PACS FoV, and the bulk of the emission is contained in the inner 3$\times$3 spaxels. This is in contrast to the work presented by \citealt{fernandez-ontiveros16} (and earlier works using the SPG data archive without any re-processing with HIPE) in which the correction for point-source losses in the 5$\times$5 extracted spectra were not included in the standard pipelines available before HIPE v.14. 

We compared the continuum level of the PACS spectra with the corresponding 40$''$ aperture fluxes of the PACS photometry maps (from the HSA, obs. IDs 1342221744 and 1342221745) at 70~\mum, 100~\mum, and 160~\mum. The PACS 40$''$ aperture fluxes are summarized in Table~\ref{tab:sed-results}. The PACS flux uncertainties include the errors estimated with an annular sky aperture and the 7\% absolute point-source flux calibration for scan maps \citep{balog14}. The final 5$\times$5 corrected, and background normalized, PACS spectra used in this work are shown in Fig.~\ref{fig:spire-pacs-corrected-spectrum}. The sections of the spectrum affected by spectral leakage are shown by gray filled bands, and they were not used in our analysis.

\subsection{The HIFI spectra}

We also have several single pointing HIFI observations of targeted lines and a few small maps of some of the key lines detected in the SPIRE and PACS spectra. We present here only the \twco, \CiPtwo, and \cii\ maps centered at coordinate R.A.(J2000) = $00^h 47^m 33.12^s$ and Dec(J2000) = $-25^{\circ} 17\arcmin 17\farcs6$ (Table~\ref{tab:obsid}). The single pointing spectra represent averages between the horizontal and vertical polarizations, while we combined both polarizations as independent pointings (due to the slight misalignment between their beams) when creating the maps using the HIPE task \textit{doGridding} on the Level 2 products.

\subsection{The SOFIA \ GREAT \& upGREAT spectra}

During Cycle 3 flight campaign of SOFIA we made single-pointed observations of the \cii~158~\mum\ fine-structure line at 1900.54~GHz as well as the high-$J$ CO $J=11\to10$ transition at 1267.01~GHz toward the nuclear region of NGC253. The observations were performed using the German Receiver for Astronomy at Terahertz Frequencies (GREAT\footnote{GREAT is a development by the MPI f\"ur Radioastronomie and the KOSMA / Universit\"at zu K\"oln, in cooperation with the MPI f\"ur Sonnensystemforschung and the DLR Institut f\"ur Planetenforschung} single-pixel, \citealt{heyminck12}, \& upGREAT seven-pixels, \citealt{Risacher16}). The front-end configuration corresponded to the low frequency array (LFA-V) and low frequency channel (L1) of upGREAT and GREAT, respectively. The fourth generation fast Fourier spectrometer (4GFFT, \citealt{Klein12}) provided 4 GHz bandwidth with 16384 channels (i.e., about 244.1 kHz of spectral resolution).
For the opacity corrections across L1 and the LFA-V, the precipitable water vapor column was obtained from a free fit to the atmospheric total power emission. The dry constituents were fixed to the standard model values. All receiver and system temperatures are on the single-sideband scale. Calibrated data products were obtained from the {\textit{KOSMA atmospheric calibration} software for SOFIA/GREAT \citep{guan12}} version January 2016. The spectral temperatures are first expressed as forward-beam Rayleigh-Jeans temperatures $T_{\rm A}^*$ using a forward efficiency ($\eta_{\rm f}=0.97$). They were later converted to $T_{\rm mb}$ scale by using the main beam coupling efficiencies (as measured toward Mars) $\eta_{\rm mb}$ (L1) = 0.69 and  $\eta_{\rm mb}$ (LFA-V) = (0.70, 0.73, 0.71, 0.69, 0.63, 0.65, 0.71) for each pixel.  The estimated main beam sizes\footnote{\url{http://www3.mpifr-bonn.mpg.de/div/submmtech/heterodyne/great/GREAT_calibration.html}} are 22\farcs7 for L1 at 1267.01~GHz and 15\farcs1 for the LFA at 1900.54~GHz.

The observations were done under the U.S.\ proposal 03\_0039 (PI: Andrew Harris). The project was meant to observe the entire nuclear region of NGC253 in six pointing positions. However, due to reduced time during the flights scheduled for these observations, there was time to perform only the south-west (SW) position along the bar.  This position likely contains the densest and most excited gas within the nucleus.  
Because the nuclear \cii\ line is broad, the SW position was observed with two tuning set-ups shifted by 100~\kms\ w.r.t.\ the systemic velocity of 250~\kms\ (P150 and P350, for +150~\kms\ and +350~\kms, respectively) in order to have sufficient baseline coverage. For each pixel, to combine the two tunings, a baseline offset was determined from the mean value of a line-free velocity interval and applied to the P350 spectrum. The spectra were then stitched together (with overlapping regions averaged together), and a zero order baseline was removed. Most of the data analysis and process of the SOFIA/upGREAT observations was done using the GILDAS\footnote{\url{http://www.iram.fr/IRAMFR/GILDAS}} 
package CLASS90 \citep{pety05}.

\begin{figure*}[htp]
\centering
\hspace{-0.6cm}\includegraphics[angle=0,width=0.90\textwidth]{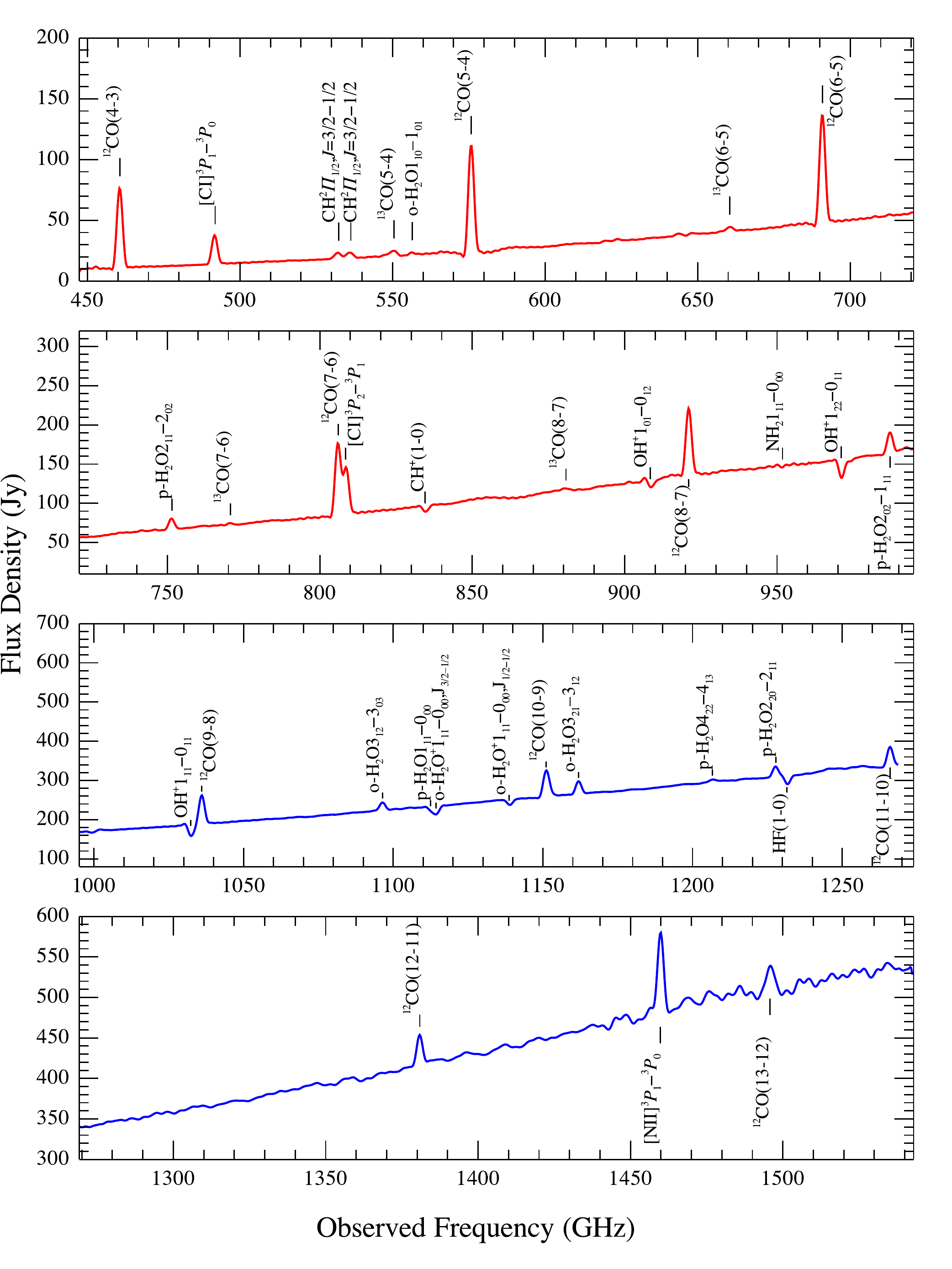}%

\caption{{\footnotesize Combined SLW and SSW bands of the SPIRE FTS spectrum of NGC253, 
corrected for a $40"$ beam size. Emission and absorption lines are indicated for detections 
with S/N$>3\sigma$.}}

\label{fig:full-spire-lines}
\end{figure*}

%
   \begin{table}[htp]
            \centering
      \caption{\footnotesize{Fluxes from the SPIRE FTS emission lines extracted from the 40\arcsec\ aperture observations of NGC~253.}}
         \label{tab:spire-emission-fluxes}
         \tabcolsep 1.5pt
         \scriptsize
         \begin{tabular}{lccc}
	    \hline\hline
	    \noalign{\smallskip}
            Line   &    $\nu_{\rm rest}^{~\mathrm{a}}$   &        Flux$^{\mathrm{b}}$       &  Luminosity$^{~\mathrm{c}}$  \\
                   &              (GHz)              &  ($10^{-16}$~\Wm) &      ($10^{4}$~\Lsun)        \\
	    \noalign{\smallskip}
	    \hline
	    \noalign{\smallskip}  

            \twco\ $J = 4\rightarrow3$  &   461.041  &  13.48$\pm$1.37  &  51.2$\pm$6.5  \\
            \twco\ $J = 5\rightarrow4$  &   576.268  &  18.44$\pm$1.86  &  70.0$\pm$8.8  \\
            \twco\ $J = 6\rightarrow5$  &   691.473  &  18.72$\pm$1.89  &  71.0$\pm$9.0  \\
            \twco\ $J = 7\rightarrow6$  &   806.652  &  19.19$\pm$1.94  &  72.8$\pm$9.2  \\
            \twco\ $J = 8\rightarrow7$  &   921.800  &  17.67$\pm$1.79  &  67.1$\pm$8.5  \\            
            \twco\ $J = 9\rightarrow8$  &  1036.912  &  16.29$\pm$1.72  &  61.8$\pm$8.0  \\
            \twco\ $J = 10\rightarrow9$  &  1151.985  &  13.41$\pm$1.45  &  50.9$\pm$6.7  \\
            \twco\ $J = 11\rightarrow10$  &  1267.014  &  10.72$\pm$1.20  &  40.7$\pm$5.5  \\
            \twco\ $J = 12\rightarrow11$  &  1381.995  &   7.79$\pm$0.95  &  29.6$\pm$4.2  \\
            \twco\ $J = 13\rightarrow12$  &  1496.923  &   5.69$\pm$0.79  &  21.6$\pm$3.4  \\

            \vspace{-0.25cm}\\

            \thco\ $J = 5\rightarrow4$  &   550.926  &   0.92$\pm$0.25  &   3.5$\pm$1.0  \\
            \thco\ $J = 6\rightarrow5$  &   661.067  &   0.74$\pm$0.24  &   2.8$\pm$0.9  \\
            \thco\ $J = 7\rightarrow6$  &   771.184  &   0.66$\pm$0.24  &   2.5$\pm$0.9  \\
            \thco\ $J = 8\rightarrow7$  &   881.273  &   0.58$\pm$0.24  &   2.2$\pm$0.9  \\

            \vspace{-0.25cm}\\

            \c18o\ $J = 5\rightarrow4$  &   548.831  &   0.34$\pm$0.22  &   1.3$\pm$0.8  \\
            
            \vspace{-0.25cm}\\

            \CiPone\  &   492.161  &   4.93$\pm$0.56  &  18.7$\pm$2.5  \\
            \CiPtwo\  &   809.342  &  12.25$\pm$1.25  &  46.5$\pm$5.9  \\

            \vspace{-0.25cm}\\

            \NiiPone\  &  1460.977  &  20.63$\pm$2.13  &  78.3$\pm$10.0  \\

            \vspace{-0.25cm}\\

            \oh2o\ $1_{10}-1_{01}$  &   556.936  &   0.37$\pm$0.22  &   1.4$\pm$0.8  \\
            \ph2o\ $2_{11}-2_{02}$  &   752.033  &   2.97$\pm$0.37  &  11.3$\pm$1.6  \\            
            \ph2o\ $2_{02}-1_{11}$  &   987.927  &   6.17$\pm$0.76  &  23.4$\pm$3.4  \\
            \oh2o\ $3_{12}-3_{03}$  &  1097.365  &   4.29$\pm$0.62  &  16.3$\pm$2.7  \\
            
            \oh2o\ $3_{12}-2_{21}$  &  1153.127  &   3.03$\pm$0.54  &  11.5$\pm$2.2  \\
            \oh2o\ $3_{21}-3_{12}$  &  1162.912  &   7.44$\pm$0.87  &  28.2$\pm$3.9  \\
            \ph2o\ $4_{22}-4_{13}$  &  1207.639  &   1.55$\pm$0.47  &   5.9$\pm$1.8  \\
            \ph2o\ $2_{20}-2_{11}$  &  1228.789  &   6.45$\pm$0.78  &  24.5$\pm$3.5  \\

            \vspace{-0.25cm}\\

            CH ${}^2\Pi_{1/2}$ $J=3/2-1/2^{~\mathrm{d}}$  &   532.730  &   0.99$\pm$0.25  &   3.8$\pm$1.0 \\
            CH ${}^2\Pi_{1/2}$ $J=3/2-1/2^{~\mathrm{d}}$  &   536.760  &   0.95$\pm$0.25  &   3.6$\pm$1.0  \\
            	   
            \vspace{-0.25cm}\\
	    
            OH$^+$ $1_{01}-0_{12}${$^{\mathrm{e}}$}   &   907.500  &   1.49$\pm$0.45  &   5.7$\pm$1.8  \\

	    \noalign{\smallskip}
	    \hline
	  \end{tabular}

\begin{list}{}{}
\item[$^{\mathrm{a}}$] Obtained from the LAMDA, CDMS, and NASA/JPL databases.
\item[$^{\mathrm{b}}$] The flux errors include the statistical uncertainty of the instrument, 6\% of the calibration uncertainty \citep{swinyard14}, and 8\% of the uncertainty in the source size used to correct the continuum levels (Sect.~\ref{sec:spire-corrected}).
\item[$^{\mathrm{c}}$] Luminosity estimated assuming a flat space cosmology ($H_0$=70~\kms\ Mpc$^{-1}$, $\Omega_{\Lambda}$=0.73, $\Omega_M$=0.27) and a distance of $3.5\pm0.2$~Mpc for NGC~253 \citep{rekola05}. The luminosity errors include the relative uncertainty of the respective fluxes and the distance of the galaxy, as well as a 5\% uncertainty for the assumed cosmology model.
\item[$^{\mathrm{d}}$] These lines are blended with the \hcn\ and \hcop\ $J = 6\rightarrow5$ lines at 531.716 GHz and 535.062 GHz, respectively, as detected in the HIFI spectra by \citet{rangwala14}.
\item[$^{\mathrm{e}}$] Emission part of the OH$^+$ P-Cygni feature. The flux in absorption centered at 909~GHz is shown in Table~\ref{tab:spire-absorption-fluxes}.

\end{list}

\end{table}
%

   \begin{table}[htp]
            \centering
      \caption{\footnotesize{Fluxes from the SPIRE FTS absorption lines extracted from the 40\arcsec\ aperture observations of NGC~253.}}
         \label{tab:spire-absorption-fluxes}
         \tabcolsep 1.5pt
         \scriptsize
         \begin{tabular}{lccc}
	    \hline\hline
	    \noalign{\smallskip}
            Line   &    $\nu_{\rm rest}^{~\mathrm{a}}$   &        Flux       &  Luminosity$^{~\mathrm{b}}$  \\
                   &              (GHz)              &  ($10^{-16}$~\Wm) &      ($10^{4}$~\Lsun)        \\
	    \noalign{\smallskip}
	    \hline
	    \noalign{\smallskip}




            CH$^+$ $J = 1\rightarrow0$  &   835.138  &  -1.71$\pm$0.25  &  -6.5$\pm$1.1  \\

	    \noalign{\smallskip}

            o-NH$_2$ $1_{11}-0_{00}$  &   952.578  &  -1.60$\pm$0.24  &  -6.1$\pm$1.0  \\
	    
	    \noalign{\smallskip}

            OH$^+$ $1_{01}-0_{12}$  &   909.159  &  -2.48$\pm$0.49  &  -9.4$\pm$2.0  \\
            OH$^+$ $1_{22}-0_{11}$  &   971.805  &  -5.63$\pm$0.69  &  -21.4$\pm$3.1  \\
            OH$^+$ $1_{12}-0_{12}$ $^{~\mathrm{c}}$  &  1033.119  &  -6.53$\pm$0.76  &  -24.8$\pm$3.4  \\

            \vspace{-0.25cm}\\


           \ph2o\ $1_{11}-0_{00}$  &  1113.343  &  -2.86$\pm$0.54  &  -10.8$\pm$2.2  \\

            \vspace{-0.25cm}\\
            

            \oh2o$^+$ $1_{11}-0_{00}$ $J_{\tfrac{3}{2}-\tfrac{1}{2}}$  &  1115.204  &  -4.57$\pm$0.65  &  -17.4$\pm$2.8  \\
            \oh2o$^+$ $1_{11}-0_{00}$ $J_{\tfrac{1}{2}-\tfrac{1}{2}}$  &  1139.654  &  -3.16$\pm$0.55  &  -12.0$\pm$2.3  \\
            
            \vspace{-0.25cm}\\


            HF $J = 1\rightarrow0$  &  1232.476  &  -4.99$\pm$0.63  &  -18.9$\pm$2.8  \\
	    
            
	    \noalign{\smallskip}
	    \hline
	  \end{tabular}
\begin{list}{}{}
\item[$^{\mathrm{a}}$] Obtained from LAMDA, CDMS and NASA/JPL databases.
\item[$^{\mathrm{b}}$] Luminosity estimated assuming a Flat Space Cosmology (H$_0$=70~\kms\ Mpc$^{-1}$, $\Omega_{\lambda}$=0.73, $\Omega_M$=0.27)
 and a distance of $3.5\pm0.2$~Mpc for NGC~253 \citep{rekola05}. The luminosity errors include the relative uncertainty of the respective fluxes and the distance of the galaxy, as well as a 5\% uncertainty for the assumed Cosmology model.
\item[$^{\mathrm{c}}$] This line is likely blended with the OH$^+$ $1_{11}-0_{11}$ line at 1032.998 GHz. Both lines have comparable Einstein-A coefficients (1.41$\times$10$^{-2}$ s$^{-1}$).

\end{list}

\end{table}
%

\begin{figure*}[!tp]

 \begin{tabular}{cccc}
  \hspace{-0.50cm}\epsfig{file=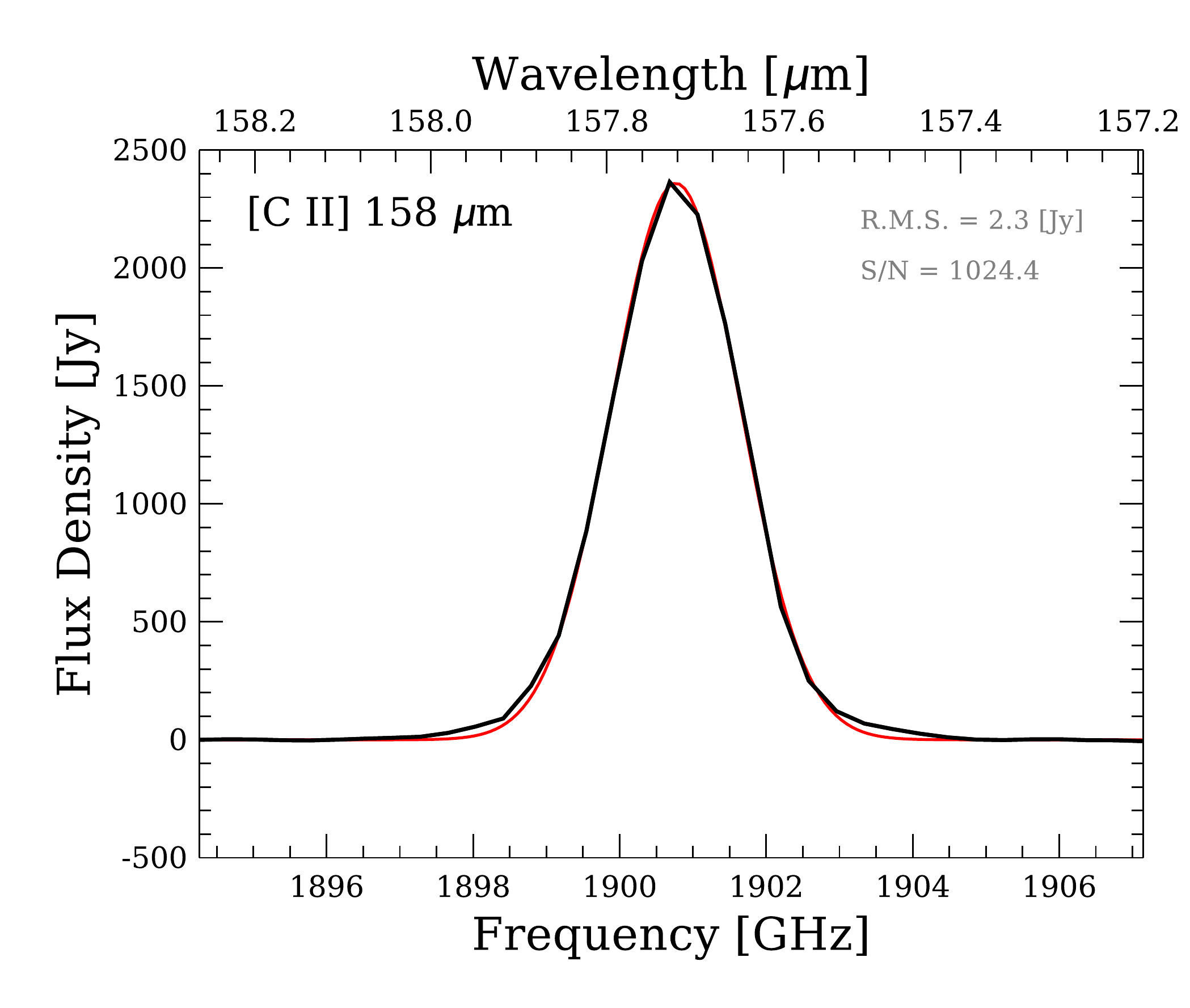,angle=0,width=0.27\linewidth} &
  \hspace{-0.50cm}\epsfig{file=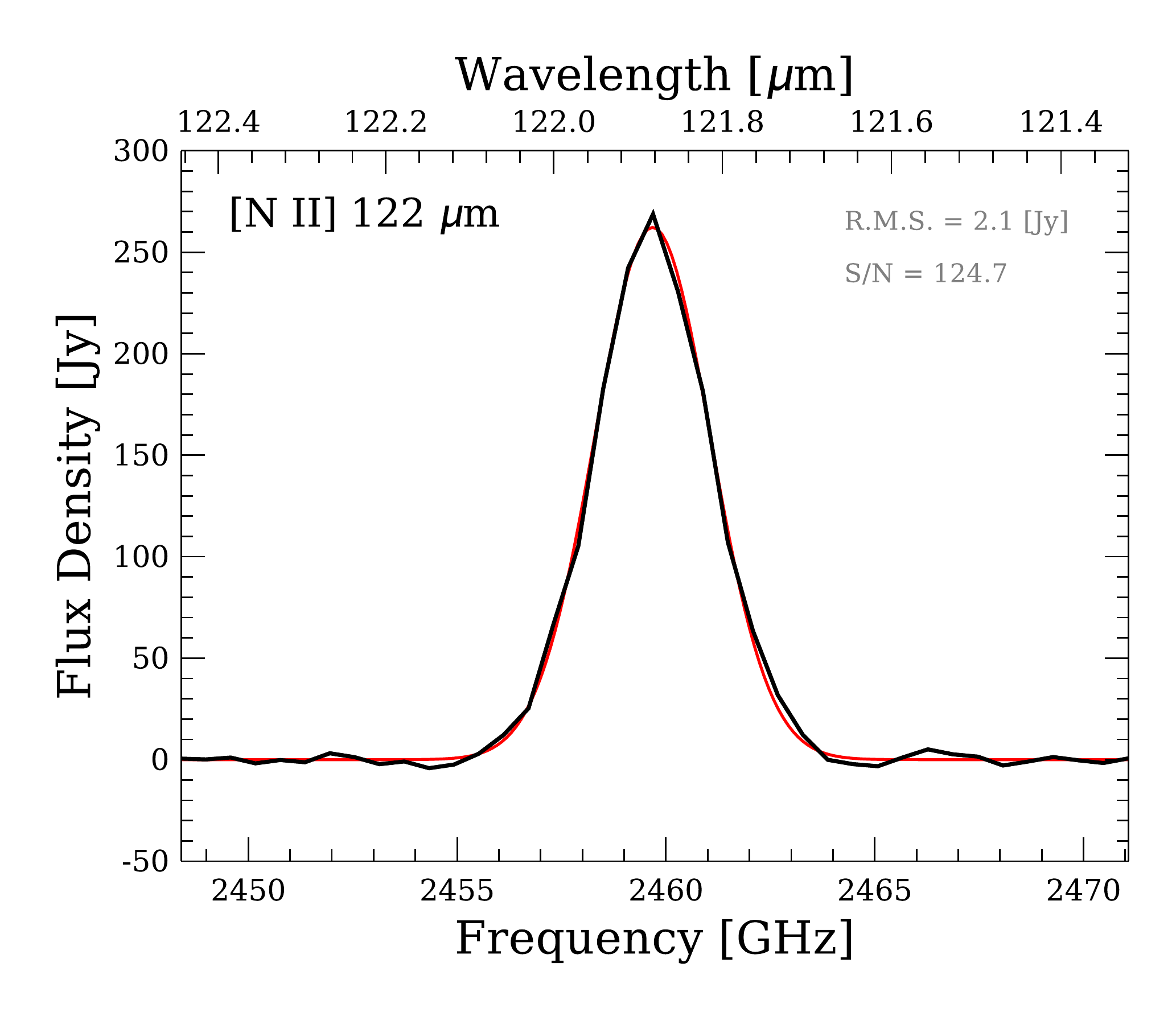,angle=0,width=0.27\linewidth} &
  \hspace{-0.50cm}\epsfig{file=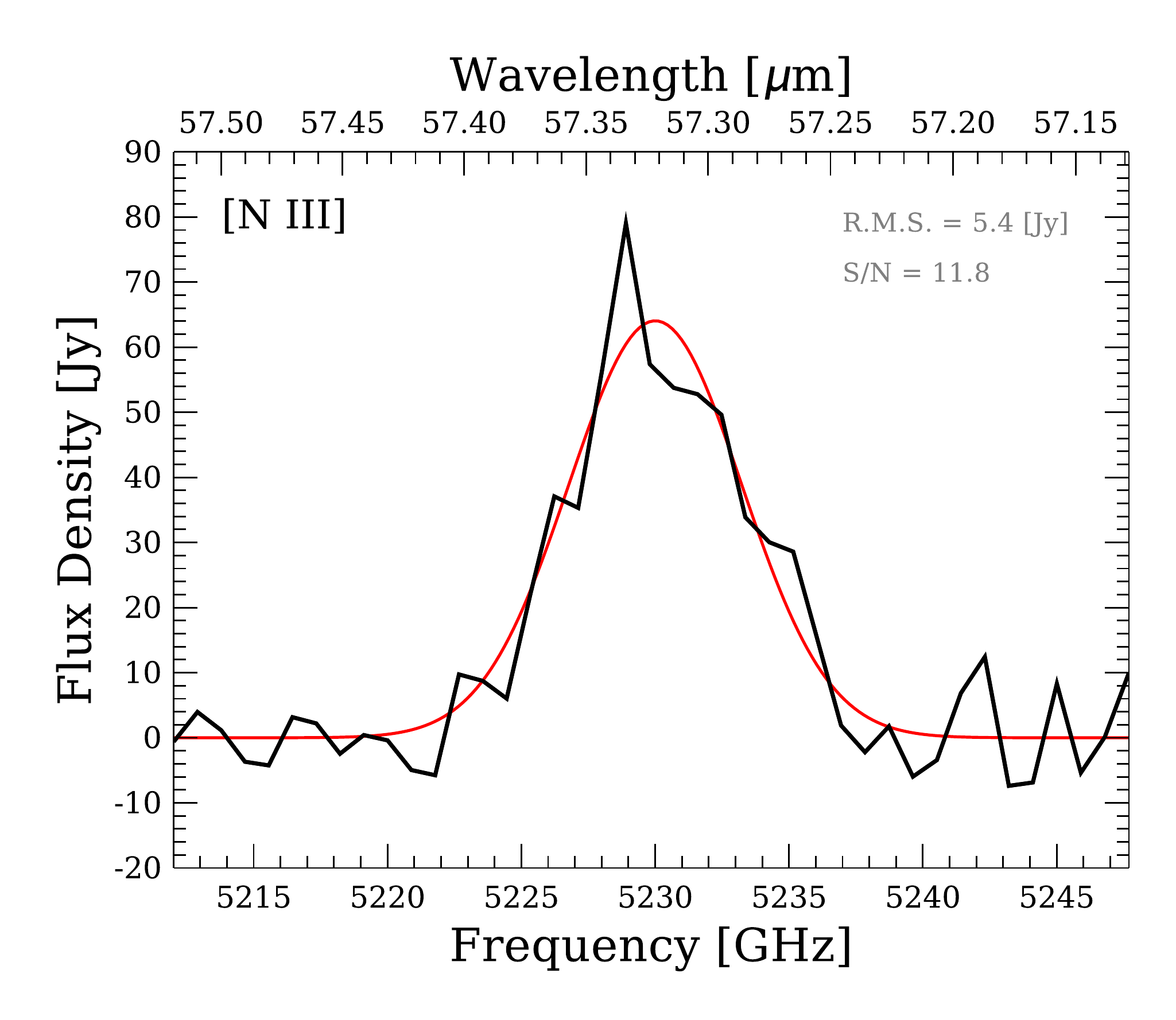,angle=0,width=0.27\linewidth}&
  \hspace{-0.50cm}\epsfig{file=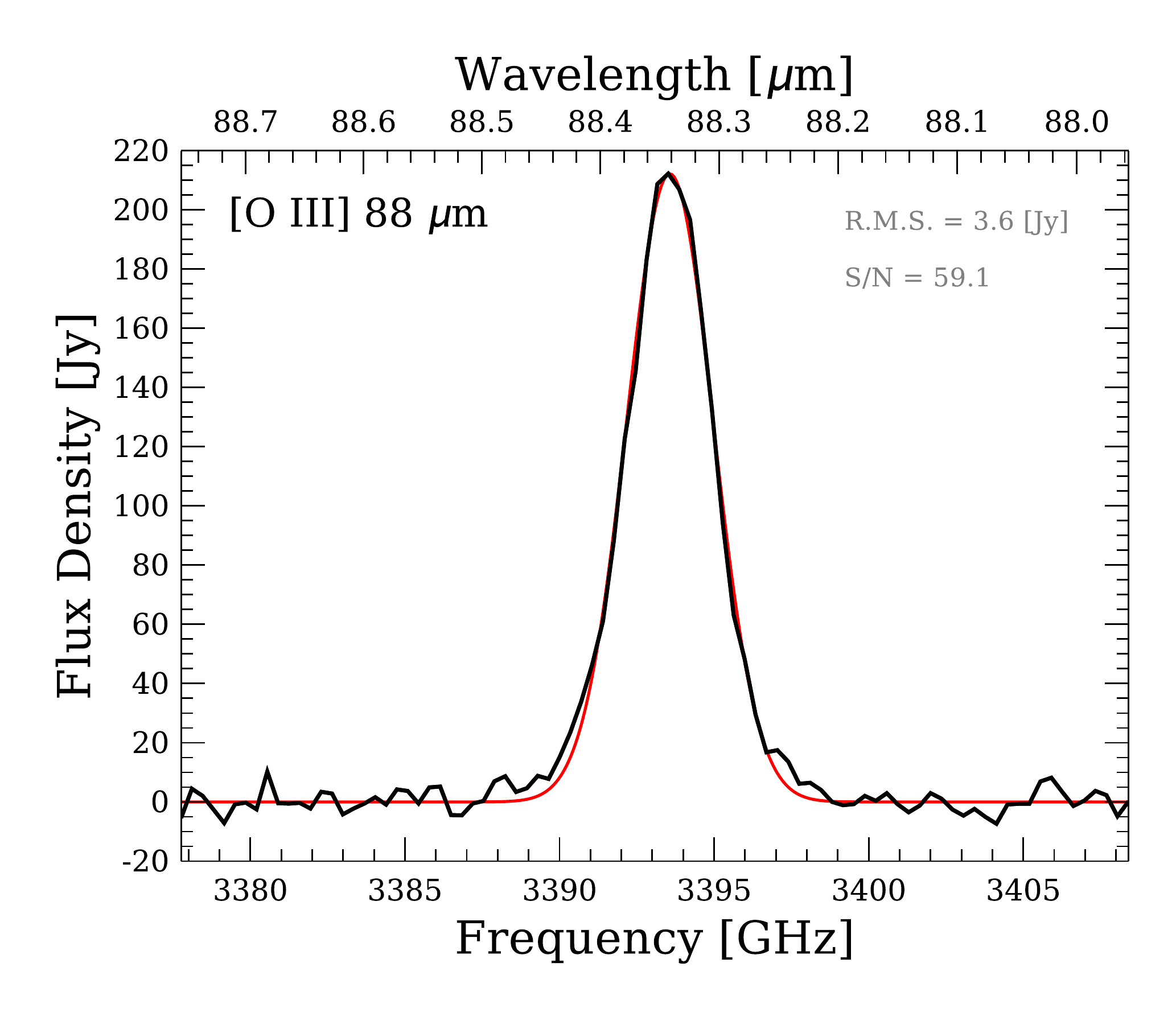,angle=0,width=0.27\linewidth}\\  
   
  \hspace{-0.50cm}\epsfig{file=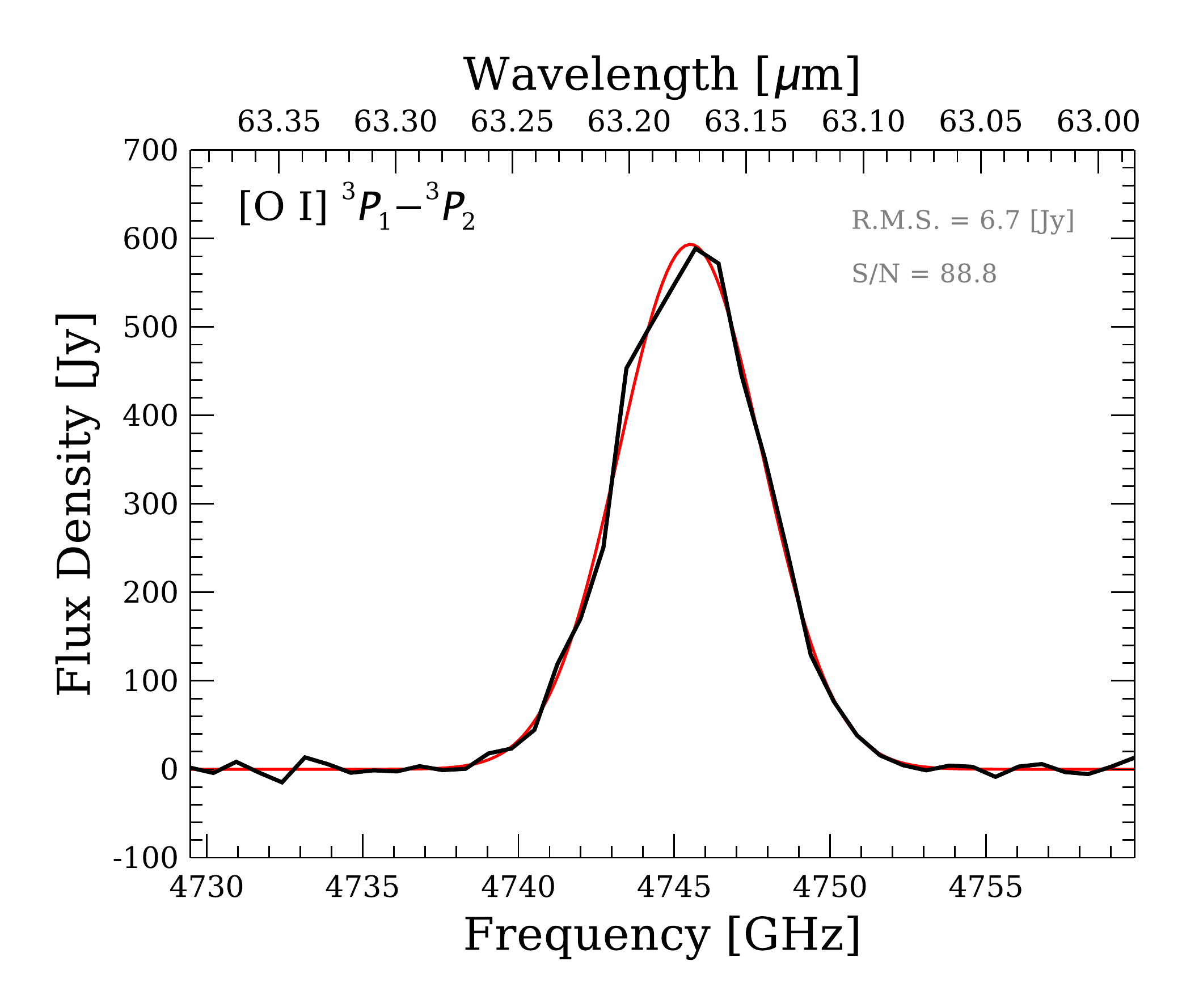,angle=0,width=0.27\linewidth} &
  \hspace{-0.50cm}\epsfig{file=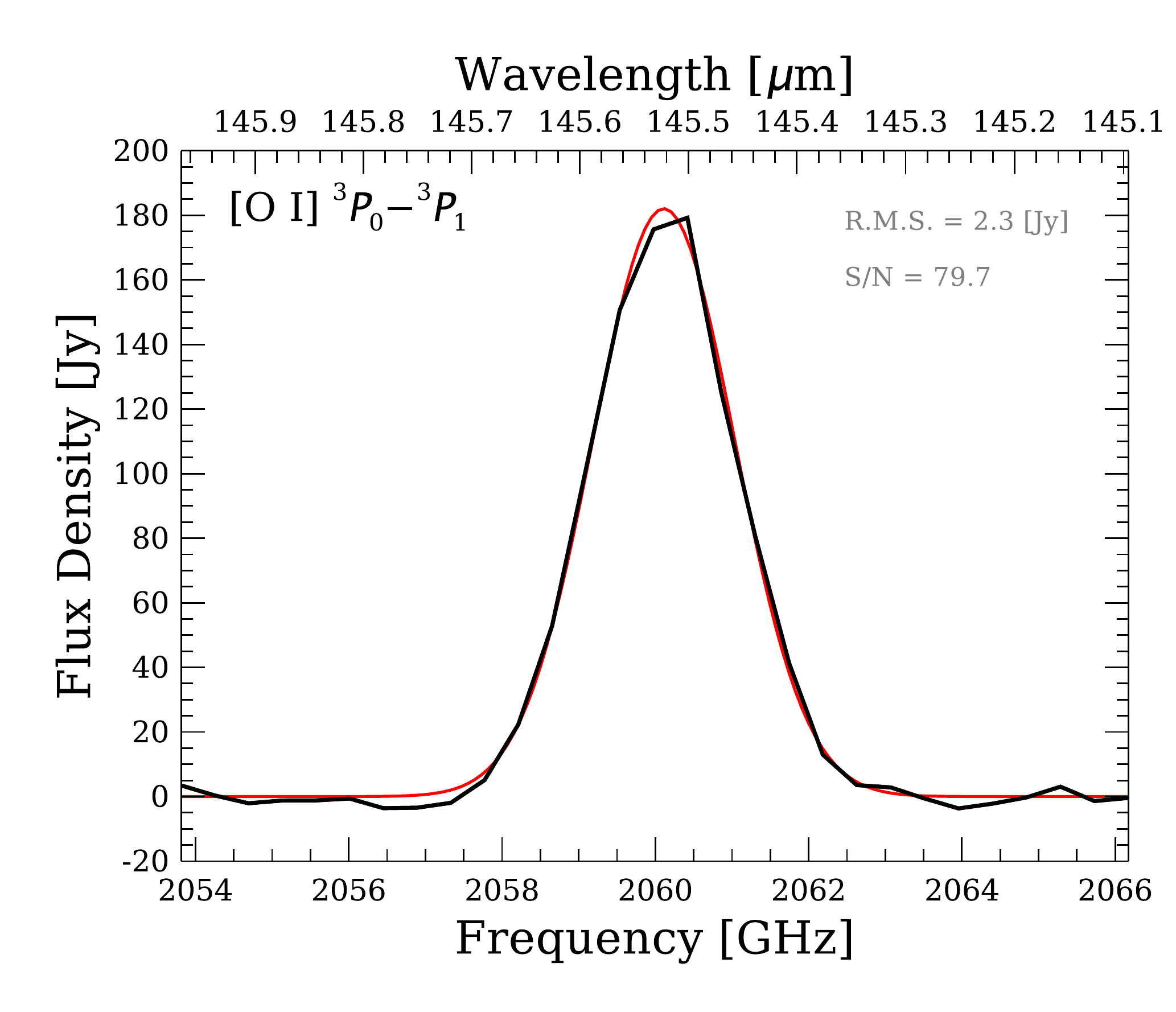,angle=0,width=0.27\linewidth} &
  \hspace{-0.50cm}\epsfig{file=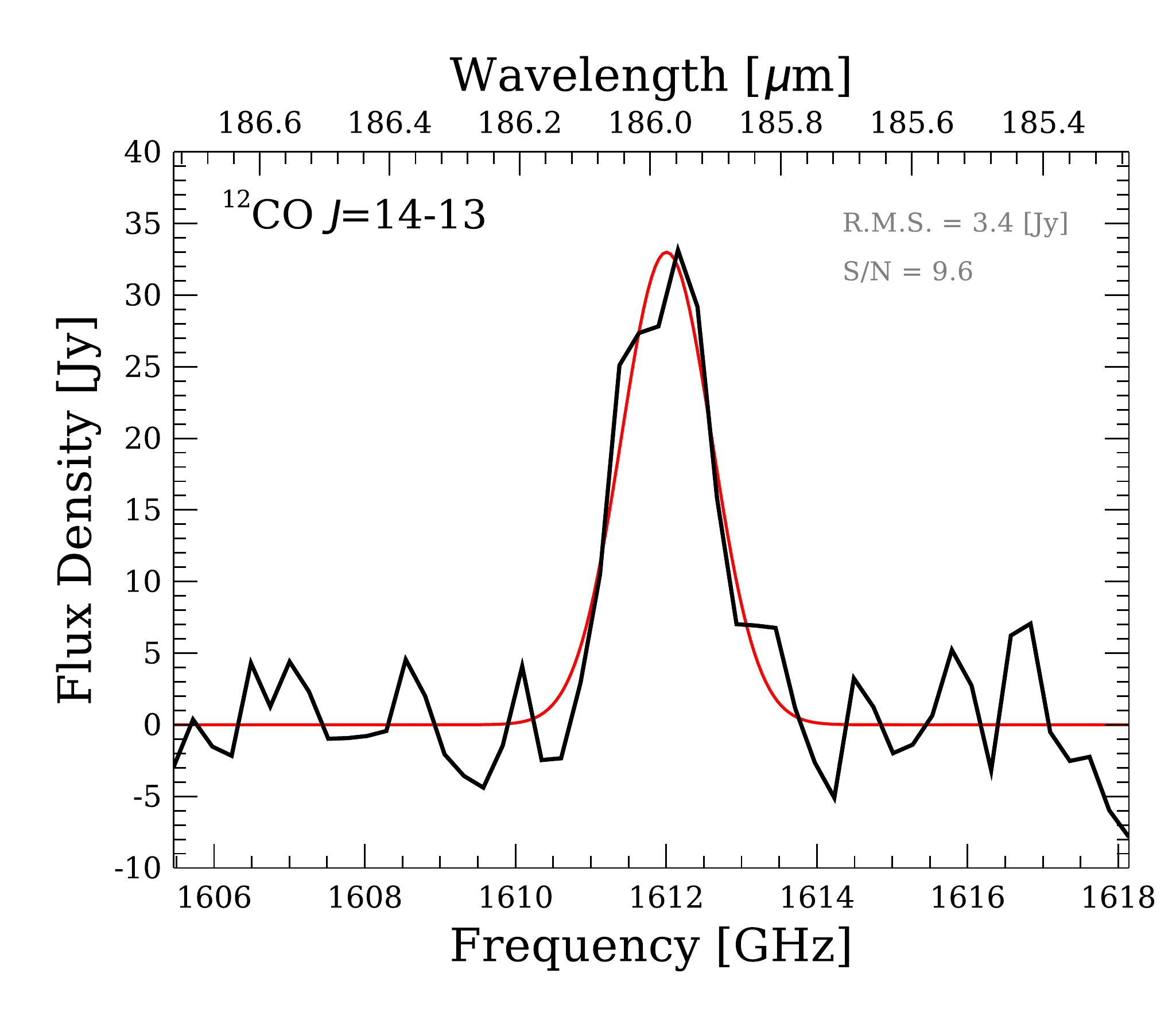,angle=0,width=0.27\linewidth} &
  \hspace{-0.50cm}\epsfig{file=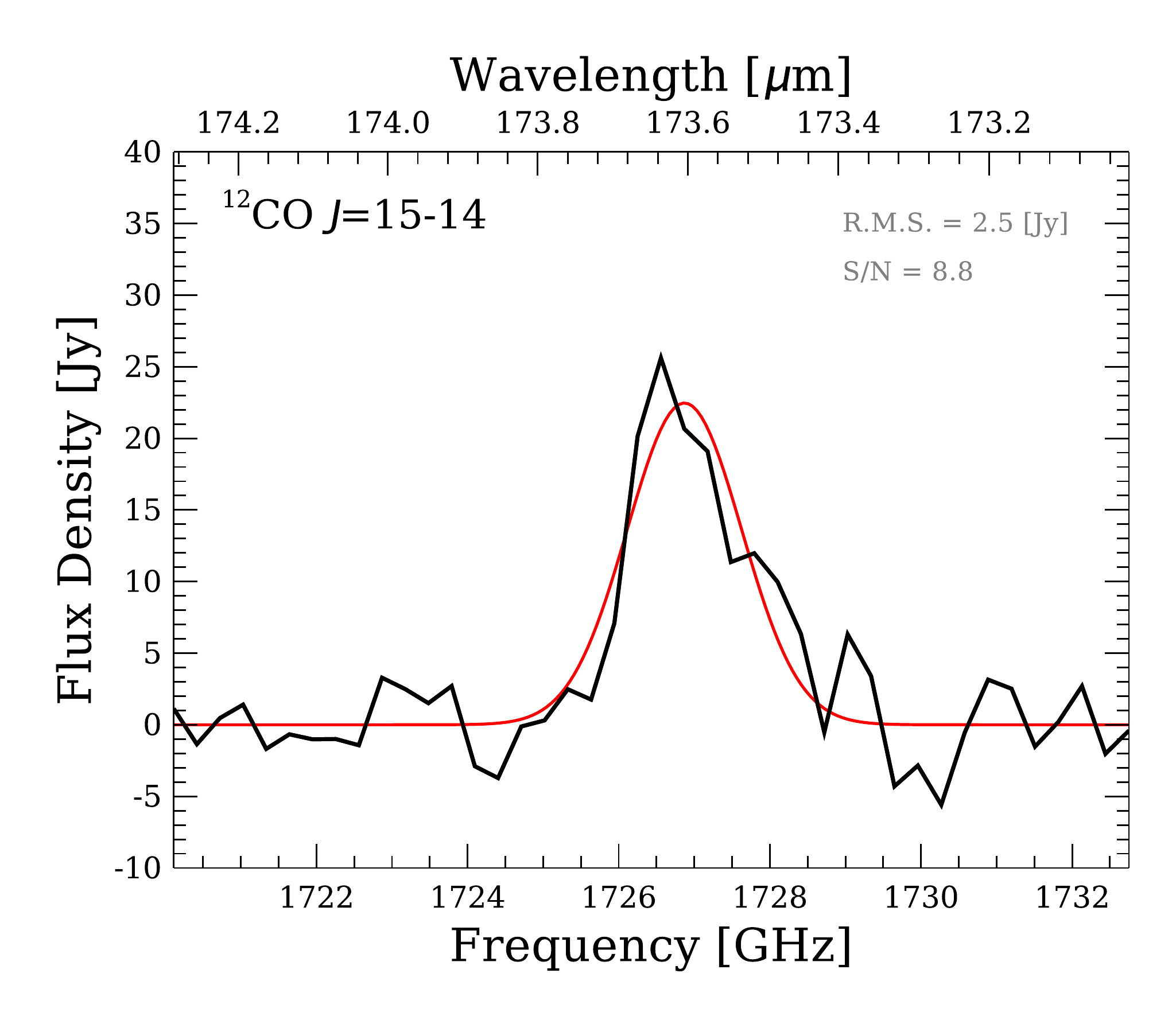,angle=0,width=0.27\linewidth} \\

  \hspace{-0.50cm}\epsfig{file=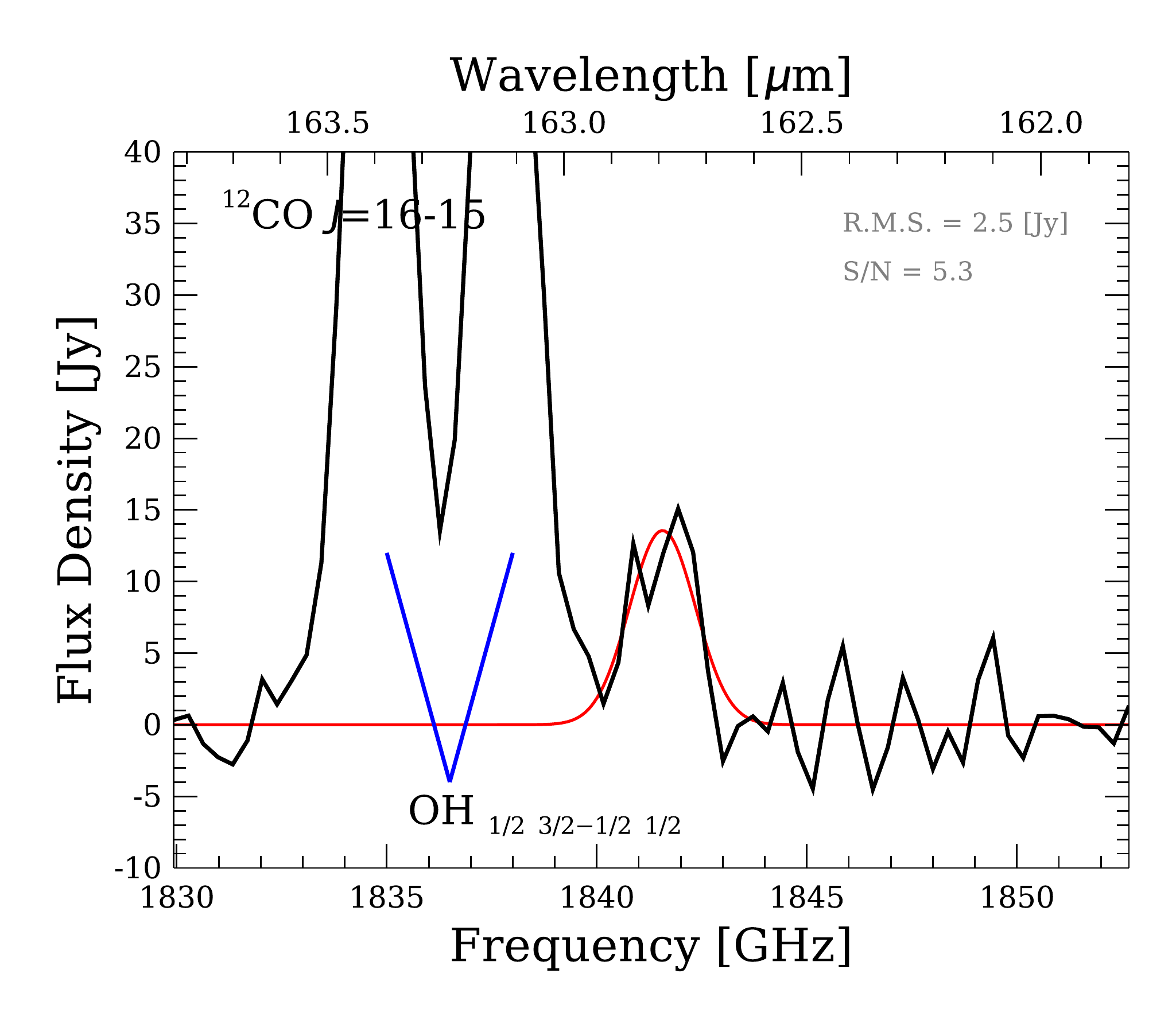,angle=0,width=0.27\linewidth} &
  \hspace{-0.50cm}\epsfig{file=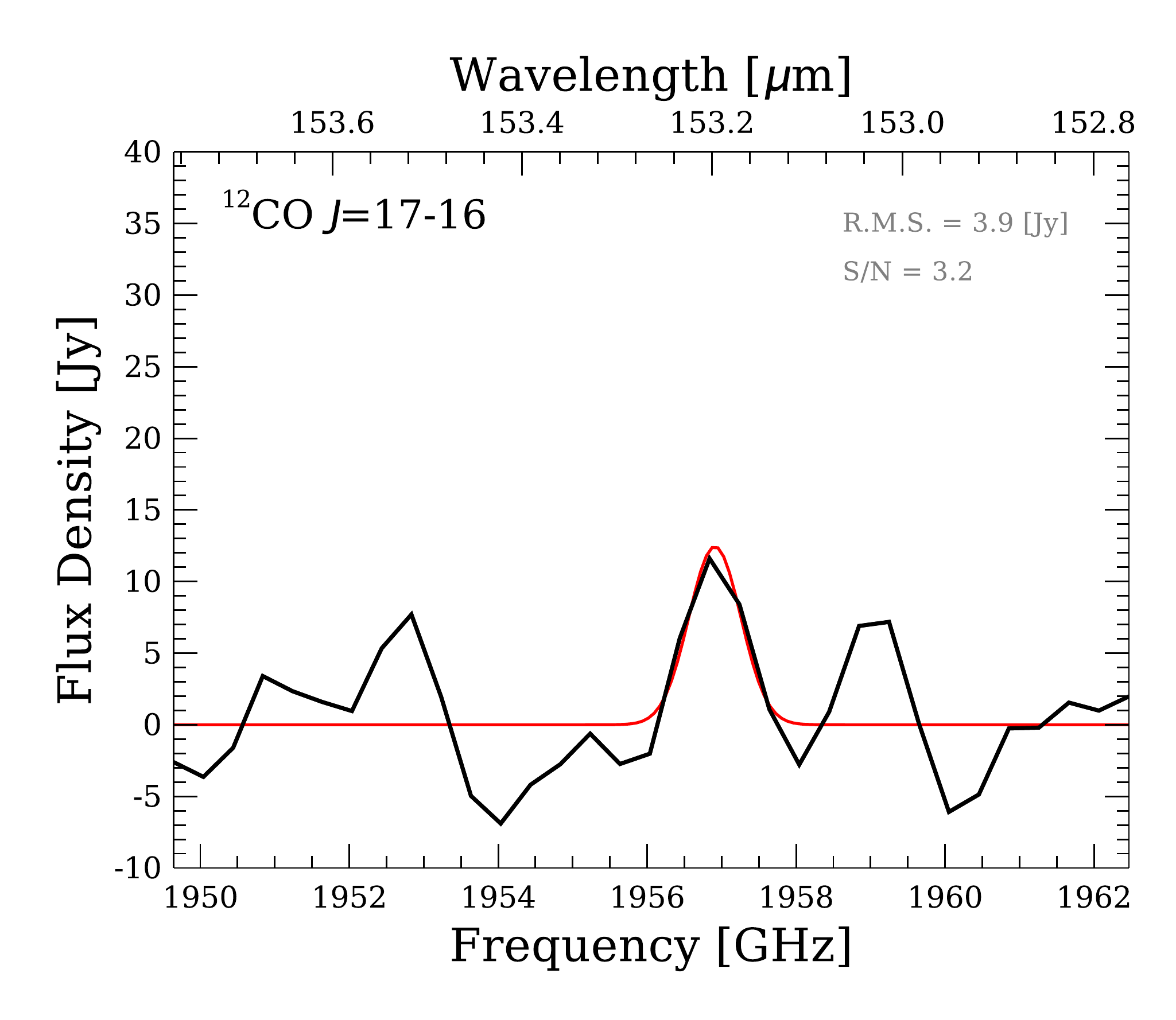,angle=0,width=0.27\linewidth} &
  \hspace{-0.50cm}\epsfig{file=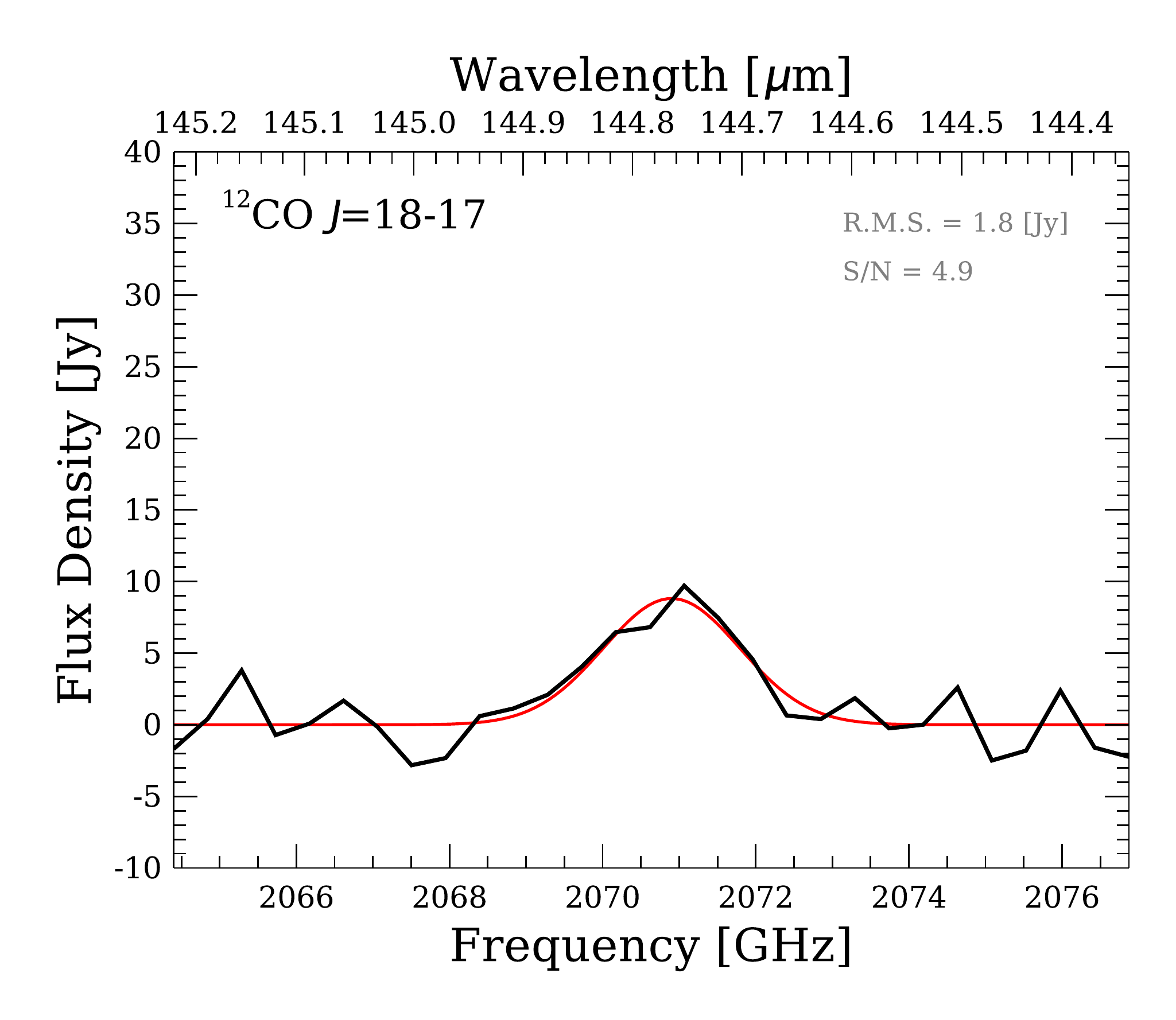,angle=0,width=0.27\linewidth} &
  \hspace{-0.50cm}\epsfig{file=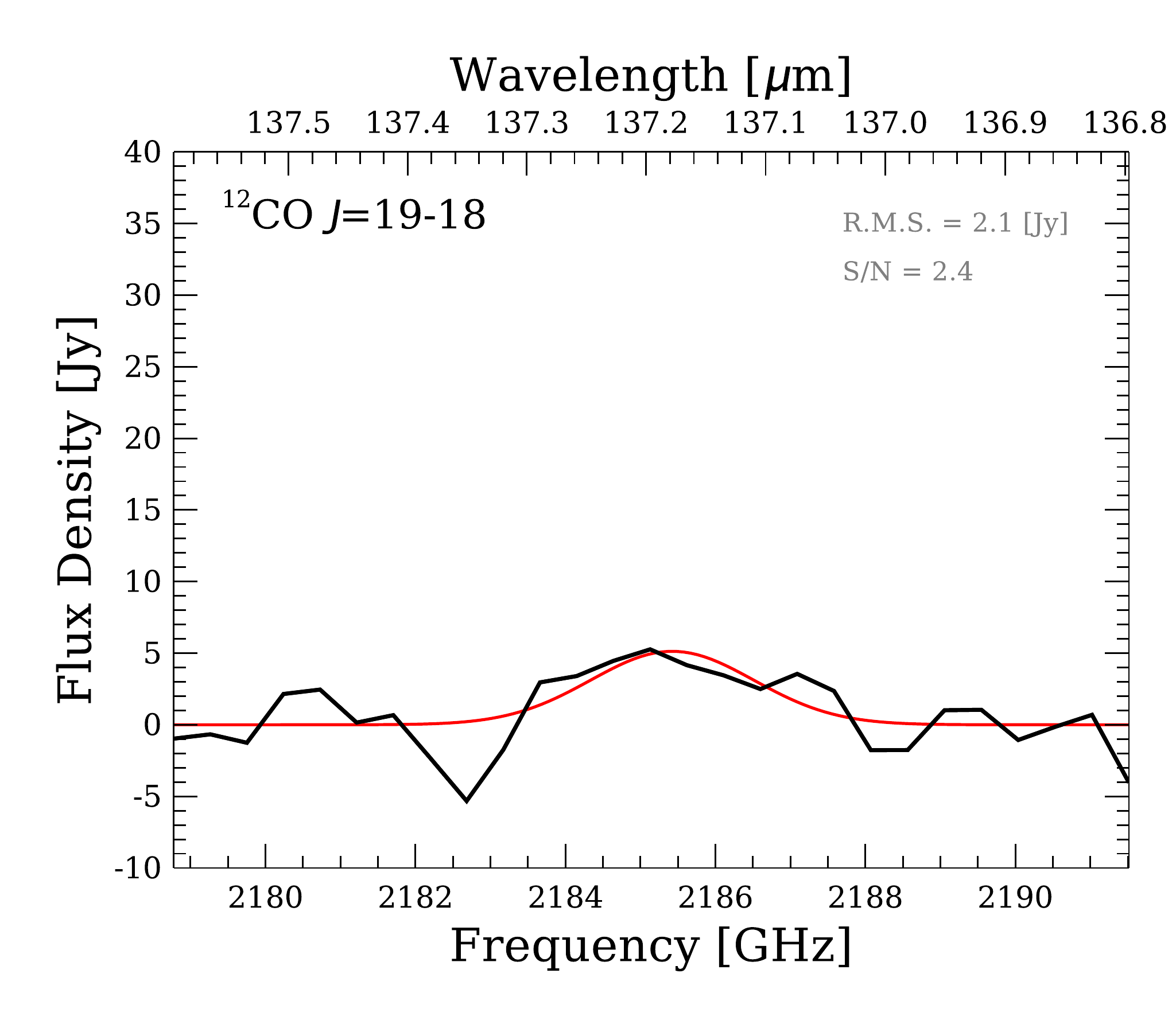,angle=0,width=0.27\linewidth} \\

  \hspace{-0.50cm}\epsfig{file=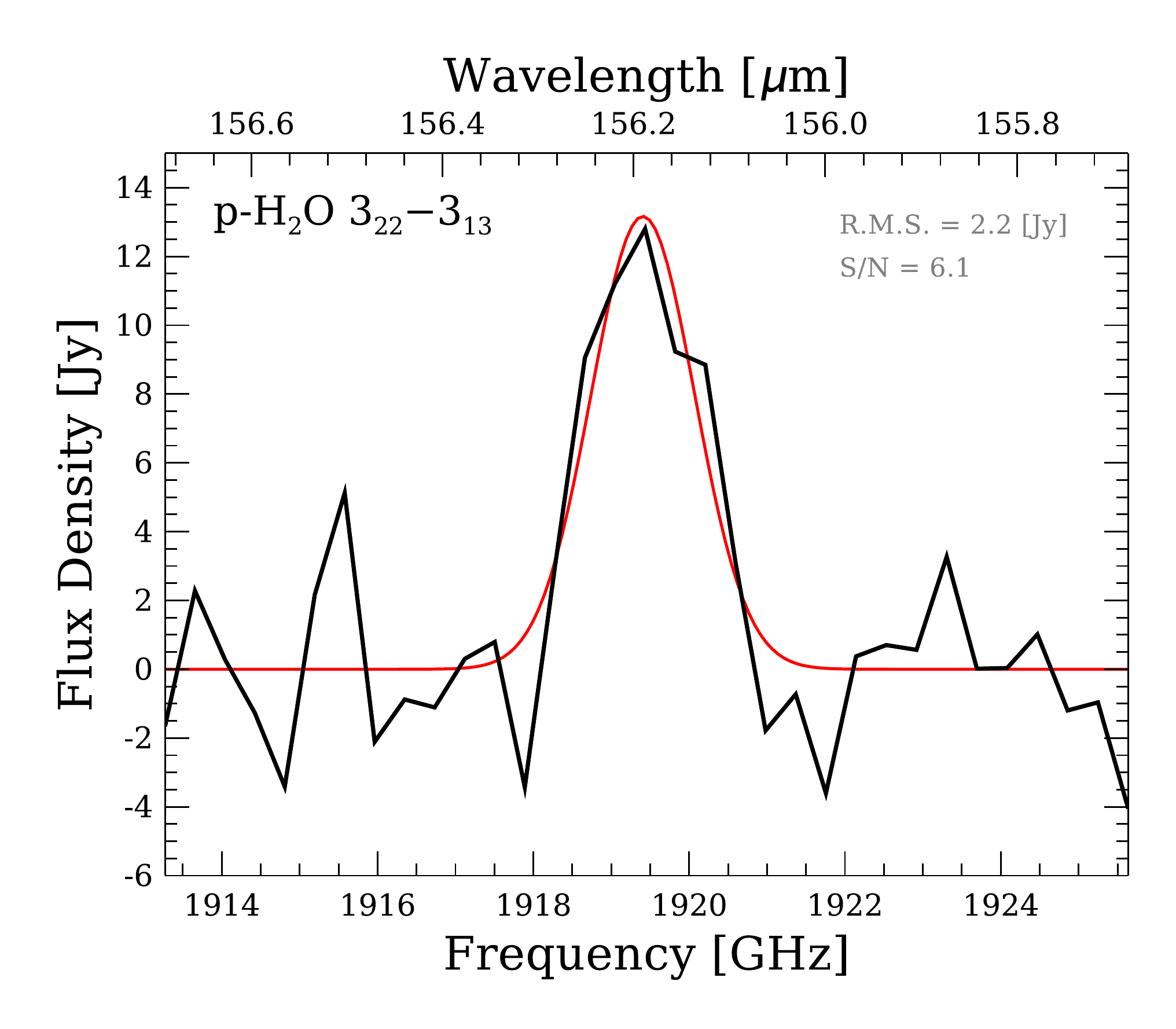,angle=0,width=0.27\linewidth} &
  \hspace{-0.50cm}\epsfig{file=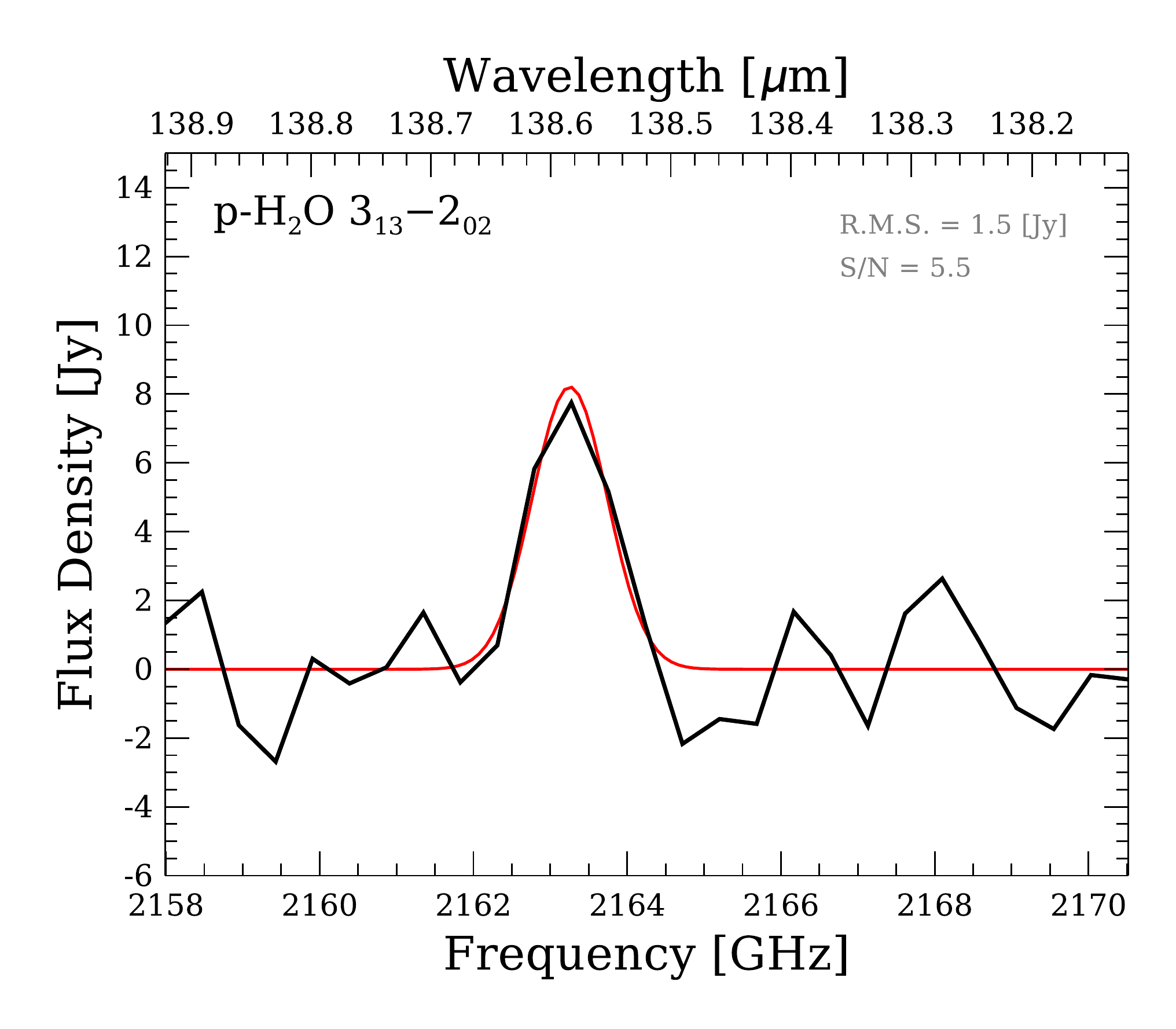,angle=0,width=0.27\linewidth} &
  \hspace{-0.50cm}\epsfig{file=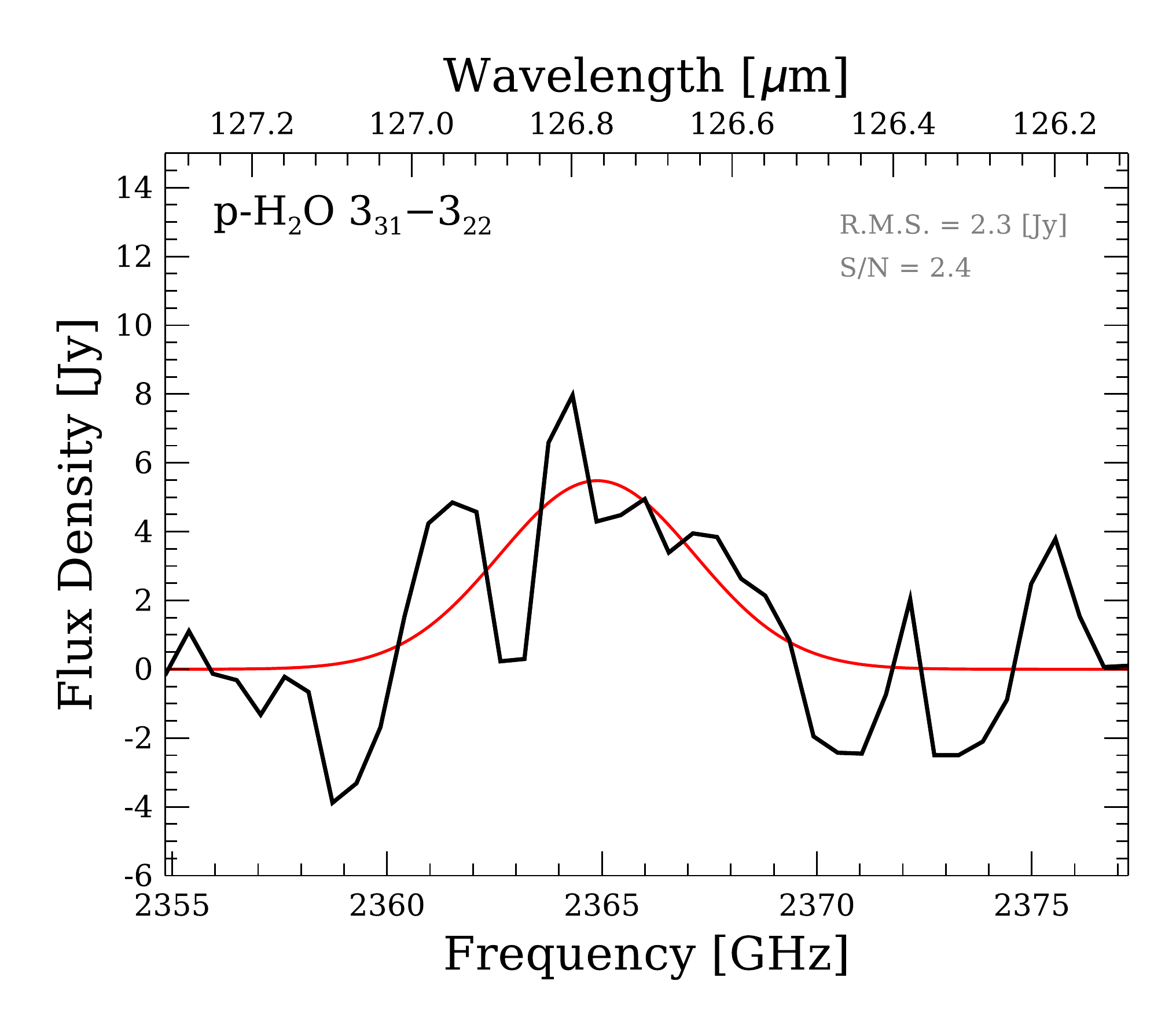,angle=0,width=0.27\linewidth} &
  \hspace{-0.50cm}\epsfig{file=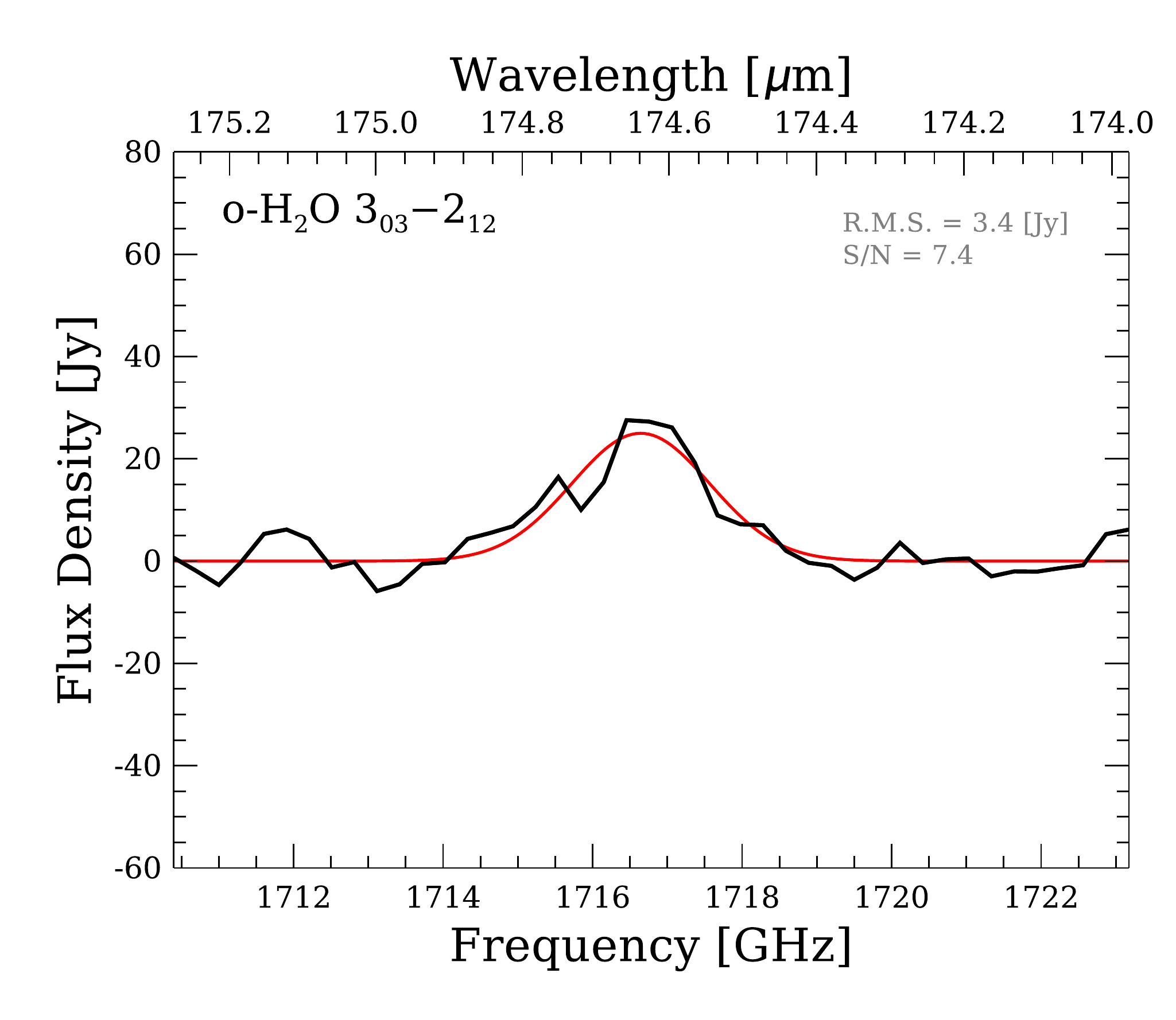,angle=0,width=0.27\linewidth} \\

  \hspace{-0.50cm}\epsfig{file=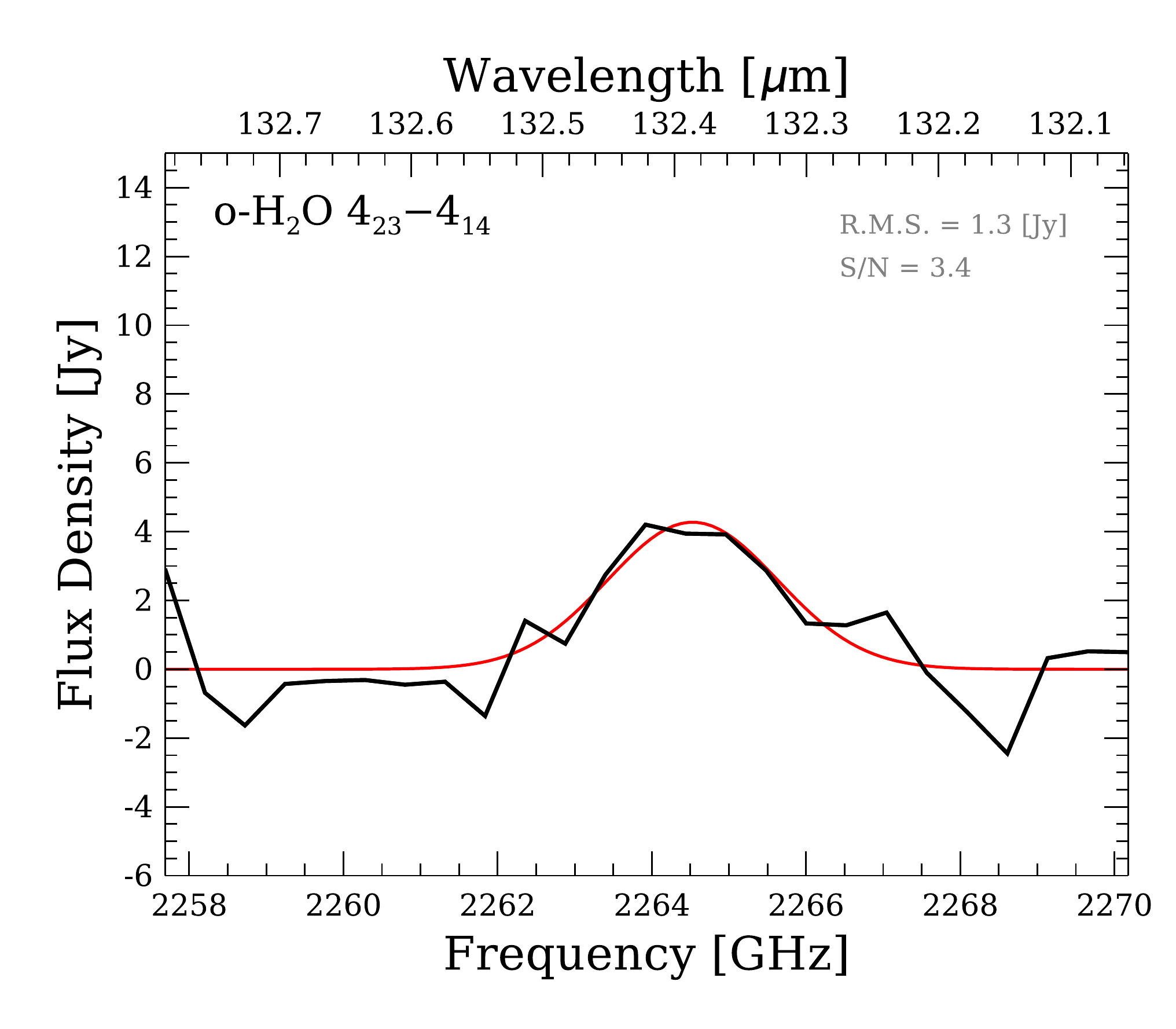,angle=0,width=0.27\linewidth} &
  \hspace{-0.50cm}\epsfig{file=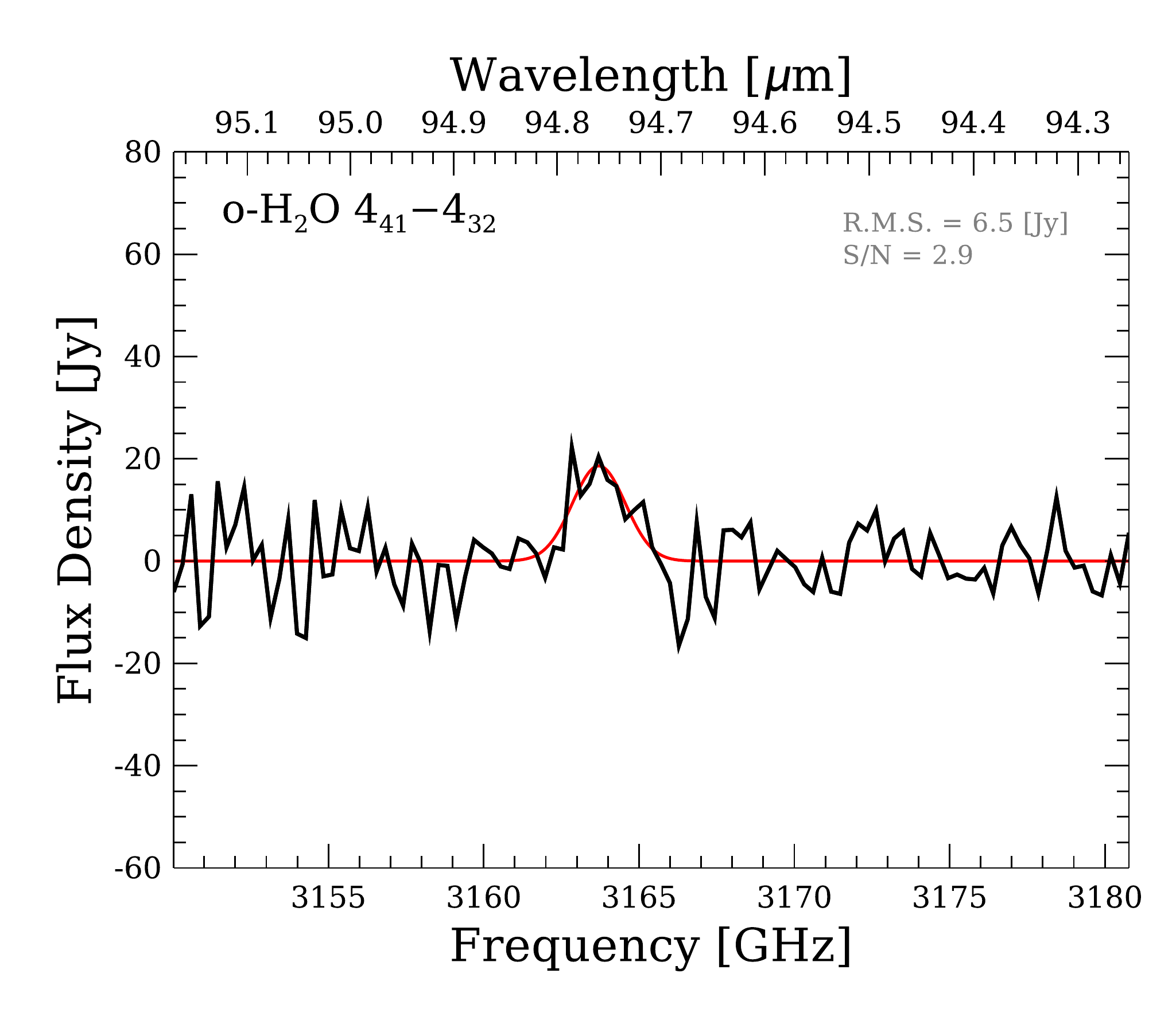,angle=0,width=0.27\linewidth} &  
  \hspace{-0.50cm}\epsfig{file=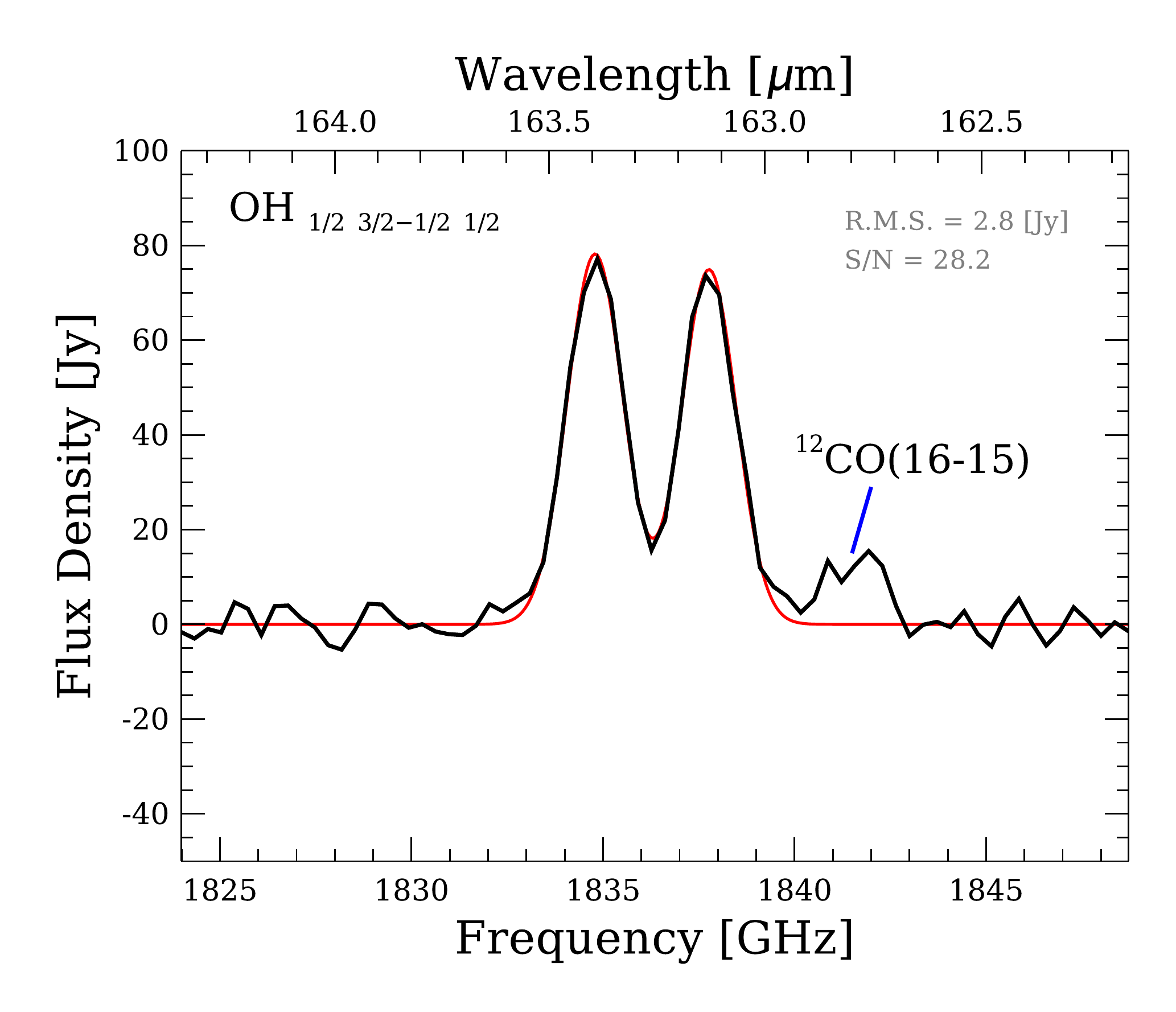,angle=0,width=0.27\linewidth} &
  \hspace{-0.50cm}\epsfig{file=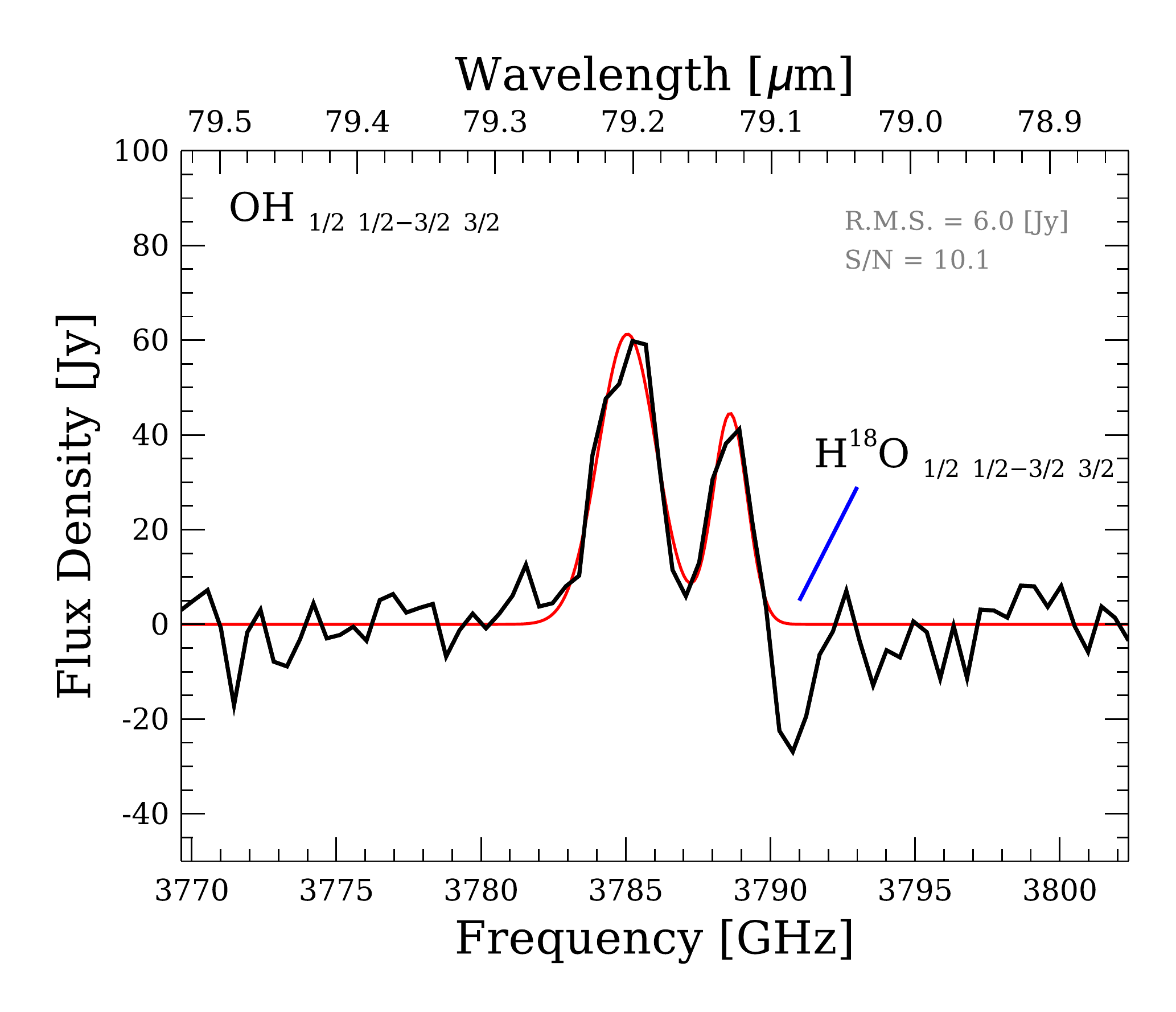,angle=0,width=0.27\linewidth}\\
    
 \end{tabular}
 \caption{\footnotesize{Detected emission lines in the PACS 5$\times$5 spaxels extracted spectrum of NGC~253.}}

  \label{fig:pacs-emission-lines}
\end{figure*}

\begin{figure*}[!tp]

 \begin{tabular}{cccc}
  \hspace{-0.50cm}\epsfig{file=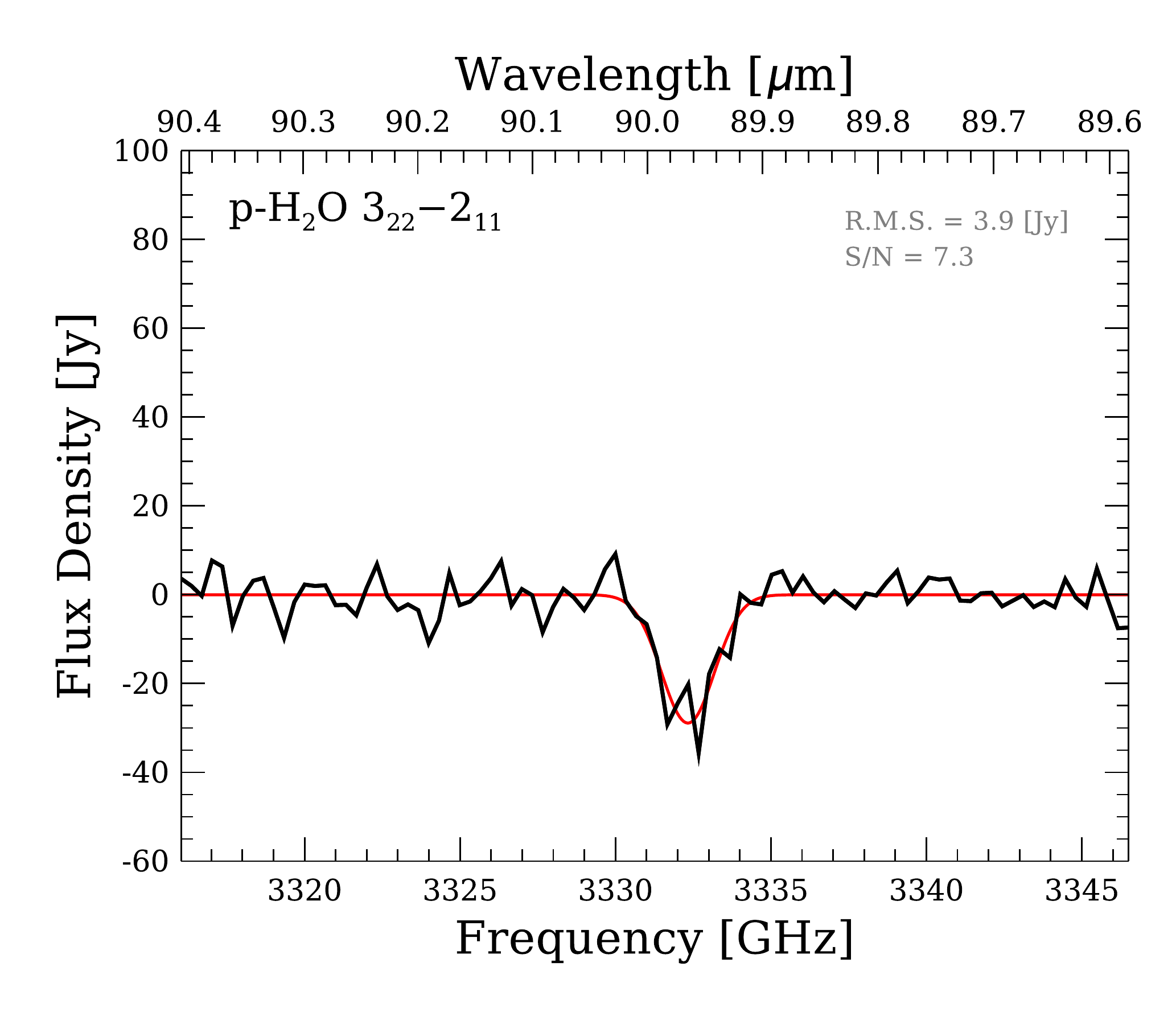,angle=0,width=0.27\linewidth} &
  \hspace{-0.50cm}\epsfig{file=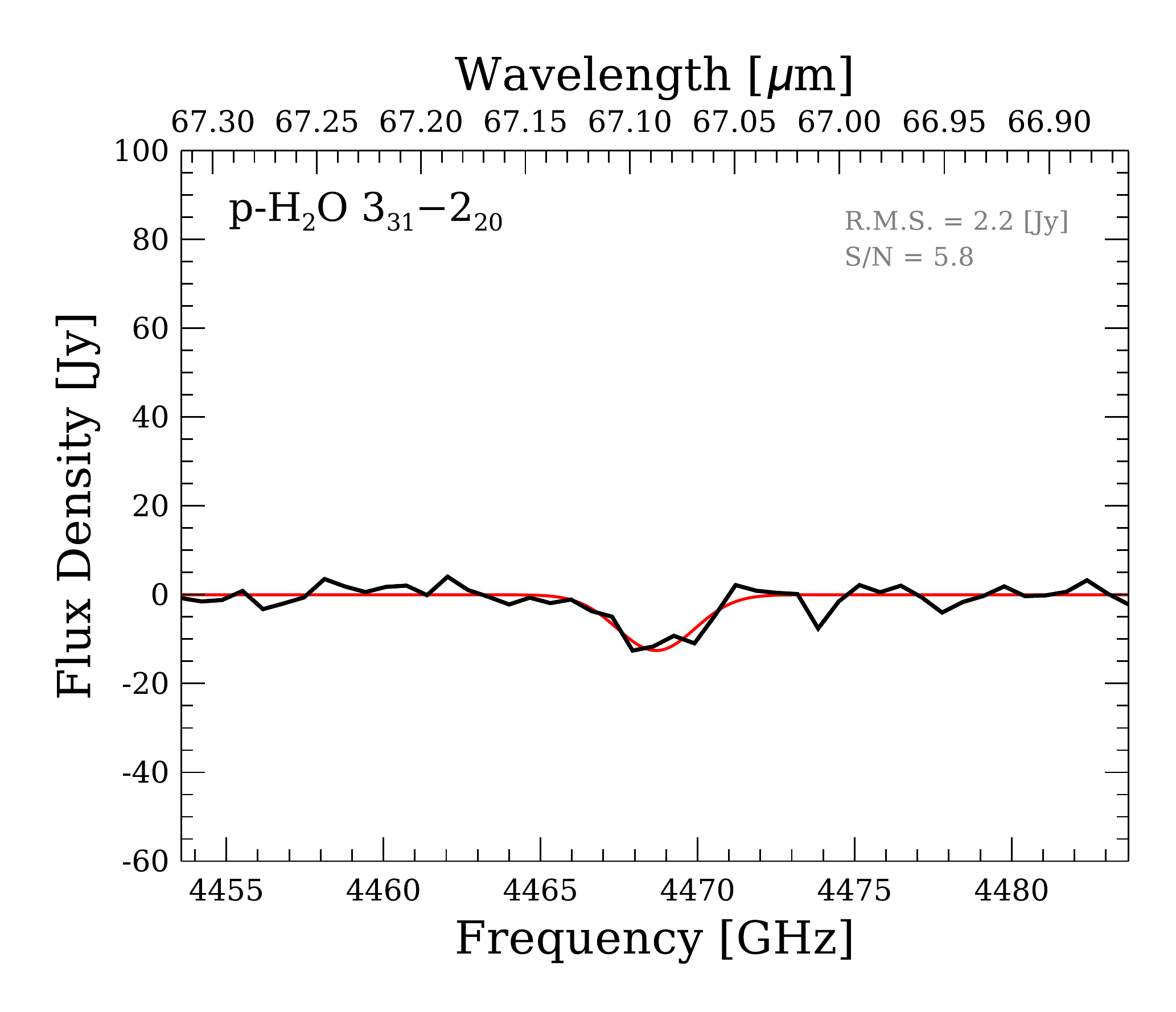,angle=0,width=0.27\linewidth} &
  \hspace{-0.50cm}\epsfig{file=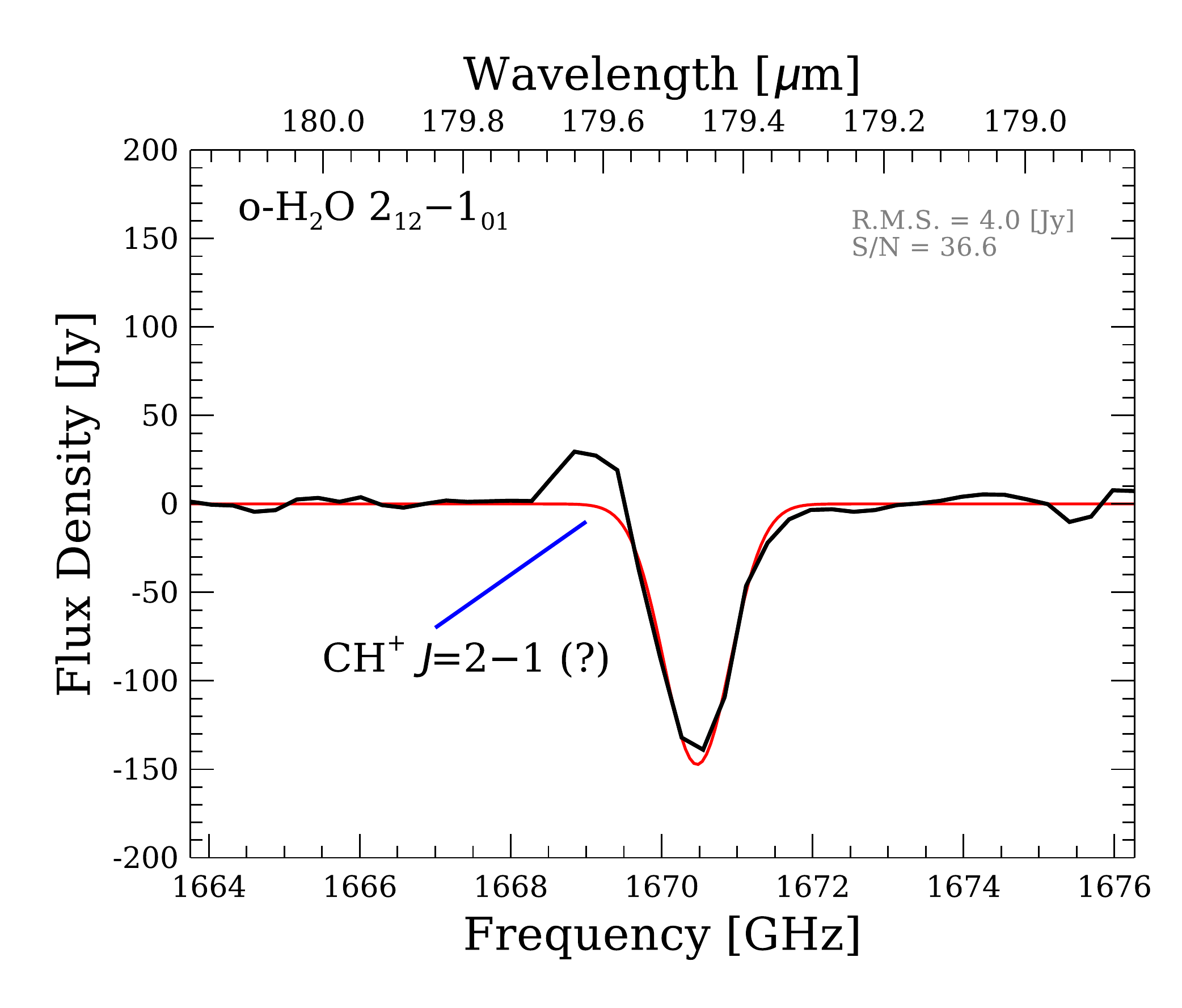,angle=0,width=0.27\linewidth} &
  \hspace{-0.50cm}\epsfig{file=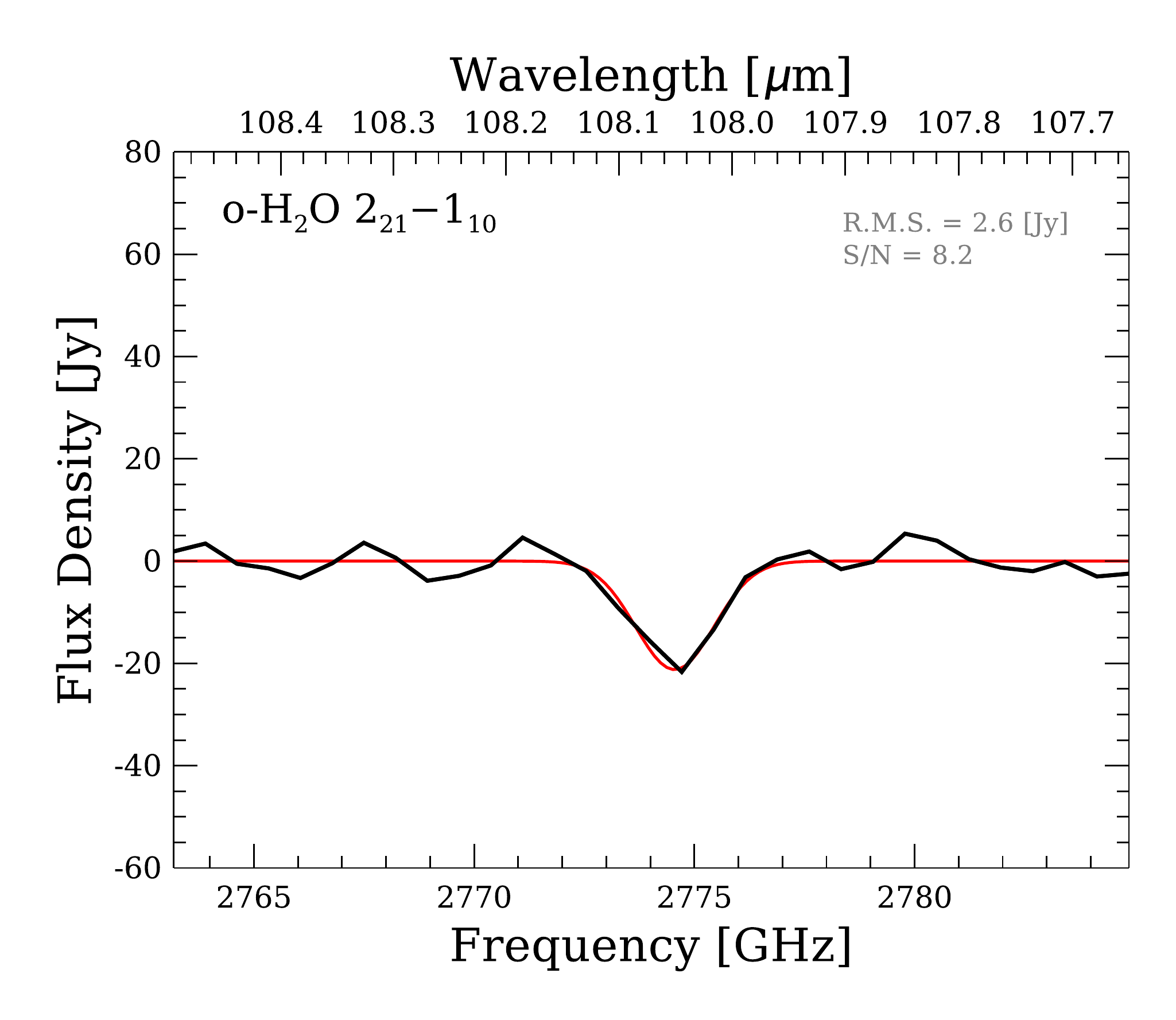,angle=0,width=0.27\linewidth}\\
  
  \hspace{-0.50cm}\epsfig{file=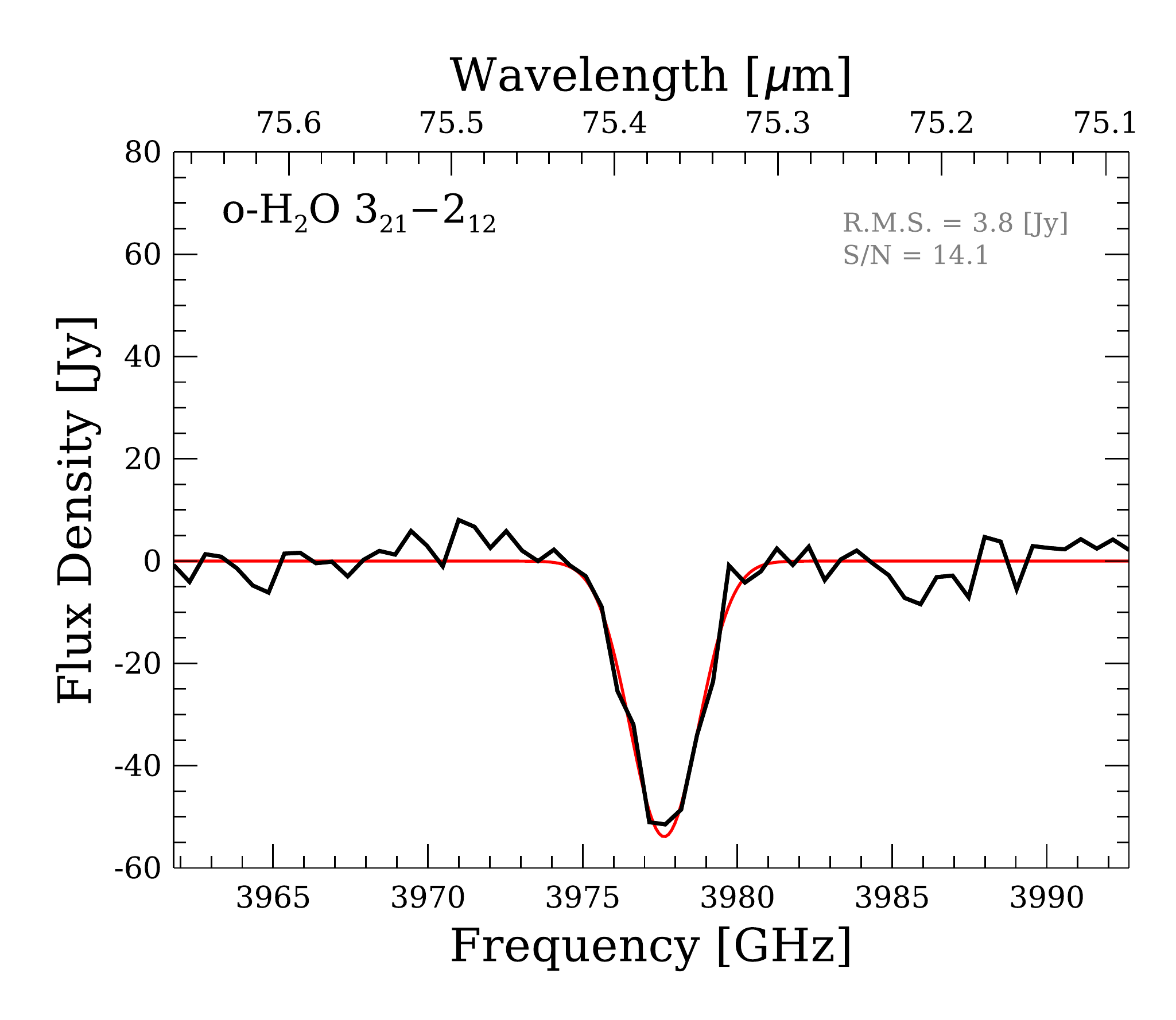,angle=0,width=0.27\linewidth} &
  \hspace{-0.50cm}\epsfig{file=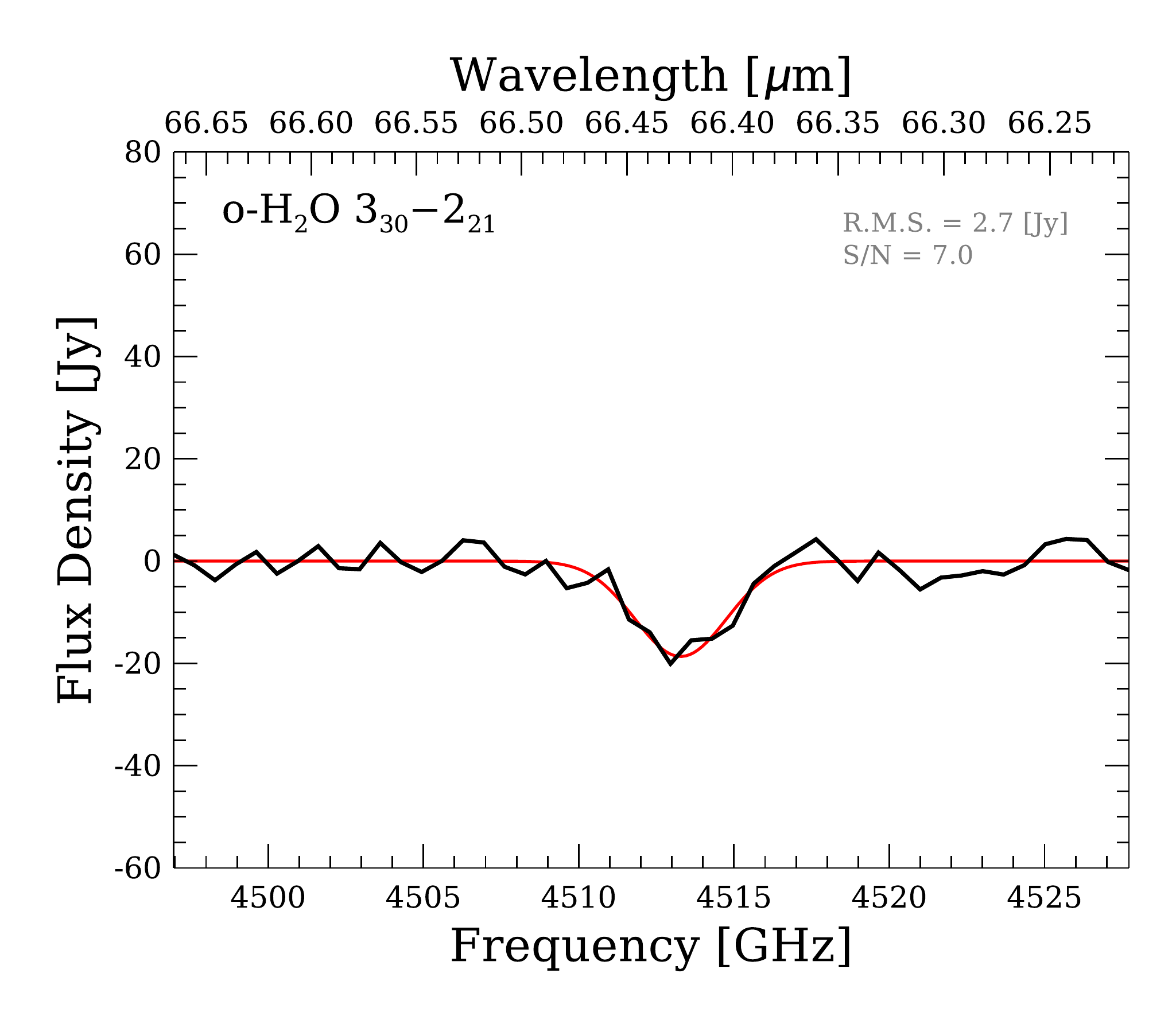,angle=0,width=0.27\linewidth} &
  \hspace{-0.50cm}\epsfig{file=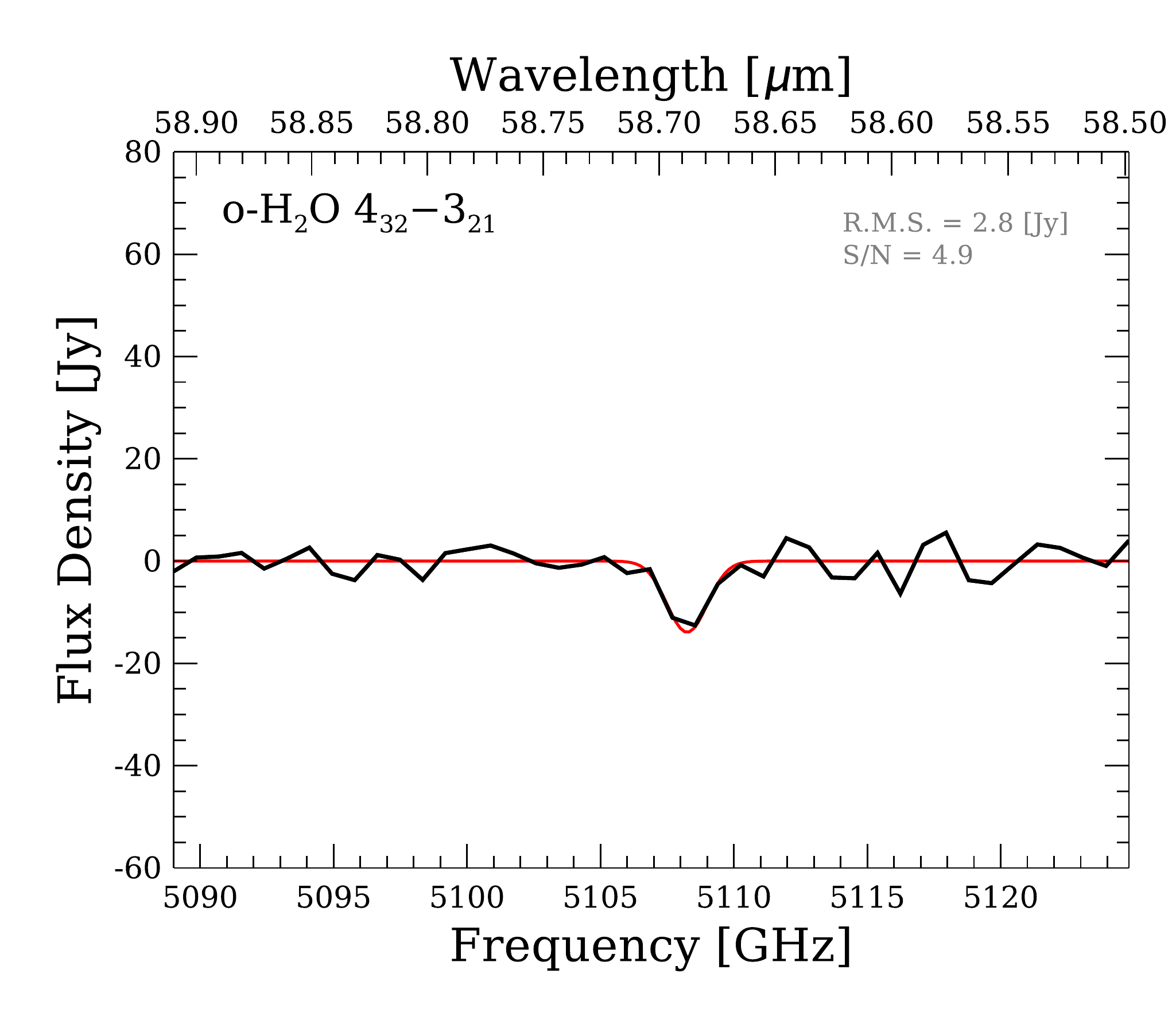,angle=0,width=0.27\linewidth} &
  \hspace{-0.50cm}\epsfig{file=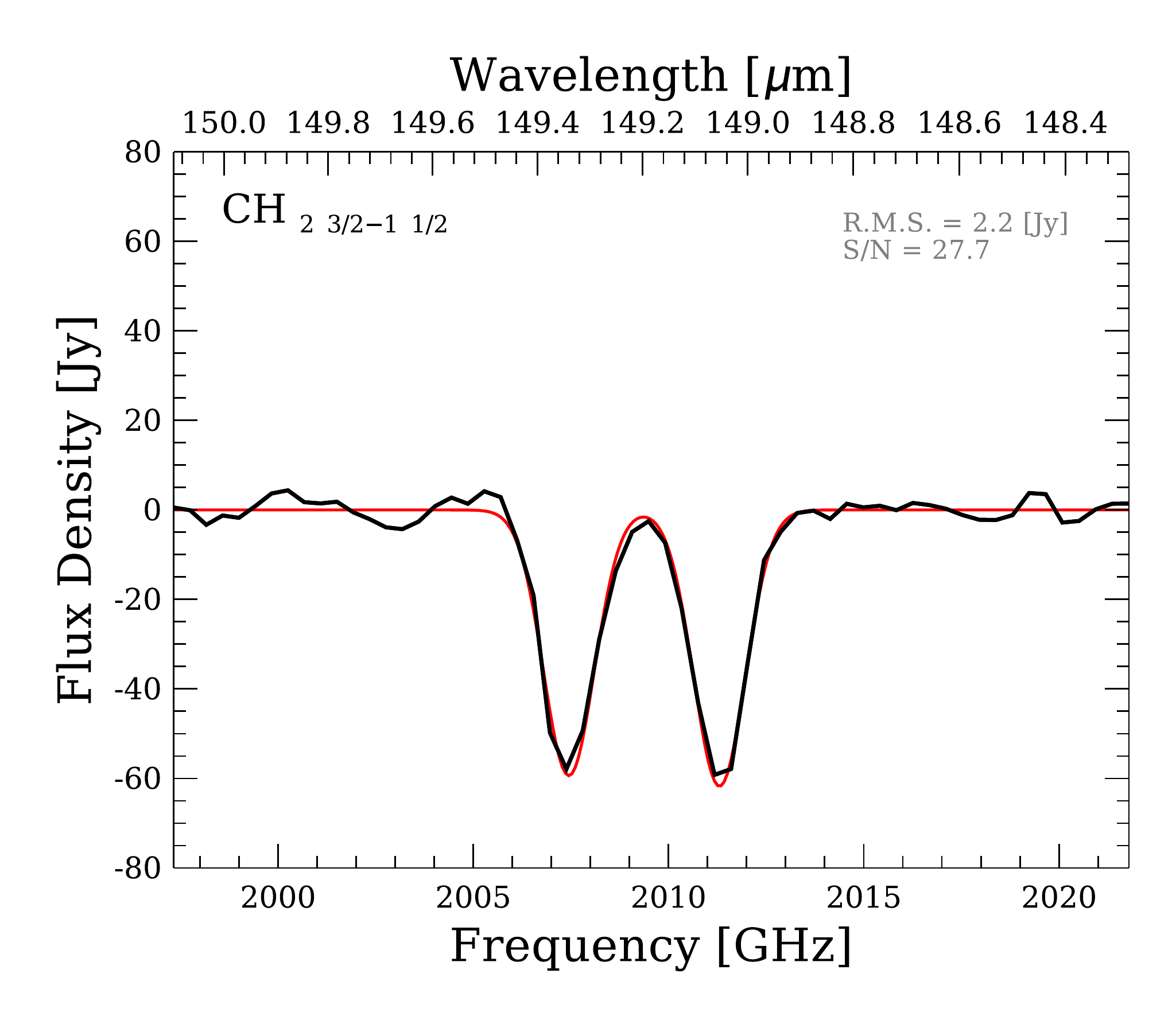,angle=0,width=0.27\linewidth} \\
    
  \hspace{-0.50cm}\epsfig{file=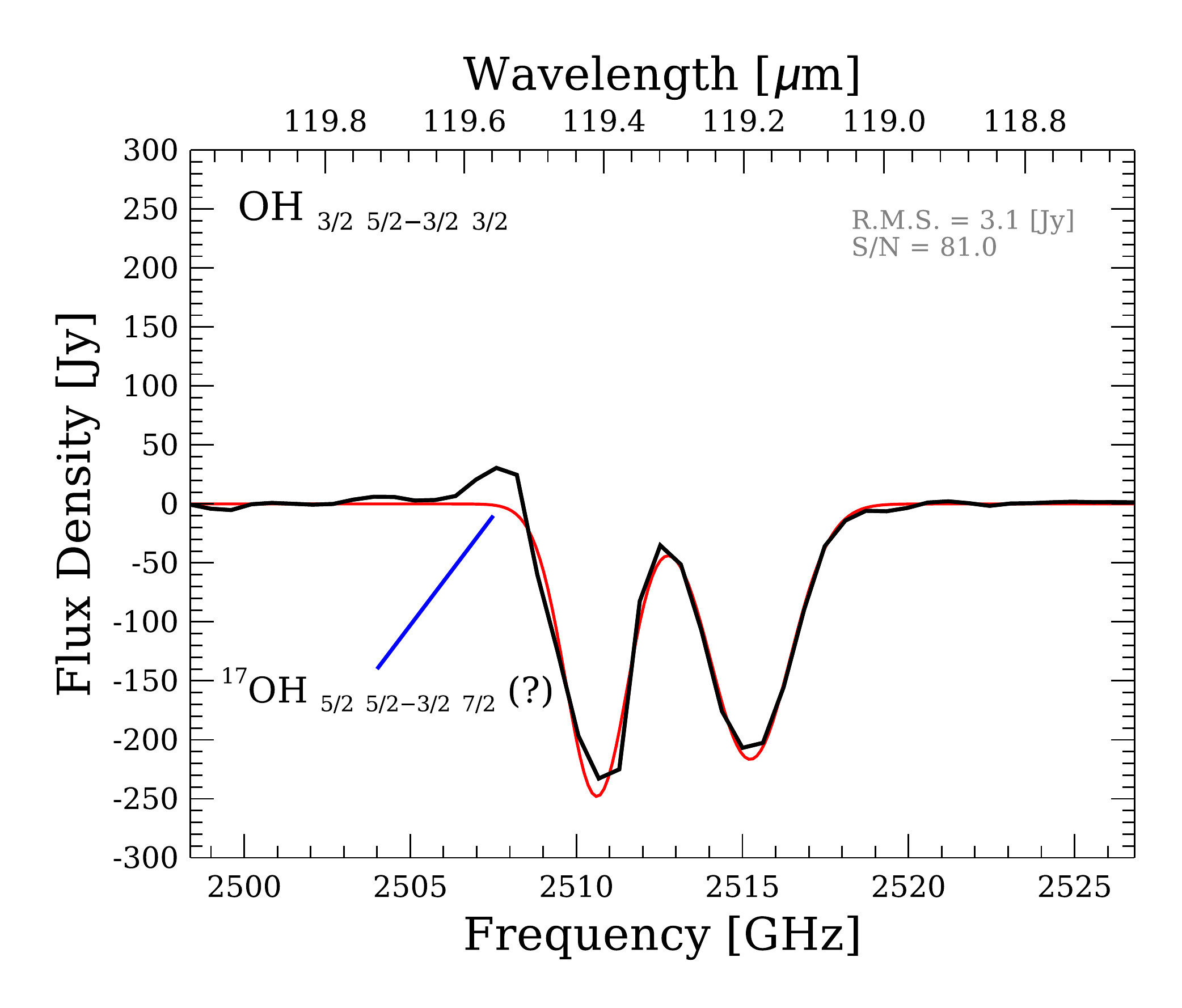,angle=0,width=0.27\linewidth} &
  \hspace{-0.50cm}\epsfig{file=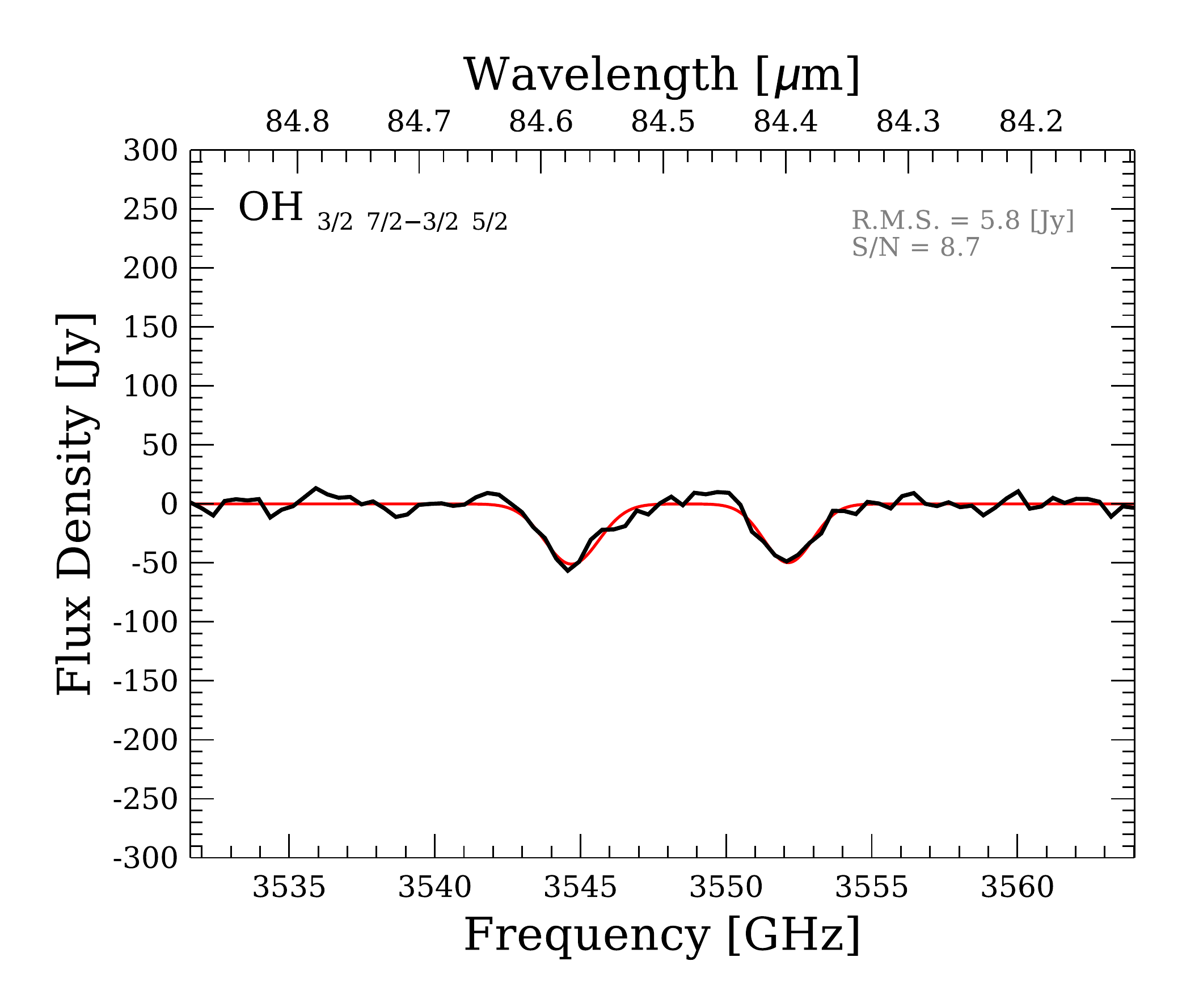,angle=0,width=0.27\linewidth} &
    \hspace{-0.50cm}\epsfig{file=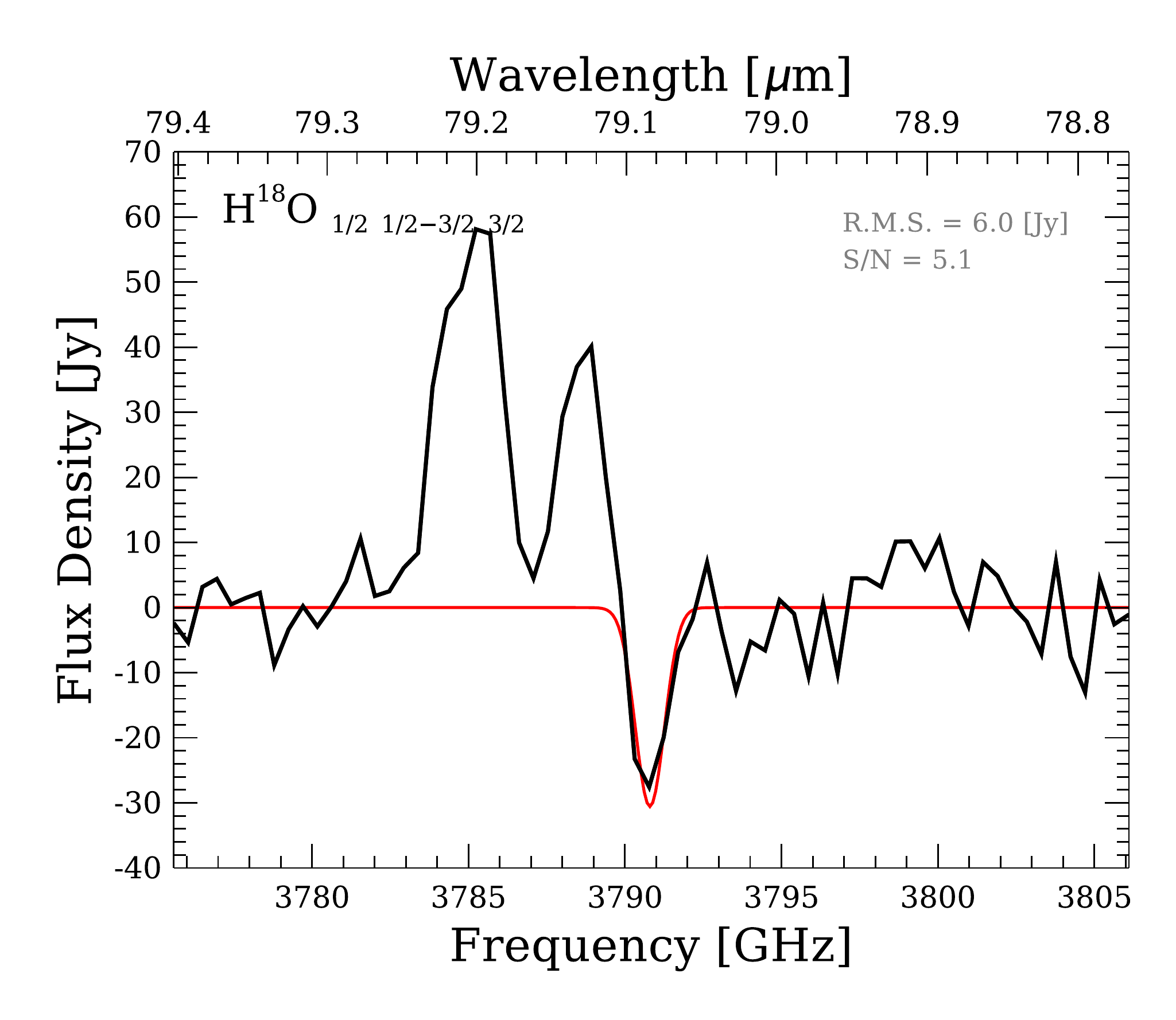,angle=0,width=0.27\linewidth}&
  \\

 \end{tabular}
 \caption{\footnotesize{Detected absorption lines in the PACS 5$\times$5 spaxels extracted spectrum of NGC~253.}}

  \label{fig:pacs-absorption-lines}
\end{figure*}

   \begin{table}[htp]
         \centering
      \caption{\footnotesize{Fluxes of the emission lines extracted from the PACS 5$\times$5 spectrum of NGC~253.}}\label{tab:pacs-emission-fluxes}
         \tabcolsep 2.0pt
         \scriptsize
         \begin{tabular}{lcccc}
	    \hline\hline
	    \noalign{\smallskip}
            Line   &    $\lambda_{\rm rest}^{~\mathrm{a}}$   &    $\nu_{\rm rest}^{~\mathrm{a}}$   &        Flux$^{~\mathrm{b}}$       &  Luminosity$^{~\mathrm{c}}$  \\
                   &              ($\mu$m)              &              (GHz)              &  ($10^{-16}$~\Wm) &      ($10^{4}$~\Lsun)        \\                   
            \noalign{\smallskip}
            \hline
            \noalign{\smallskip}
 
 
 
            [N III] 57 $\mu$m  &  57.320  &  5230.155  &  54.75$\pm$9.00  &  207.8$\pm$37.6  \\
            \nii\ 122 $\mu$m  &  121.888  &  2459.566  &  95.40$\pm$14.36  &  362.1$\pm$61.1  \\
 
            \noalign{}\\
 
            \cii\  &  157.741  &  1900.537  &  479.87$\pm$72.08  &  1821.2$\pm$306.5  \\
 
            \noalign{}\\
 
            [O III] 88 $\mu$m  &  88.356  &  3392.991  &  73.32$\pm$11.05  &  278.3$\pm$47.0  \\
            \PzeroOi\  &  145.525  &  2060.070  &  42.51$\pm$6.41  &  161.3$\pm$27.2  \\
            \PoneOi\  &  63.184  &  4744.775  &  347.60$\pm$52.53  &  1319.2$\pm$223.1  \\
 
            \noalign{}\\
 
            \twco\ $J = 14\rightarrow13$  &  186.000  &  1611.787  &   5.64$\pm$1.01  &  21.4$\pm$4.2  \\
            \twco\ $J = 15\rightarrow14$  &  173.630  &  1726.617  &   3.95$\pm$0.75  &  15.0$\pm$3.1  \\
            \twco\ $J = 16\rightarrow15$  &  162.812  &  1841.346  &   2.64$\pm$0.58  &  10.0$\pm$2.3  \\
            \twco\ $J = 17\rightarrow16$  &  153.267  &  1956.018  &   1.06$\pm$0.49  &   4.0$\pm$1.9  \\
            \twco\ $J = 18\rightarrow17$  &  144.780  &  2070.676  &   2.00$\pm$0.48  &   7.6$\pm$1.9  \\
            \twco\ $J = 19\rightarrow18$  &  137.200  &  2185.076  &   1.45$\pm$0.58  &   5.5$\pm$2.2  \\
 
            \noalign{}\\
 
            \ph2o\ $3_{22}-3_{13}$  &  156.190  &  1919.409  &   2.09$\pm$0.48  &   7.9$\pm$1.9  \\
            \ph2o\ $3_{13}-2_{02}$  &  138.530  &  2164.098  &   1.07$\pm$0.30  &   4.0$\pm$1.2  \\
            \ph2o\ $3_{31}-3_{22}$  &  126.710  &  2365.973  &   3.21$\pm$1.00  &  12.2$\pm$3.9  \\
 
            \noalign{}\\
 
            \oh2o\ $3_{03}-2_{12}$  &  174.630  &  1716.729  &   5.37$\pm$1.00  &  20.4$\pm$4.1  \\
            \oh2o\ $4_{23}-4_{14}$  &  132.410  &  2264.122  &   1.24$\pm$0.36  &   4.7$\pm$1.4  \\
            \oh2o\ $4_{41}-4_{32}$  &  94.705  &  3165.533  &   3.85$\pm$1.02  &  14.6$\pm$4.0  \\
 
            \noalign{}\\
 
            OH $_{\tfrac{1}{2}~\tfrac{3}{2}-\tfrac{1}{2}~\tfrac{1}{2}}$  &  163.400  &  1834.715  &  13.30$\pm$2.14  &  50.5$\pm$9.0  \\
            OH $_{\tfrac{1}{2}~\tfrac{3}{2}-\tfrac{1}{2}~\tfrac{1}{2}}$  &  163.120  &  1837.865  &  12.51$\pm$2.03  &  47.5$\pm$8.5  \\
            OH $_{\tfrac{1}{2}~\tfrac{1}{2}-\tfrac{3}{2}~\tfrac{3}{2}}$  &  79.179  &  3786.257  &  14.57$\pm$2.87  &  55.3$\pm$11.7  \\
            OH $_{\tfrac{1}{2}~\tfrac{1}{2}-\tfrac{3}{2}~\tfrac{3}{2}}$  &  79.115  &  3789.301  &   6.68$\pm$1.75  &  25.3$\pm$6.9  \\

	    \noalign{\smallskip}
	    \hline
	  \end{tabular}

\begin{list}{}{}
\scriptsize
\item[$^{\mathrm{a}}$] Obtained from LAMDA, CDMS and NASA/JPL databases.
\item[$^{\mathrm{b}}$] The flux errors include the statistical uncertainties of the instrument and a 15\% of calibration uncertainty, adopted for the 5$\times$5 spaxel spectra.
\item[$^{\mathrm{c}}$] Luminosity estimated assuming a Flat Space Cosmology (H$_0$=70~\kms\ Mpc$^{-1}$, $\Omega_{\lambda}$=0.73, $\Omega_M$=0.27)
  and a distance of $3.5\pm0.2$~Mpc for NGC~253 \citep{rekola05}. The luminosity errors include the relative uncertainty of the respective fluxes and the distance of the galaxy, as well as a 5\% uncertainty for the assumed Cosmology model.
\end{list}

\end{table}
%

   \begin{table}[htp]
         \centering
      \caption{\footnotesize{Fluxes of the absorption lines extracted from the PACS 5$\times$5 spectrum of NGC~253.}}\label{tab:pacs-absorption-fluxes}
         \tabcolsep 2.0pt
         \scriptsize
         \begin{tabular}{lcccc}
	    \hline\hline
	    \noalign{\smallskip}
            Line   &    $\lambda_{\rm rest}^{~\mathrm{a}}$   &    $\nu_{\rm rest}^{~\mathrm{a}}$   &        Flux$^{~\mathrm{b}}$       &  Luminosity$^{~\mathrm{c}}$  \\
                   &              ($\mu$m)              &              (GHz)              &  ($10^{-16}$~\Wm) &      ($10^{4}$~\Lsun)        \\                   
            \noalign{\smallskip}
            \hline
            \noalign{\smallskip}
            
 
 
 
            \ph2o\ $3_{22}-2_{11}$  &  89.988  &  3331.479  &  -6.04$\pm$1.19  &  -22.9$\pm$4.8  \\
            \ph2o\ $3_{31}-2_{20}$  &  67.090  &  4468.512  &  -4.20$\pm$0.92  &  -15.9$\pm$3.7  \\
 
            \noalign{}\\
 
            \oh2o\ $2_{12}-1_{01}$  &  179.526  &  1669.906  &  -15.40$\pm$2.54  &  -58.4$\pm$10.6  \\
            \oh2o\ $2_{21}-1_{10}$  &  108.070  &  2774.058  &  -4.85$\pm$1.00  &  -18.4$\pm$4.1  \\
            \oh2o\ $3_{21}-2_{12}$  &  75.380  &  3977.082  &  -15.05$\pm$2.47  &  -57.1$\pm$10.3  \\
            \oh2o\ $3_{30}-2_{21}$  &  66.440  &  4512.228  &  -6.87$\pm$1.43  &  -26.1$\pm$5.8  \\
            \oh2o\ $4_{32}-3_{21}$  &  58.700  &  5107.197  &  -2.72$\pm$0.92  &  -10.3$\pm$3.6  \\
 
            \noalign{}\\
 
            OH $_{\tfrac{3}{2}~\tfrac{5}{2}-\tfrac{3}{2}~\tfrac{3}{2}}$  &  119.440  &  2509.984  &  -60.74$\pm$9.69  &  -230.5$\pm$40.7  \\
            OH $_{\tfrac{3}{2}~\tfrac{5}{2}-\tfrac{3}{2}~\tfrac{3}{2}}$  &  119.230  &  2514.405  &  -67.72$\pm$10.84  &  -257.0$\pm$45.5  \\
            OH $_{\tfrac{3}{2}~\tfrac{7}{2}-\tfrac{3}{2}~\tfrac{5}{2}}$  &  84.600  &  3543.646  &  -10.24$\pm$3.60  &  -38.9$\pm$14.0  \\
            OH $_{\tfrac{3}{2}~\tfrac{7}{2}-\tfrac{3}{2}~\tfrac{5}{2}}$  &  84.420  &  3551.202  &  -10.57$\pm$2.01  &  -40.1$\pm$8.2  \\
 
            \noalign{}\\
 
            H$^{18}$O $_{\tfrac{1}{2}~\tfrac{1}{2}-\tfrac{3}{2}~\tfrac{3}{2}}$  &  79.080  &  3791.002  &  -3.56$\pm$2.90  &  -13.5$\pm$11.1  \\
            CH $2_{\tfrac{3}{2}~2-}-1_{\tfrac{1}{2}~1+}$  &  149.390  &  2006.771  &  -9.66$\pm$1.52  &  -36.7$\pm$6.4  \\
            CH $2_{\tfrac{3}{2}~2+}-1_{\tfrac{1}{2}~1-}$  &  149.092  &  2010.787  &  -9.82$\pm$1.54  &  -37.3$\pm$6.5  \\

            
	    \noalign{\smallskip}
	    \hline
	  \end{tabular}

\begin{list}{}{}
\scriptsize
\item[$^{\mathrm{a}}$] Obtained from LAMDA, CDMS and NASA/JPL databases.
\item[$^{\mathrm{b}}$] The flux errors include the statistical uncertainties of the instrument and a 15\% of calibration uncertainty, adopted for the 5$\times$5 spaxel spectra.
\item[$^{\mathrm{c}}$] Luminosity estimated assuming a Flat Space Cosmology (H$_0$=70~\kms\ Mpc$^{-1}$, $\Omega_{\lambda}$=0.73, $\Omega_M$=0.27)
  and a distance of $3.5\pm0.2$~Mpc for NGC~253 \citep{rekola05}. The luminosity errors include the relative uncertainty of the respective fluxes and the distance of the galaxy, as well as a 5\% uncertainty for the assumed Cosmology model.
\end{list}

\end{table}
%

   \begin{table*}[htp]
            \centering
      \caption{\footnotesize{Line widths and fluxes$^{\mathrm{a}}$ from HIFI spectra of NGC~253.}}
         \label{tab:hifi-fluxes}
         \tabcolsep 5.8pt
         \scriptsize
         \begin{tabular}{lccccc}
	    \hline\hline
	    \noalign{\smallskip}
            Line   &    FWHM$^{~\mathrm{b}}$   &    FWHM$^{~\mathrm{c}}$   &       Intensity$^{~\mathrm{d}}$         &        Flux$^{~\mathrm{c}}$         & Instruments  \\
                   &         (\kms)            &         (\kms)            &           (\Kkms)         &           ($10^{-16}$~\Wm)          &  Ratio   \\
	    \noalign{\smallskip}
	    \hline
	    \noalign{\smallskip}

            \twco\ $J = 5\rightarrow4$  &  195.8$\pm$24.6  &                     &  215.9$\pm$27.1   &    \\
            \twco\ $J = 6\rightarrow5$  &  188.3$\pm$23.7  &                     &  201.4$\pm$25.3   &    \\
            \twco\ $J = 9\rightarrow8$  &  157.8$\pm$19.8  &  164.3$\pm$20.7  &  88.4$\pm$11.1   &  14.6$\pm$1.8 &  0.94$^{~\mathrm{e}}$ \\

            \noalign{\smallskip}

            \thco\ $J = 5\rightarrow4$  &  171.9$\pm$21.6 &                     &  12.0$\pm$1.5    &   \\
            \thco\ $J = 6\rightarrow5$  &  191.8$\pm$24.3  &                     &  12.9$\pm$1.6   &    \\
            \thco\ $J = 9\rightarrow8$$^{~\mathrm{h}}$  &  --  &                     &  --   &    \\

            \noalign{\smallskip}

            \CiPone\                    &  193.7$\pm$24.4   &                     &  71.7$\pm$9.0  &                     &  \\
            \CiPtwo\                    &  174.0$\pm$21.9   &  184.5$\pm$23.2  &  98.9$\pm$12.4  &  10.7$\pm$1.3  & 0.86$^{~\mathrm{e}}$ \\
            
            \noalign{\smallskip}

            \cii\                    &  195.7$\pm$24.6   &  205.7$\pm$25.9  &  637.5$\pm$80.2  &  406.7$\pm$51.2   & 0.79$^{~\mathrm{f}}$  \\

            \noalign{\smallskip}
            
\cii\ & & & 264.9$\pm$33.0 & 56.6$\pm$7.0 & 1.45$^{~\mathrm{g}}$ \\

	    \noalign{\smallskip}
	    \hline
	  \end{tabular}

\begin{list}{}{}
\scriptsize
\item[$^{\mathrm{a}}$] The errors quoted include the r.m.s. obtained from the baseline subtraction and uncertainties of 6\% in the side band ratio, 3\% in the planetary model, 10\% in the beam efficiency, 2\% in the pointing, and 3\% in the correction for standing waves.
\item[$^{\mathrm{b}}$] FWHM obtained from a single component Gaussian fit of the corresponding HIFI spectrum.
\item[$^{\mathrm{c}}$] FWHM and Flux convolved to a 40$"$ HPBW from the \twco~$J$=9-8 and \CiPtwo\ HIFI maps.
\item[$^{\mathrm{d}}$] Velocity integrated temperatures obtained from the Gaussian fit of the HIFI spectra at their respective beam sizes.
\item[$^{\mathrm{e}}$] Ratio between the HIFI and SPIRE fluxes for the 40$"$ HPBW spectra.
\item[$^{\mathrm{f}}$] Ratio between the HIFI and PACS fluxes for the 40$"$ HPBW spectra.
\item[$^{\mathrm{g}}$] Ratio between the HIFI and upGREAT fluxes observed at about offset position (-11\farcs5, -8\farcs2) for the 15\farcs1 HPBW spectra.
\item[$^{\mathrm{h}}$] The \thco~$J = 9\rightarrow8$ line was observed but not clearly detected since the spectrum is severely affected by standing waves.
\end{list}

\end{table*}

\begin{figure*}[htp]
\centering
\hspace{-0.0cm}\includegraphics[angle=0,width=0.33\textwidth]{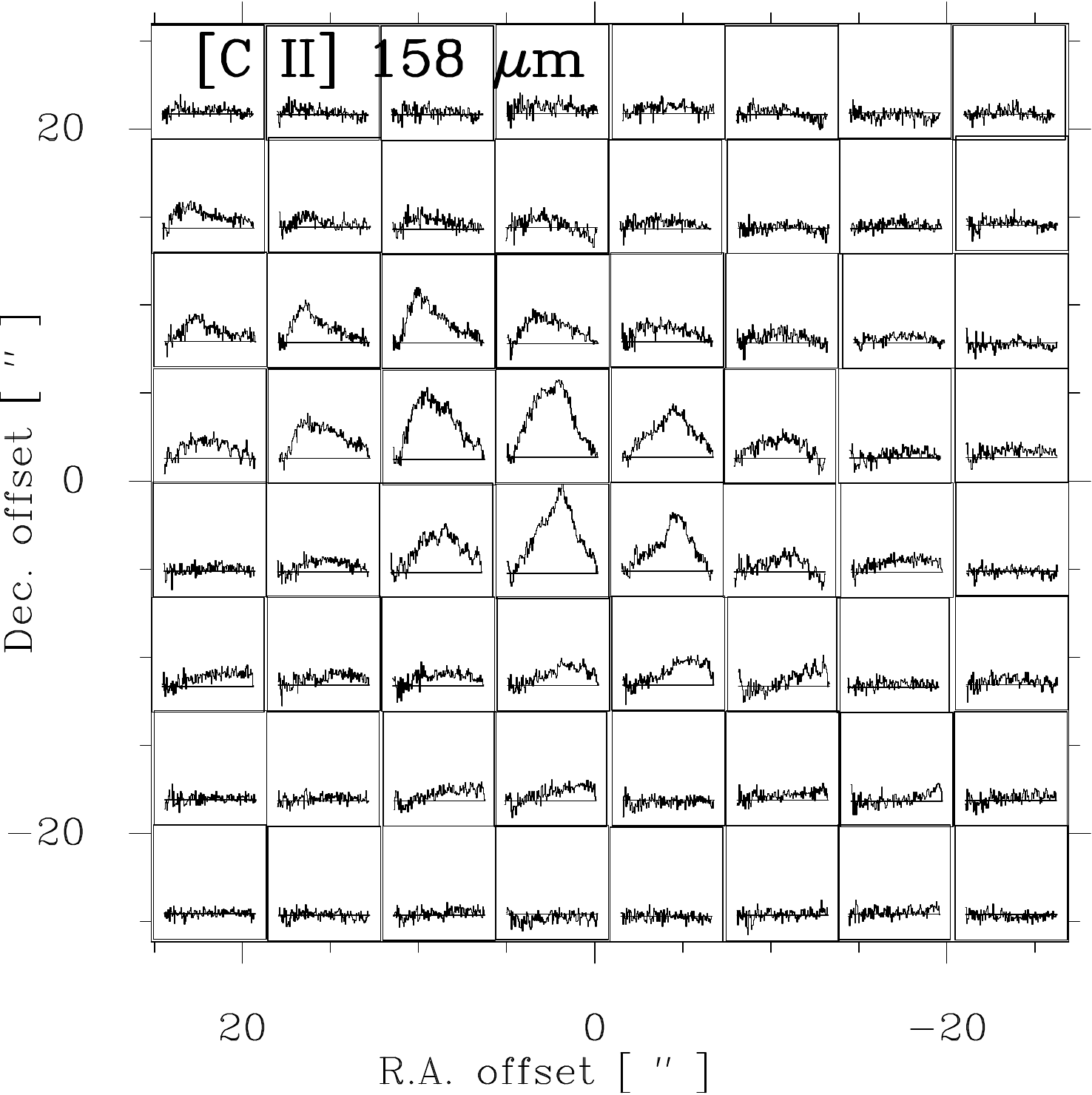}%
\hspace{-0.0cm}\includegraphics[angle=0,width=0.33\textwidth]{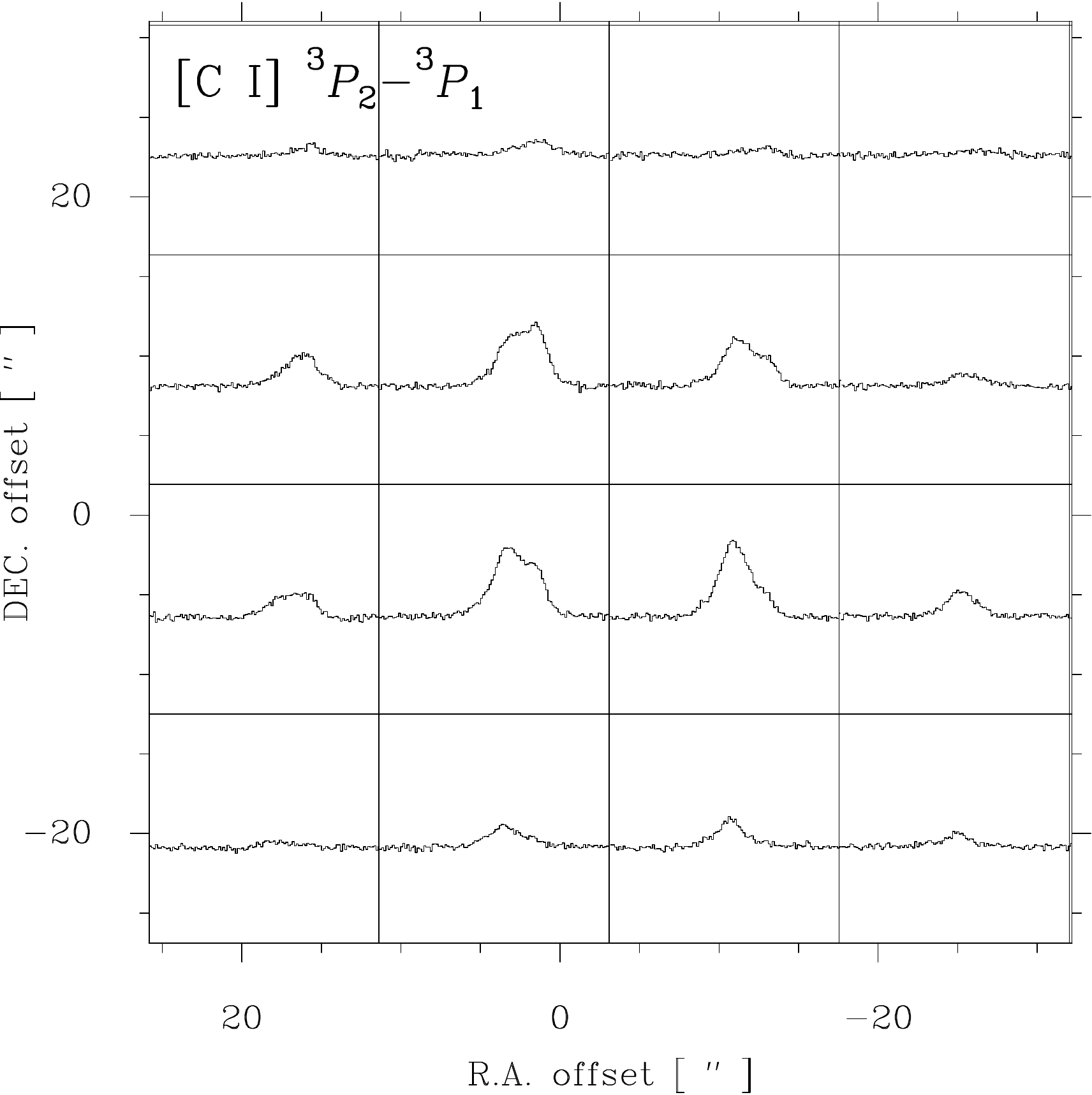}%
\hspace{-0.0cm}\includegraphics[angle=0,width=0.33\textwidth]{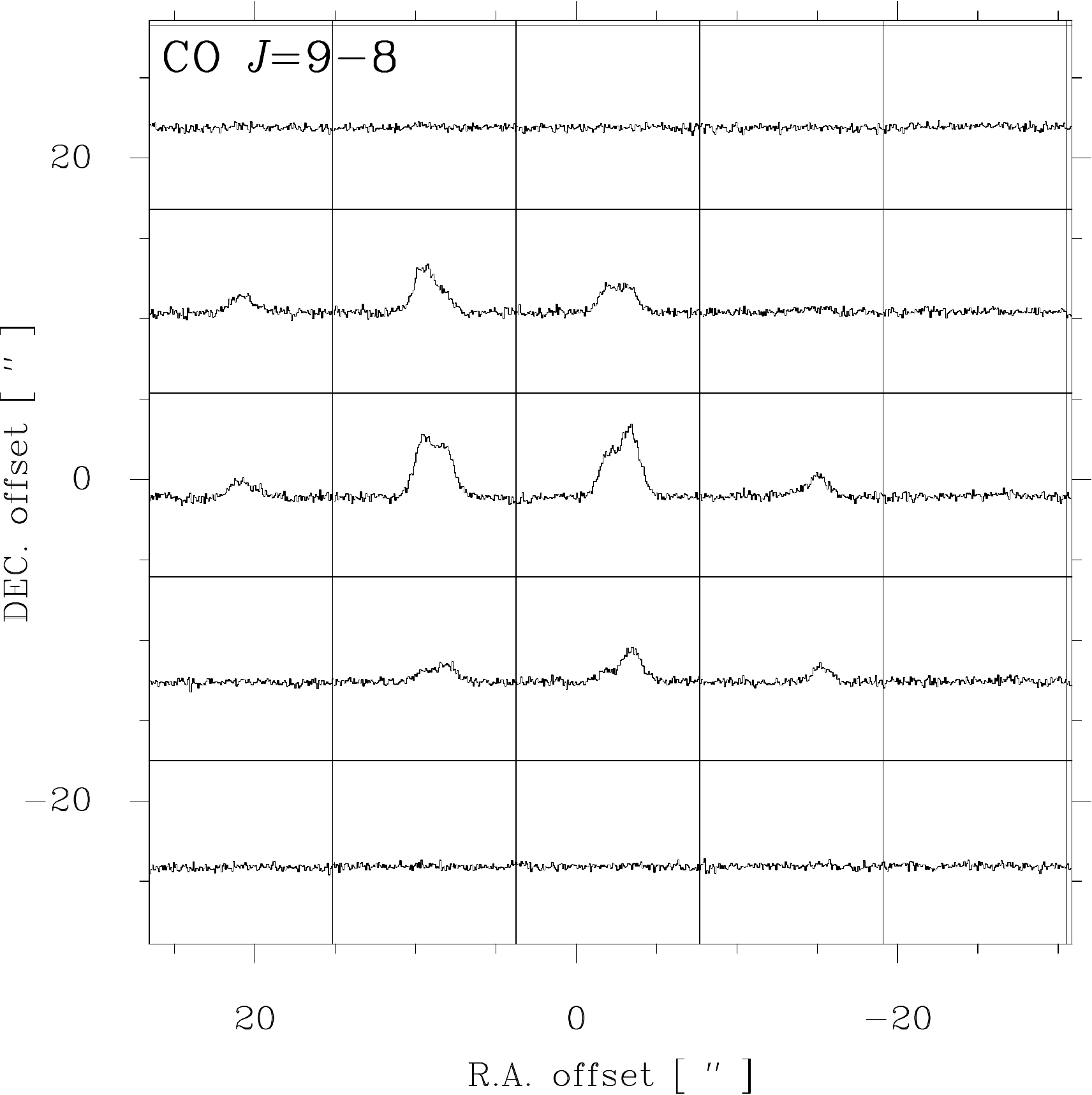}%

\hspace{-0.0cm}\includegraphics[angle=0,width=0.33\textwidth]{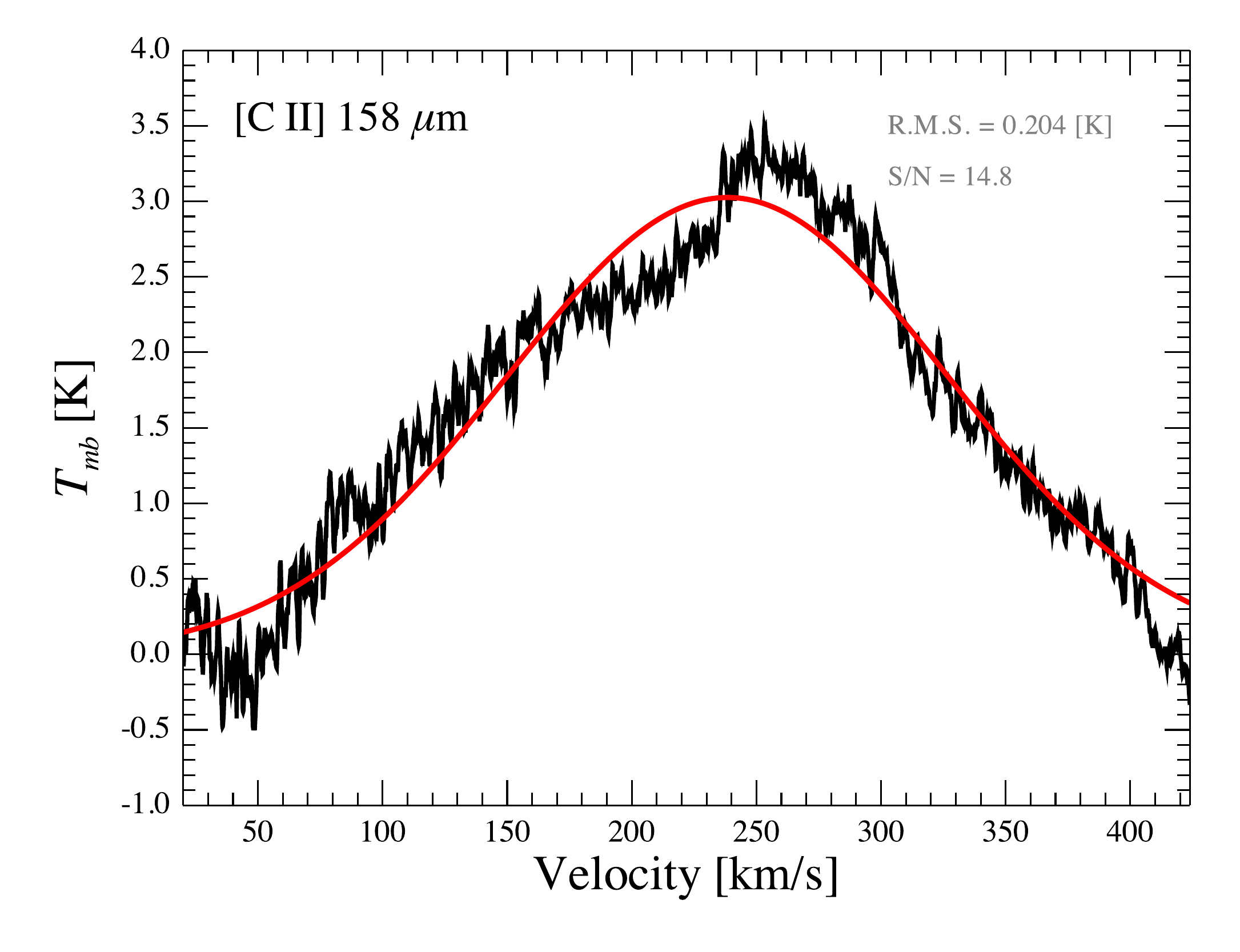}%
\hspace{-0.0cm}\includegraphics[angle=0,width=0.33\textwidth]{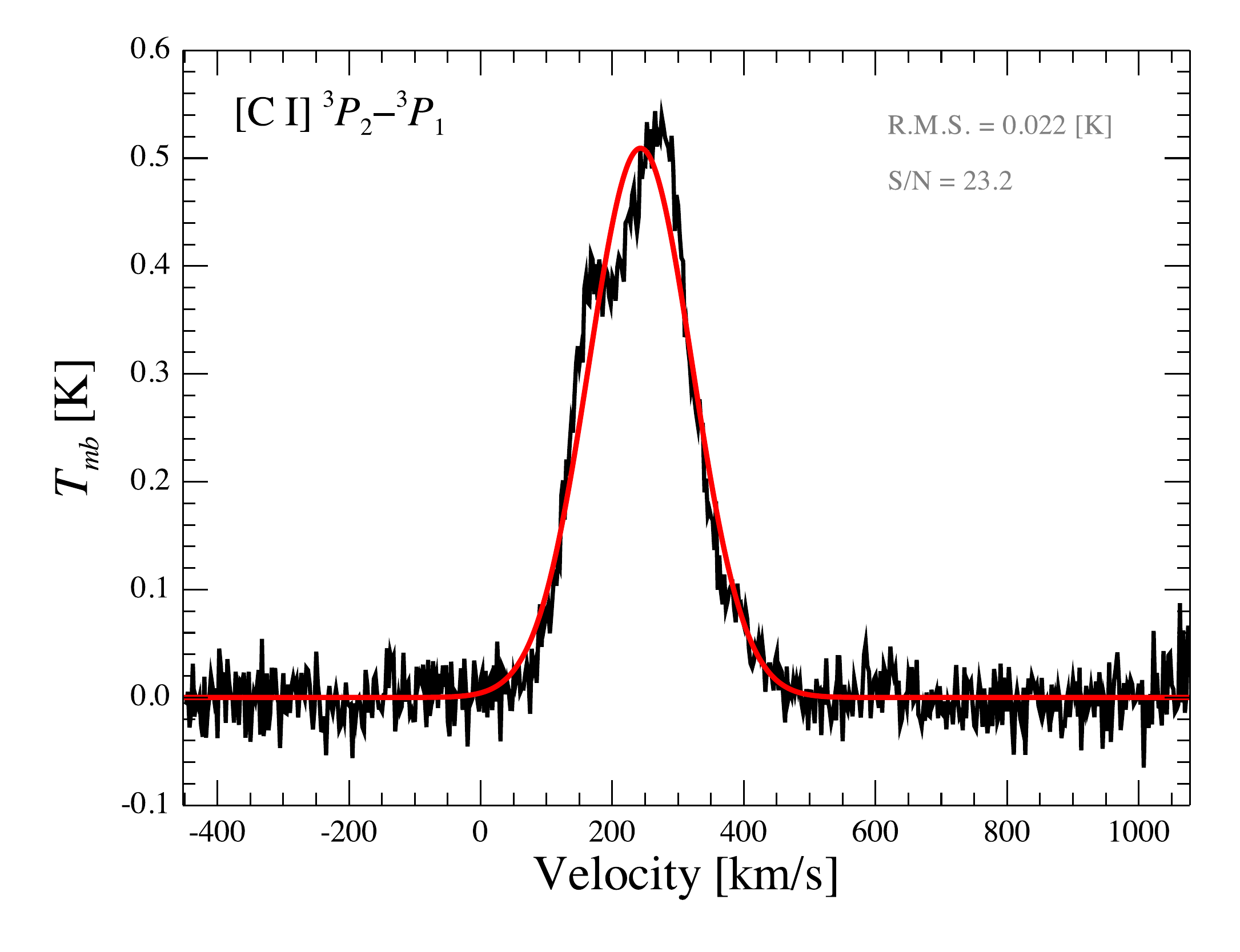}%
\hspace{-0.0cm}\includegraphics[angle=0,width=0.33\textwidth]{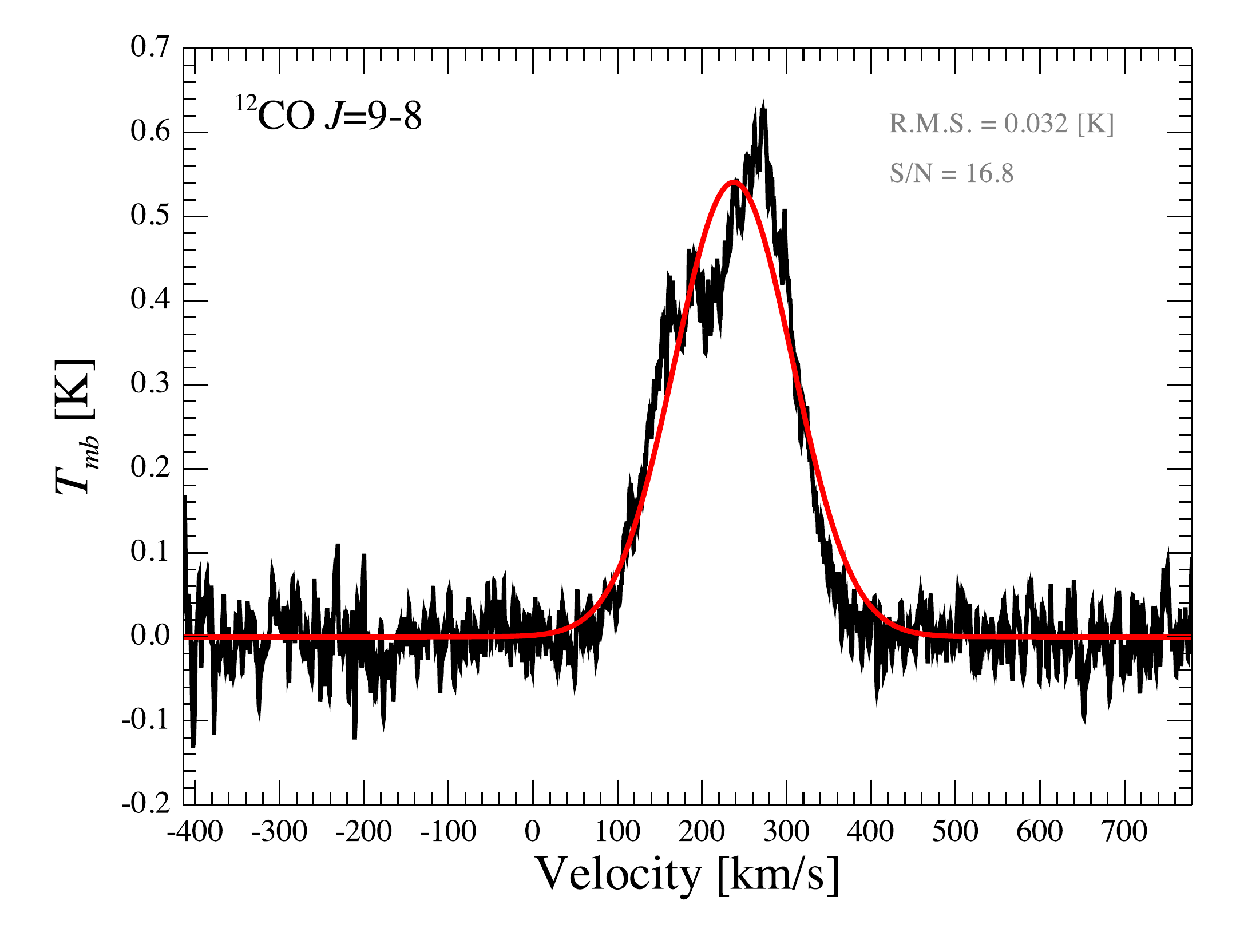}%

\vspace{-0.4cm}

\caption{{\footnotesize \textit{Top panels} - HIFI maps of \cii, \CiPtwo\ and \twco\ $J=9\rightarrow8$ in NGC253. For clarity each grid cell show the average spectrum between the vertical and horizontal polarizations. The central (0\arcsec, 0\arcsec) position correspond to the R.A.(J2000) = $00^h 47^m 33.12^s$ and Dec(J2000) = $-25^{\circ} 17\arcmin 17\farcs6$ coordinates.  \textit{Bottom panels} - Spectra obtained convolving the maps above with a 40$''$ HPBW. }}
\label{fig:hifi-spectral-maps}
\end{figure*}

\begin{figure*}[htp]
\centering
\hspace{-0.4cm}\includegraphics[angle=0,width=0.56\textwidth]{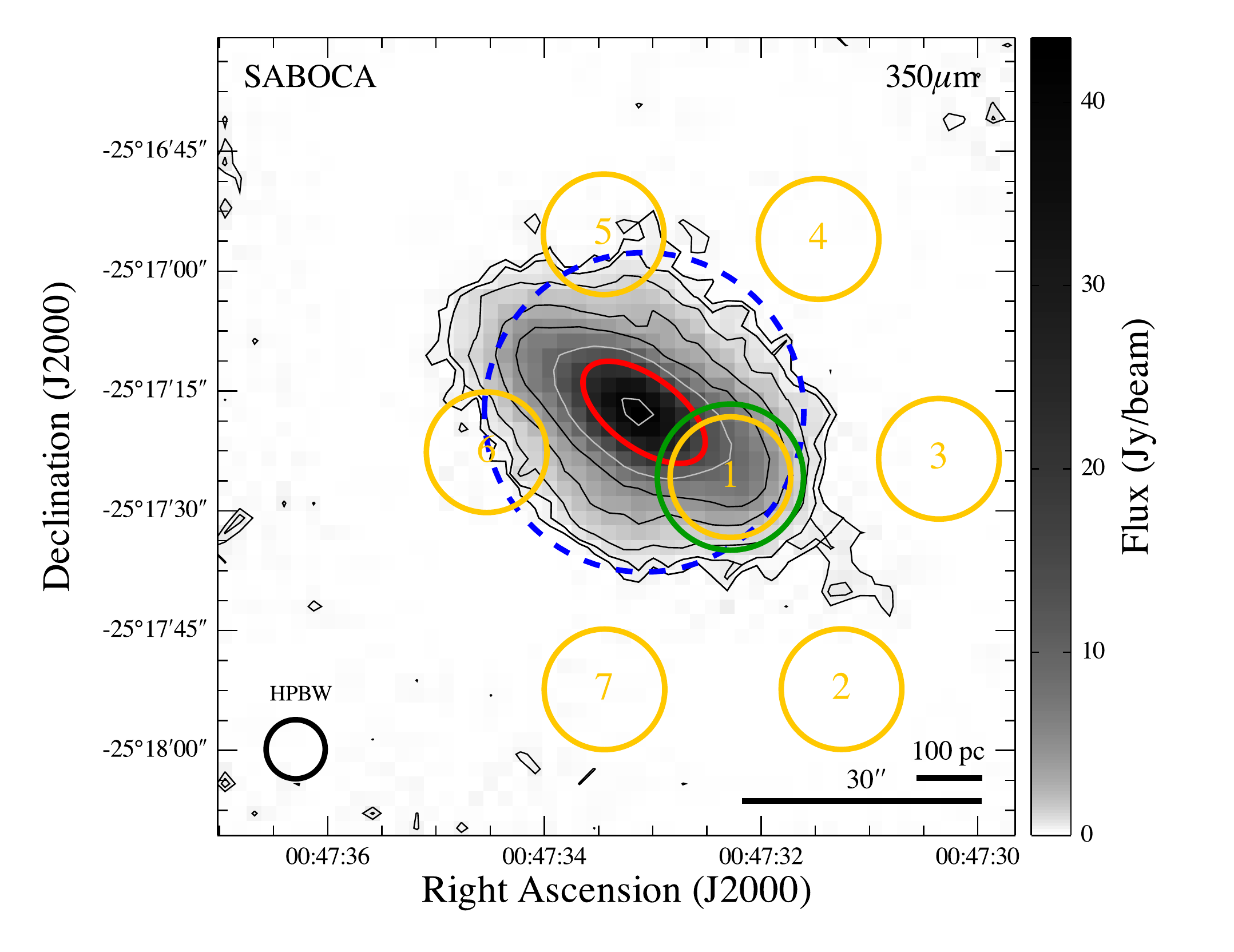}%
\hspace{+0.0cm}\includegraphics[angle=0,width=0.44\textwidth]{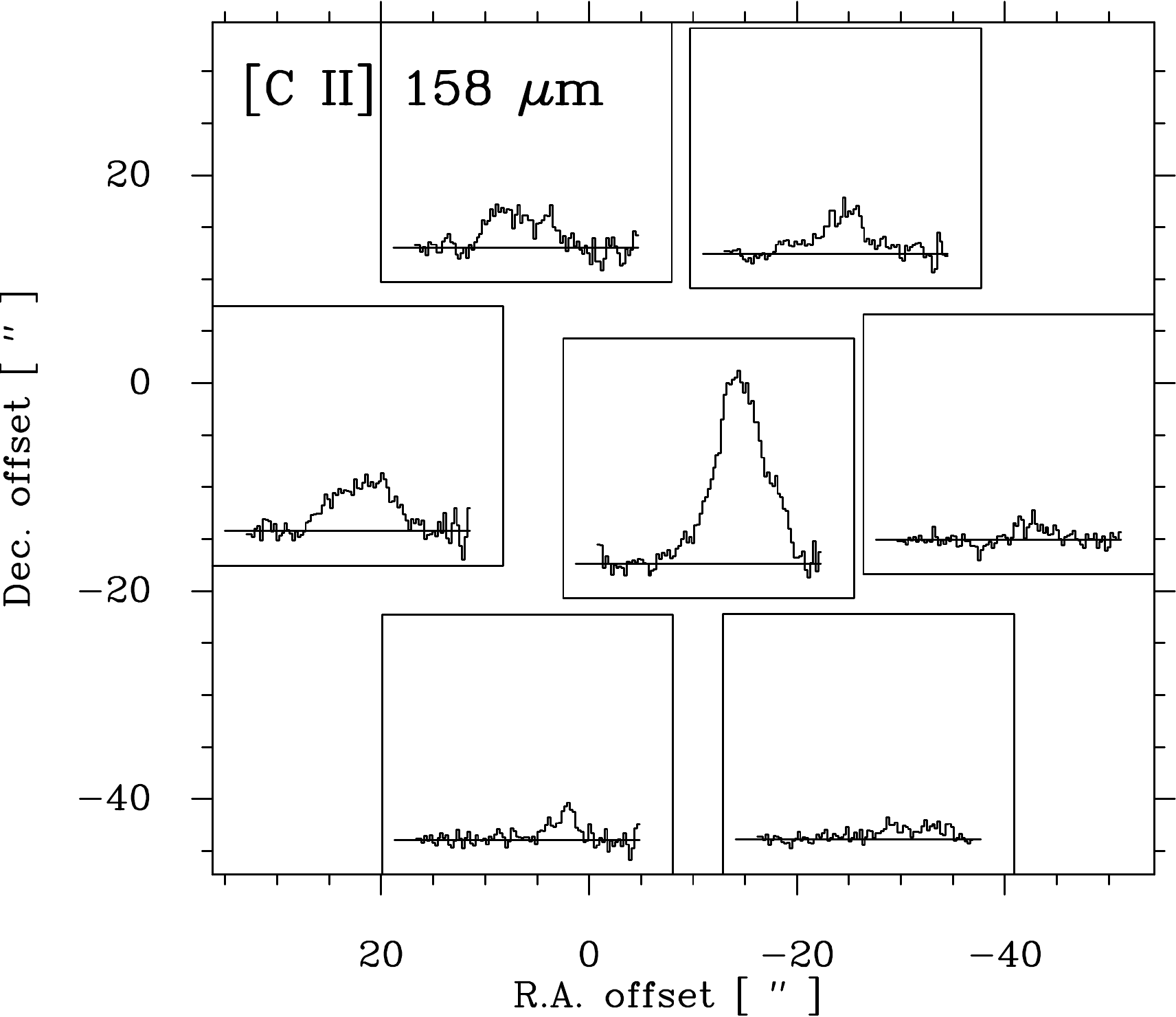}%

\vspace{-0.2cm}

\caption{{\footnotesize Footprint of the upGREAT multi-beam (seven pixels) receiver on SOFIA ({\it{left}} ) overlaid on the dust continuum emission obtained with SABOCA on APEX at 350~\mum, towards the nuclear region of NGC253. The beam size of upGREAT is 15\farcs1 at 1900.5 GHz. The \cii~158~\mum\ spectra of the seven pixels is shown on the {\it{right}} panel at the corresponding relative offset positions. The central position, i.e., the (0\arcsec, 0\arcsec) offset, is that used for the SABOCA map at R.A.(J2000) = $00^h 47^m 33.40^s$ and Dec(J2000) = $-25^{\circ} 17\arcmin 20\farcs7$. The spectra on the right panel expand the velocity range [-200~\kms, 600~\kms] and a $T_{\rm mb}$ scale from -0.2 K to 1.3 K.}}
\label{fig:upGREAT-CII}
\end{figure*}

\begin{figure}[!ht]
\centering
\hspace{0.0cm}\includegraphics[angle=0,width=0.45\textwidth]{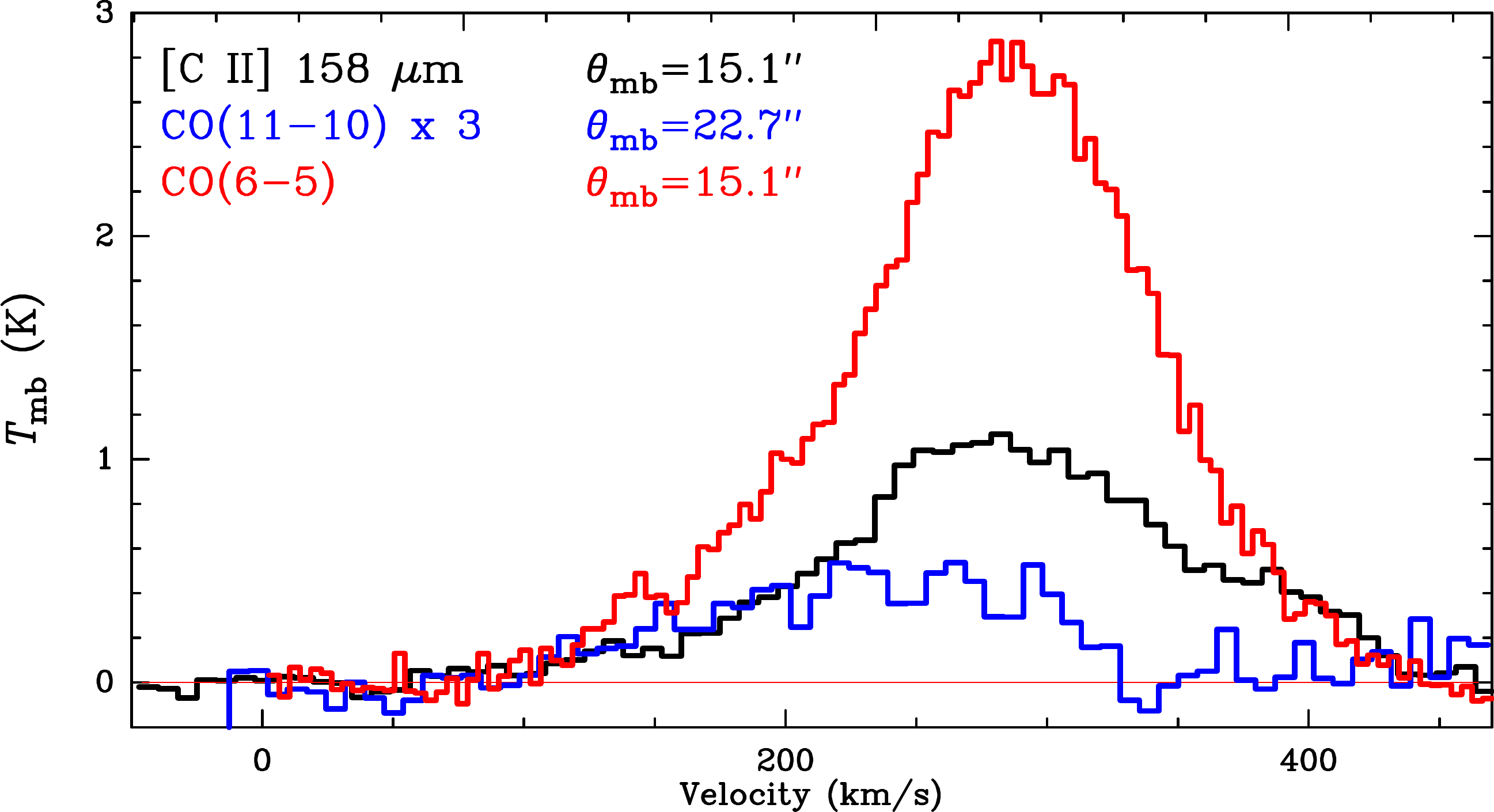}\\

\hspace{0.0cm}\includegraphics[angle=0,width=0.45\textwidth]{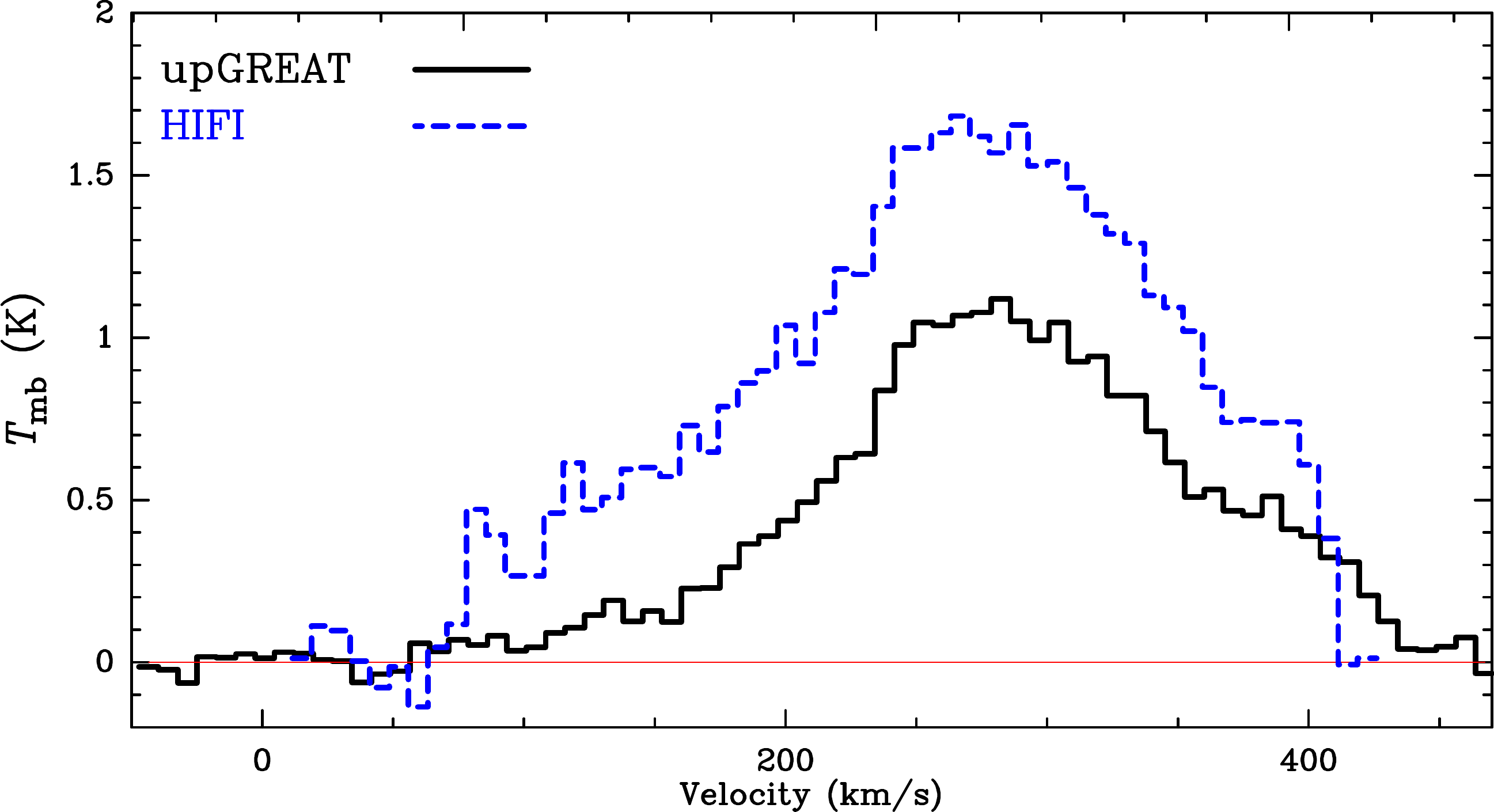}%

\vspace{-0.1cm}

\caption{{\footnotesize \textit{Top} - SOFIA/upGREAT spectra of the \cii~158~\mum\ fine-structure and \twco~$J=11\to10$ lines, compared with the \twco~$J=6\to5$ line obtained with APEX/CHAMP$^+$, as observed toward the offset position (-11\farcs5, -8\farcs2) south-west from the nuclear region (central pixel in Fig.~\ref{fig:upGREAT-CII}). The corresponding HPBWs are indicated. The CHAMP$^+$ map was convolved to the same resolution of the \cii\ beam of 15\farcs1. The \twco~$J=6\to5$ spectrum was extracted from the same position (within $\pm$2\arcsec) of the central pixel of upGREAT. For better visibility the fainter \twco~$J=11\to10$ spectrum was multiplied by a factor 3. \textit{Bottom} - SOFIA/upGREAT and HIFI spectra of the \cii~158~\mum\ fine-structure line. The HIFI spectrum is slightly closer to the central position at about (-10\farcs0, -7\farcs6), after regridding and convolving the HIFI \cii\ map of Fig.~\ref{fig:hifi-spectral-maps} to the 15\farcs1 HPBW resolution of SOFIA/upGREAT.}}
\label{fig:upGREAT-CHAMP}
\end{figure}

   \begin{table}[htp]
            \centering
      \caption{\footnotesize{Line fluxes from SOFIA/upGREAT and APEX/CHAMP+ toward the south-west position in the nuclear region of NGC~253.}}
         \label{tab:upGREAT-fluxes}
         \scriptsize
         \begin{tabular}{lcc}
	    \hline\hline
	    \noalign{\smallskip}
            Line   &   Intensity$^{~\mathrm{a}}$ &  Flux$^{~\mathrm{b}}$     \\
                     &     (K \kms)  & ($10^{-16}$~\Wm)        \\
	    \noalign{\smallskip}
	    \hline
	    \noalign{\smallskip}

\twco\ $J = 6\rightarrow5$ & 398.91$\pm$39.89 &  4.10$\pm$0.41 \\
\twco\ $J = 11\rightarrow10$ & 24.29$\pm$2.55 & 3.47$\pm$0.36 \\
\cii~158~\mum\ & 182.35$\pm$20.18 & 38.95$\pm$4.31 \\

	    \noalign{\smallskip}
	    \hline
	  \end{tabular}

\begin{list}{}{}
\scriptsize
\item[$^{\mathrm{a}}$] The errors quoted include the r.m.s. obtained from the baseline subtraction and 10\% accounting for calibration and pointing uncertainties.

\item[$^{\mathrm{b}}$] Fluxes were estimated using the corresponding beam sizes of 15\farcs1 for \cii\ and \twco~$J=6\to5$, and 22\farcs7 for \twco~$J=11\to10$.

\item[$^{\mathrm{b}}$] The \twco~$J=11\to10$ flux would be about 56\% smaller if a 15\farcs1 is considered instead.

\end{list}

\end{table}

\section{Results}\label{sec:results}

More than 60 lines (in absorption and emission) were detected and identified in the wavelength range (57~\mum --671~\mum) covered by both SPIRE and PACS.
The corrected SPIRE apodized spectrum of NGC~253 is shown in Fig.~\ref{fig:full-spire-lines}. All 
the line fitting and analysis, however, was done using the {\it unapodized} spectrum due to its 
higher spectral resolution, the less blended lines, and the more accurate fluxes obtained from 
fitting Sinc functions compared to the about 5\% less flux obtained when fitting Gaussians to the 
apodized spectrum (cf., SPIRE data reduction guide, Sect. 7.10.6 in version 3).
We detected 35 lines in the SPIRE spectra, including few unidentified lines not reported here. 
We detected 8 \hho\ lines in emission, and CH$^+$ $J=1\rightarrow0$, three OH$^+$ and two \ohhop\ lines in absorption, among others. The emission part of the OH$^+$ P-Cygni feature is more evident at 907 GHz than the $N=1-0$ line observed at 971~GHz, also observed and velocity resolved with HIFI \citep{vdtak16}.
The fluxes (in units of \Wm)  and the equivalent luminosities are summarized in Tables~\ref{tab:spire-emission-fluxes} and \ref{tab:spire-absorption-fluxes}.

We detected more than 30 lines in the PACS long range SED spectrum. The individual emission lines (including their corresponding 
Gaussian fit, R.M.S. and S/N ratio) are shown in Fig.~\ref{fig:pacs-emission-lines}. The absorption 
lines detected are shown in Fig.~\ref{fig:pacs-absorption-lines}.
In addition to several ionized species (\cii, \nii, \niii, and \oiii) we also detected five more 
\twco\ lines, extending the ladder observed with the SPIRE-FTS. We only detected an upper limit 
for the \twco\ $J=19\rightarrow18$, and the $J=17\rightarrow16$ is in a spectral region with very 
low S/N (with an uncertainty of 45\%), so we do not trust in the flux obtained for this transition.
We also detected two OH doublet lines in emission and two doublets in absorption, as well as 
H$^{18}$O in absorption.
The fluxes and luminosities of all the detected (and identified) PACS lines are listed in Tables~\ref{tab:pacs-emission-fluxes} and \ref{tab:pacs-absorption-fluxes}. There are 30 lines currently identified in the PACS spectra of NGC~253.

The HIFI maps of \twco\ $J=9\rightarrow8$, \CiPtwo\ and \cii\ are shown in Fig.~\ref{fig:hifi-spectral-maps}. The spectra obtained convolving the maps with an equivalent (HPBW) 40\arcsec\ beam is also shown in order to compare them with the corresponding SPIRE and PACS data. 
Although the horizontal and vertical polarization spectra were used independently to create the final maps, the grid maps of Fig.~\ref{fig:hifi-spectral-maps} show the average spectrum of the two polarizations for clarity. 
The spectra of the single pointing observations, at their respective beam resolutions, can be found in Fig.~\ref{fig:hifi-single-point-spectra} (Appendix~\ref{sec:appendix-HIFI-single}).

Even though we clearly see more than one component in the velocity resolved HIFI lines, we fit a single Gaussian component to the spectra, since this fit is good enough to extract the total flux and the width (FWHM) of the lines. In Table~\ref{tab:hifi-fluxes} we list the FWHM and velocity integrated temperatures (in $T_{mb}$ scale). For the HIFI maps we also list the FWHM and total flux (\Wm) of the spectra convolved with an equivalent (HPBW) 40$''$ beam, as well as the flux ratio for the lines that were observed with other instruments.

\begin{figure}[tp]
\centering
\hspace{-0.0cm}\includegraphics[angle=0,width=0.5\textwidth]{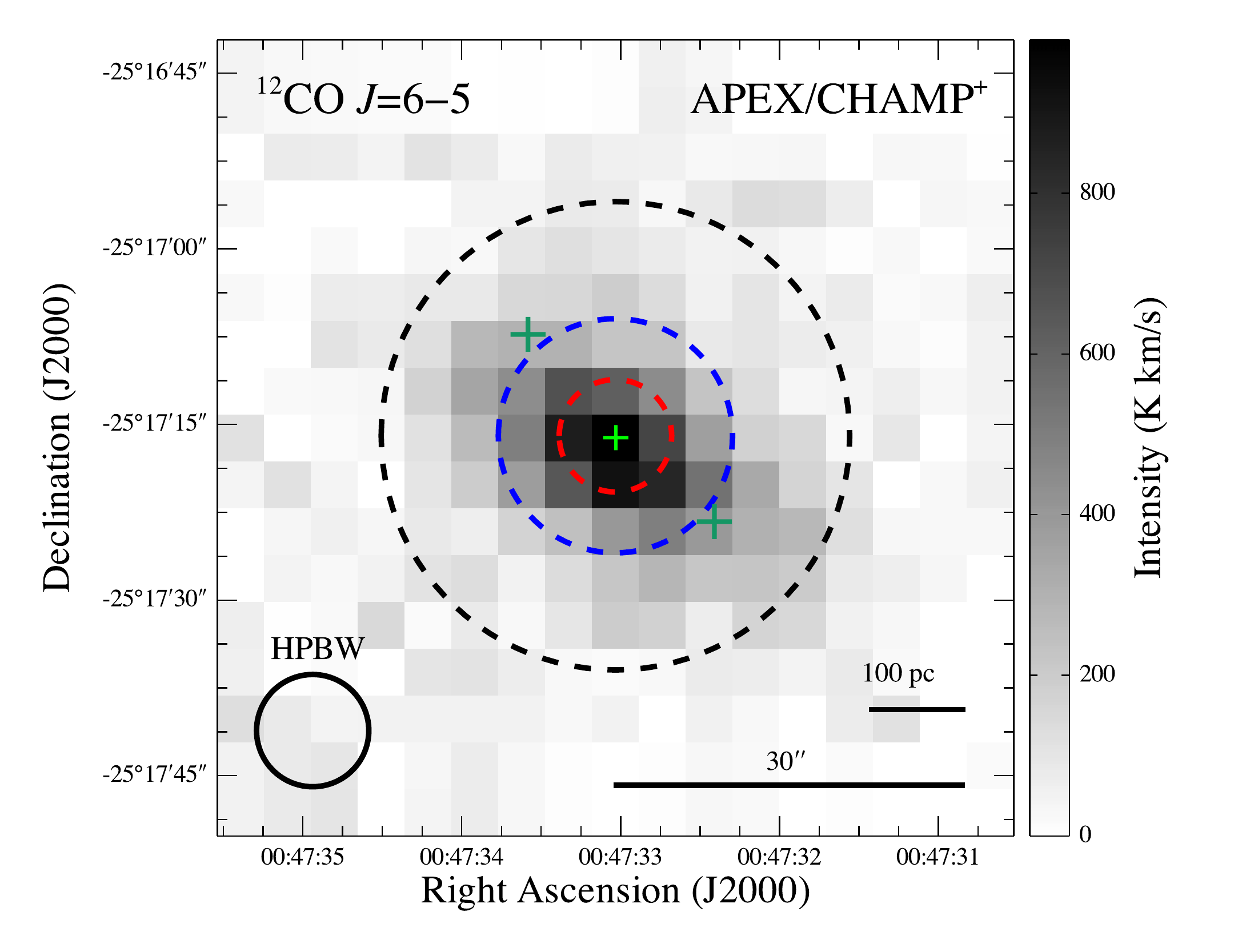}%

\vspace{-0.3cm}

\hspace{-0.0cm}\includegraphics[angle=0,width=0.45\textwidth]{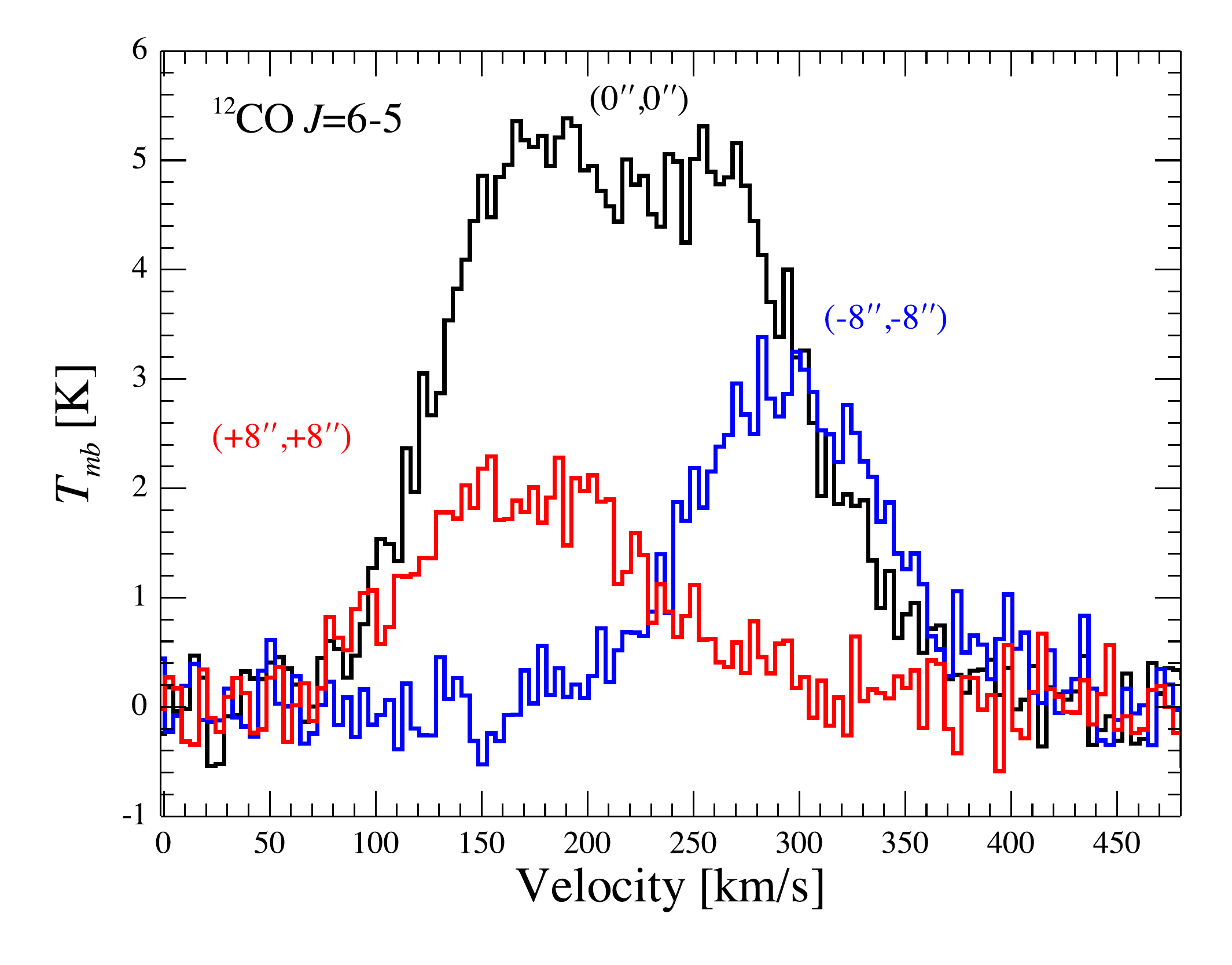}%

\vspace{-0.4cm}

\hspace{-0.0cm}\includegraphics[angle=0,width=0.45\textwidth]{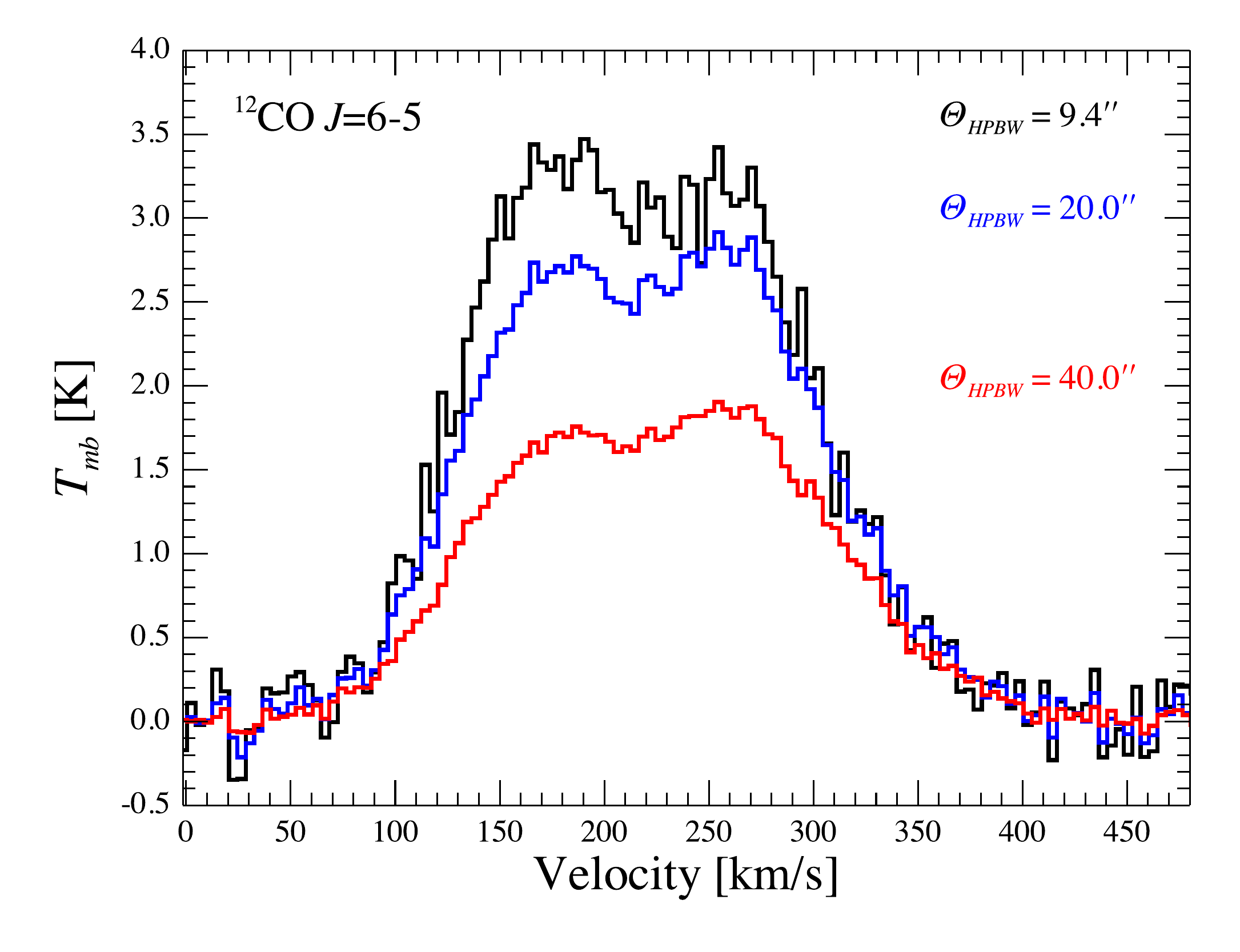}%

\vspace{-0.4cm}

\caption{{\footnotesize \textit{Top panel} - APEX/CHAMP$^+$ map of the \twco\ $J=6\rightarrow5$ 
emission in NGC253 at the original resolution (HPBW) of $\sim$9$''$.4. 
\textit{Middle panel} - Line profiles at offsets $\Delta\alpha=\Delta\delta=$+8$''$,0$''$ and 
-8$''$ (shown with crosses in the map, from top-left to bottom-right, respectively). \textit{Bottom 
panel} - Line profiles of the spectrum at offset (0$''$,0$''$) at the original resolution and after 
convolving the \twco\ $J=6\rightarrow5$ map with equivalent beams of HPBW 20$''$ and 40$''$ (shown in the map with the dashed circles, from the smallest to the largest, respectively).}
\label{fig:CO6-5_champ}}
\end{figure}

For the \cii\ map we obtained a source size of about 27$''$.9$\times$18$''$.3 (average of 
$\sim$23$''$.1) from a 2-D Gaussian fit. Unfortunately, the \ci\ and \twco\ maps are too small 
(only 5$\times$5 pixels) to fit a 2D-Gaussian (the fitting procedure does not converge). 
So we assumed that the \ci\ emitting region has the same size as that of \cii. 
In the case of line \twco\ $J=9\rightarrow8$, instead, we used the \twco\ $J=6\rightarrow5$ map, with a 
resolution (HPBW) of 9$''$.4, obtained with CHAMP$^+$ \citep{kasemann06, gusten08} on APEX 
(Fig.~\ref{fig:CO6-5_champ}), assuming the emitting region of these two lines have similar 
sizes (a discussion about the sizes can be found in the next section). A 2-D Gaussian fit of 
the $J=6\rightarrow5$ map gives a CO source size of 20$''$.8$\times$12$''$.5 
(or $\sim$16$''$.7 on average), which is consistent with the source size 
found from the SABOCA map (c.f. Fig.~\ref{fig:continuum-maps}) when considering a 
10\% uncertainty in the estimates.

   \begin{table}[tp]
            \centering
      \caption{\footnotesize{Line parameters of the APEX/CHAMP$^+$ \twco\ $J=6\rightarrow5$ spectrum.}}
         \label{tab:champ-CO6-5}
         \tabcolsep 5.8pt
         \scriptsize
         \begin{tabular}{cccc}
	    \hline\hline
	    \noalign{\smallskip}
          HPBW   &  FWHM$^{~\mathrm{a}}$  &  Intensity$^{~\mathrm{c}}$  &  Flux$^{~\mathrm{b}}$ \\\relax
        [${''}$]  &  (\kms)  &  (\Kkms)  &  ($10^{-16}$~\Wm)  \\
	    \noalign{\smallskip}
	    \hline
	    \noalign{\smallskip}
	    
 
	    
          9.4  &   173.4$\pm$20.4  &  1019.4$\pm$120.1  &     4.1$\pm$0.8 \\
         20.0  &   184.5$\pm$21.7  &   458.5$\pm$53.9  &     8.3$\pm$1.6 \\
         36.7  &   190.9$\pm$22.4  &   254.5$\pm$29.9  &    15.5$\pm$3.0 \\
         40.0  &   191.7$\pm$22.5  &   236.9$\pm$27.9  &    17.1$\pm$3.3 \\        
         
	    
	    
	    
         
	    \noalign{\smallskip}
	    \hline
	  \end{tabular}
	  
\begin{list}{}{}
\scriptsize
\item[$^{\mathrm{a}}$] The errors quoted include the r.m.s. obtained from the baseline subtraction in the original spectra, and uncertainties of 5\% in the calibration, 3\% in the planetary model, 10\% in the beam efficiency, and 2\% in the pointing. For the error in the flux, an additional 10\% of uncertainty was considered for the source size used to compute the flux in units of \Wm.
\end{list}	  

\end{table}
%

The fluxes reported in Table~\ref{tab:hifi-fluxes} were obtained using the average source sizes estimated for \twco\ $J=9\rightarrow8$, \ci, and \cii. 
These fluxes are, respectively, 6\%, 14\%, and 21\% smaller than the corresponding fluxes obtained with SPIRE and PACS. The larger difference in the PACS 
line may be because the PACS calibration could be affected by the bright \cii\ line and the 
continuum level around it (c.f. inset in Fig.~\ref{fig:dust-sed-fit}). On the other hand, the 
HIFI \twco\ $J=6\rightarrow5$ integrated intensity is only about 10 \Kkms\ (5\%) higher than the 
intensity obtained from the APEX/CHAMP$^+$ map convolved with the same equivalent beam (HPBW=36$''$.7) of 
HIFI at the \twco\ $J=6\rightarrow5$ frequency. Considering the uncertainties of the data from all the instruments, there is not significant difference between the fluxes of, for instance, \twco~$J=6\to5$ from APEX/CHAMP$^+$ and Heschel/SPIRE (Tables~\ref{tab:champ-CO6-5} and \ref{tab:spire-emission-fluxes}, respectively). 
Similarly, the fluxes of \twco~$J=9\to8$ obtained with SPIRE and HIFI (Tables~\ref{tab:spire-emission-fluxes} and \ref{tab:hifi-fluxes}, respectively) are practically the same, given the uncertainties. On the other hand, the \cii\ flux obtained with PACS is 21\% than that obtained with HIFI. Since the uncertainties of both instruments is similar (15\% and 13\% for PACS and HIFI, respectively), and given that the HIFI spectra of \cii\ do not have a fully covered baseline (due to the relatively short bandwidth of HIFI at $\sim$1.9~THz), we conclude that the \cii\ flux obtained with PACS is more reliable. Unfortunately we cannot yet compare the PACS flux with the SOFIA/upGREAT flux because we did not manage to map the full central region of NGC~253 in the \cii\ line with upGREAT. But we can compare the spectrum of the central pixel of the latter with the associated HIFI spectrum, as discussed below.

The CHAMP$^+$ fluxes, convolved with different beam sizes, are listed in Table~\ref{tab:champ-CO6-5}. 
The effect of the convolution on the line profiles is shown in the bottom panel of 
Fig.~\ref{fig:CO6-5_champ}. Contrary to what could be expected, the dynamical range (or full 
width at zero intensity) of the line remains the same ($\sim$420~\kms), while the FWHM widens 
with larger beams, not because the beams cover emitting regions with exceeding kinematical 
components not seen at the central region (as shown in Fig.~\ref{fig:CO6-5_champ}, middle 
panel), but just because the peak temperature of the line decreases (due to the beam smearing effect). 
This is an effect that needs to be taken into account when interpreting the line shapes from extragalactic observations.

The footprint of the SOFIA/GREAT/upGREAT observations and the spectra obtained are shown in Fig.~\ref{fig:upGREAT-CII}. The central pixel of the upGREAT array correspond to the \cii\ line observed at the offset position (-11\farcs5, -8\farcs2) south-west (SW) from the nuclear region of NGC~253. This location encloses part of the densest gas observed in the \hcn~$J=1\to0$ high resolution map reported by \citealt{paglione04}. The rotation of the gas can be seen in the (shifted to higher velocities) shape of the velocity resolved \cii, CO~$J=11\to10$ and CO~$J=6\to5$ lines shown in Fig.~\ref{fig:upGREAT-CHAMP}. The CO~$J=6\to5$ line observed with APEX/CHAMP$^+$ was convolved to the same 15\farcs1 resolution of the \cii\ line. We convolved the \cii\ map of HIFI to the 15\farcs1 HPBW of SOFIA/upGREAT for comparison. The obtained HIFI flux of \cii\ is about 45\% brighter than the value obtained from upGREAT. Such large difference may be due to the different calibration schemes, relative pointing errors (after regridding the HIFI spectrum is about (1\farcs5,0\farcs6) closer to the central region than the upGREAT spectrum), but mostly due to the difficulty and uncertainty of fitting a good baseline to the HIFI spectra due to its narrower instant bandwidth. We also note that the beam coupling efficiencies of these instruments differ significantly. While for SOFIA/upGREAT we estimated a 70\% efficiency, Herschel/HIFI achieved only 57\% efficiency when the \cii\ map was observed.


\section{The dust continuum properties}\label{sec:continuum}

The dust properties in NGC~253 have been investigated by \citet{radovich01},  \citet{melo02}, and \citet{weiss08}, using the mid and far-IR data from ISOPHOT, IRAS, and 
the submillimiter data from LABOCA on APEX \citep{siringo09}, respectively. We have 
reanalyzed the dust temperatures, mass, optical depths, and column densities, using the 
the SPIRE and PACS photometry fluxes, complemented at the shorter wavelengths by archival data from Spitzer/MIPS (24~\mum, AOR: 22610432) and MSX (21~\mum, 15~\mum, 12~\mum\ and 8~\mum, bands E, D, C and A, respectively; only these four images are available for NGC~253 in the MSX data archive\footnote{\url{http://irsa.ipac.caltech.edu/data/MSX/}}). 

Following \citet{weiss08}, and \citet{vlahakis05}, the dust emission was modeled using the grey body formulation
\begin{equation}\label{eq:dust-sed}
 S_{\nu} = \Big( 1-e^{-\tau_{\nu}} \Big) \Big(  B_{\nu}(T_i) - B_{\nu}(T_{\rm cmb})  \Big) \Omega_s \Phi_i,
\end{equation}

\noindent
where $B_{\nu}$ is the Planck function, $\tau_{\nu}$ the dust optical depth, $\Omega_s$ the source solid angle, $T_{\rm cmb}=2.73$~K the cosmic microwave background temperature, and $T_i$ and $\Phi_i$ the dust temperature and beam area filling factor of each component. 
The dust optical depth was computed as
\begin{equation}\label{eq:tau-dust}
 \tau_{\nu} = k_d({\nu})M_{dust}/ \left( D^2 \Omega_s \Phi_c \right),
\end{equation}

\noindent
where $M_{dust}$ is the dust mass, $D$ the distance to the source, $\Phi_c$ the filling factor of the coldest component, and following \citet{weiss08} and \citet{krugel94}, the adopted dust absorption coefficient $k_d({\nu})$ was
\begin{equation}\label{eq:dust-absorption-coefficient}
 k_d({\nu}) = 0.04(\nu/250 {\rm GHz})^{\beta}~~{\rm [m^2/kg]},
\end{equation}

\noindent
with $\nu$ in GHz and $\beta$=2 \citep{priddey01}. 
We used the flux observed at 500~\mum\ (the most optically thin emission in our data set) to compute the dust mass for each component
\begin{equation}\label{eq:mass-dust}
 M_{dust,i} = \frac{F_{500} D^2 }{k_{d,500}} \Big(  B_{500}(T_i) - B_{500}(T_{\rm cmb})  \Big)^{-1}.
\end{equation}

The total dust mass was estimated to be $M_{dust} = 3.0\pm 0.9\times 10^6$~\Msun, while the total gas mass is $M_{gas} = 4.5\pm1.3\times10^8$, assuming the same gas-to-dust mass ratio of 150 used by \citet{weiss08}.

and assuming a shorter distance of 2.5~Mpc. 


\begin{figure*}[!ht]
\centering
\hspace{-0.62cm}\includegraphics[angle=0,width=0.7\textwidth]{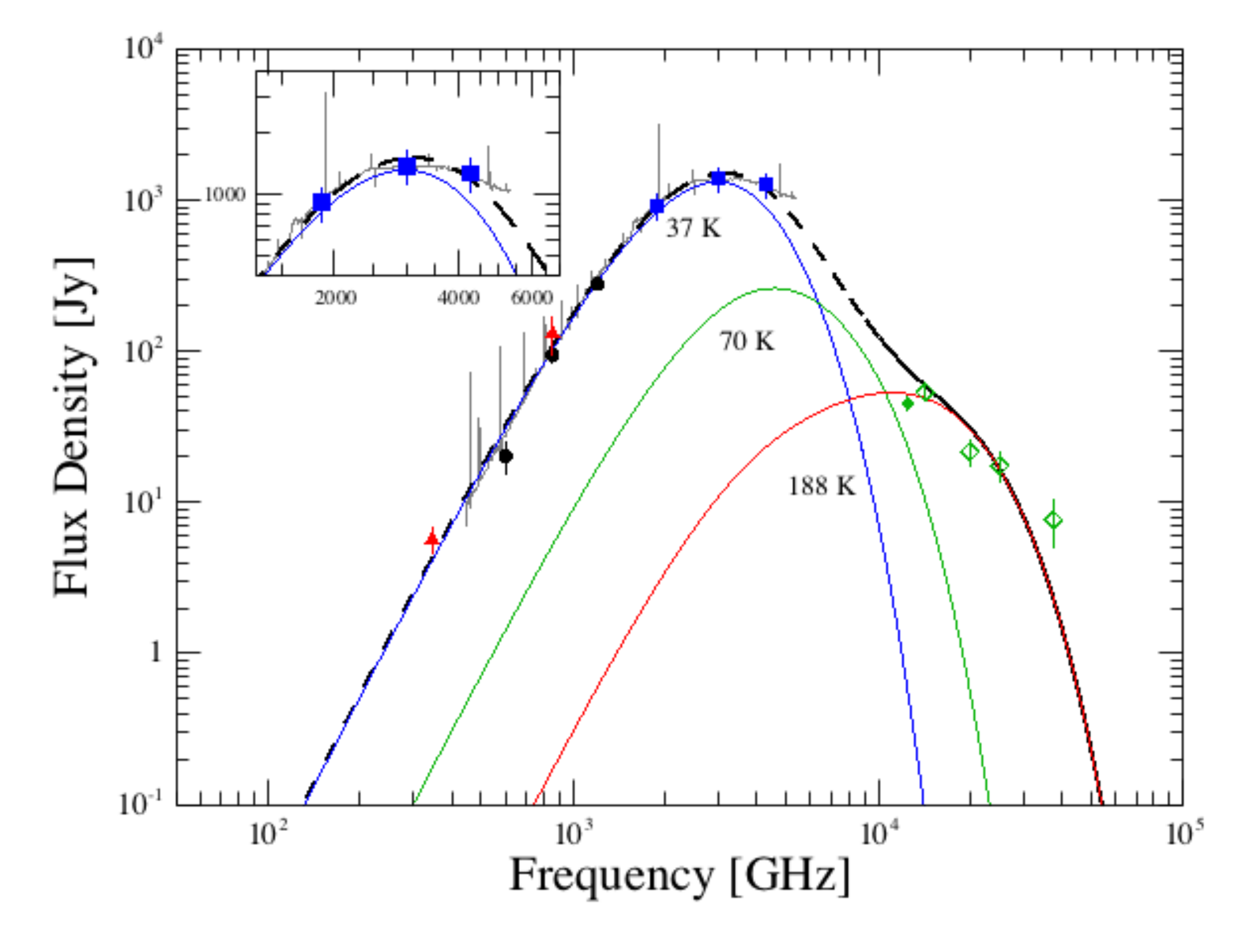}%

\vspace{-0.4cm}

\caption{{\footnotesize Spectral energy distribution of NGC253, obtained from the combined SPIRE 
and PACS spectra, corrected to a 40$''$ beam and from the 5$\times$5 spaxels corrected for point source 
losses, respectively. The continuum level of the combined spectra matches the equivalent 40$''$ 
aperture photometric fluxes at 870~\mum\ and 350~\mum\ of LABOCA and SABOCA (triangles), at 
500~\mum, 350~\mum\ and 250~\mum\ of SPIRE (circles), at 160~\mum, 100~\mum\ and 70~\mum\ of PACS 
(squares), and the total MIPS 24~\mum\ flux (diamond). The inset shows a zoom into the PACS spectra, where the continuum level do not follow the expected gray body signature around 1.9 THz ($\sim$160~\mum).}}
\label{fig:dust-sed-fit}
\end{figure*}

The dust temperatures and masses depend on the underlying source solid angle and the area filling factor of each component. 
We used the source size of 17$''$.3$\times$9$''$.2 (deconvolved from the SABOCA map). This is about half the size (30$''\times$17$''$) derived by \citet{weiss08} from a 80\arcsec\ beam. Note that Weiss \etal\ adopted a distance of 2.5~Mpc estimated assuming that the observed stars in NGC~253 were similar to the asymptotic giant branch (AGB) stars in Galactic globular clusters   \citep{davidge90, davidge91, houghton97}. Instead we used a more recent estimate of 3.5$\pm$0.2~Mpc based on models of planetary nebula accounting for dust \citep{rekola05}, which is consistent with estimates based on measurements of the magnitude of the tip of the red giant branch \citep{mouhcine05}. 


\begin{table}[!ht]
      \centering
   \caption{\footnotesize{Dust continuum SED fit parameters of NGC~253.}}
      \label{tab:sed-fit}
      \tabcolsep 5.8pt
      \scriptsize
      \begin{tabular}{lccc}
         \hline\hline
         \noalign{\smallskip}
            
	       & \multicolumn{3}{c} { Component Parameters } \\
          \cline{2-4}
          Quantity   &  Cold  &  Warm  &  Hot \\
                      
         \noalign{\smallskip}
         \hline
         \noalign{\smallskip}

         $\Phi^{\mathrm{a}}$   &  $5\times10^{-1}$  &  $1\times10^{-2}$  &  $1\times10^{-4}$  \\
         $T_{dust}$ [K]  &  36.6$\pm$3.7 &  70.0$\pm$7.0  &  187.7$\pm$37.5  \\
         $G_0$   &  3.4$\times10^{5}$  &  8.7$\times10^{6}$ &  1.2$\times10^{9}$ \\
         $M_{dust}$ [$10^6~M_{\odot}$]$^{\mathrm{b}}$  &  1.0$\pm$0.3  &  0.4$\pm$0.1  &  0.14$\pm$0.04 \\
    	    
         \noalign{\smallskip}
         \hline
	  \end{tabular}

\begin{list}{}{}
\scriptsize
\item[$^{\mathrm{a}}$] Uncertainties in the filling factors are of the order of 10\%.


\item[$^{\mathrm{b}}$] Dust mass obtained using a source size solid angle of $\Omega_s=17.3\times9.2$~arcsec$^2$ as obtained from a two-dimensional Gaussian intensity distribution fit of the SABOCA map.

\end{list}

\end{table}
%

\begin{table}[htp]
      \centering
   \caption{\footnotesize{Flux and dust properties at the observed wavelengths.}}
      \label{tab:sed-results}

      \tabcolsep 5.8pt
      \scriptsize
      \begin{tabular}{ccccccccc}
         \hline\hline
         \noalign{\smallskip}

         Wavelength  &  Observed Flux   &   $k_{\nu}^{\mathrm{a}}$    &   $\tau_{\nu}^{\mathrm{b}}$   \\\relax
         [$\mu \rm m$]  &       [Jy]  & [m$^2$ kg$^{-1}$] &                  \\
         \noalign{\smallskip}
         \hline
         \noalign{\smallskip}

         500  &    20.1$\pm$ 5.1 &  0.23  & 0.06$\pm$0.02  \\
         350  &    93.6$\pm$10.8 &  0.47  & 0.11$\pm$0.03  \\
         250  &   278.7$\pm$19.0 &  0.92  & 0.22$\pm$0.06  \\
         160  &   914.1$\pm$69.5 &  2.25  & 0.54$\pm$0.15  \\
         100  &  1383.4$\pm$102.7 &  5.75  & 1.39$\pm$0.39  \\
         70  &  1271.0$\pm$94.9 &  11.74  & 2.84$\pm$0.80  \\
         24  &    45.0$\pm$ 5.3 &  99.86  & 24.19$\pm$7.11  \\
         21  &    53.7$\pm$ 7.3 &  130.43  & 31.60$\pm$9.54  \\
         15  &    21.7$\pm$ 4.7 &  255.65  & 61.93$\pm$21.35  \\
         12  &    17.6$\pm$ 4.2 &  399.45  & 96.77$\pm$34.83  \\
          8  &     7.7$\pm$ 2.8 &  898.76  & 217.72$\pm$98.03  \\

         \noalign{\smallskip}
         \hline
	  \end{tabular}

\begin{list}{}{}
\scriptsize
\item[$^{\mathrm{a}}$] The uncertainties in the absorption coefficients are assumed to be of the order of 10\%.

\item[$^{\mathrm{b}}$] The errors in the optical depths consider the uncertainties of the temperatures of the three components and 10\% uncertainties in the source size solid angle and the corresponding uncertainty of the distance to NGC~253 $d=3.52\pm0.18$~Mpc \citep{rekola05}. 

\end{list}

\end{table}

The dust SED fit is shown in Fig~\ref{fig:dust-sed-fit} and the parameters are summarized in Table~\ref{tab:sed-fit}. The uncertainties of the dust 
temperatures consider a total of 10\% error for the source size and filling factors adopted for 
each component. Given the uncertainties, the temperature $\sim$37 K of the cold component is 
practically the same as that (30-35 K) found by \citet{weiss08}. Our second component, on the other hand, 
is considerably ($\sim$16 K) higher than the one found before. This may be due to the fact that we include fluxes at shorter wavelengths that were not used by \citet{weiss08} in their SED fit. Our third component, however, has the highest flux uncertainties of 30\% from the MSX data. This is because the flux at 21~\mum\ does not follow the trend of the MIPS 24~\mum\ flux (indicating a different flux scale between these instruments), and because the flux at 8~\mum\ and (to a lesser extend at 12~\mum) contain emission from PAHs \citep[e.g.][and references therein]{povich07}, which we did not correct for since they are difficult to asses in an unresolved source. Hence, the dust temperature of the third component should be consider an upper limit. Besides, a spectral index $\beta=2$, as assumed above, may not be appropriate for the warmest dust. Leaving it as free parameter only for the third component would lead to a poorly constrained value anyway because of the contamination by PAHs. Hence we did not investigate further on this matter.

Assuming FUV heating, we also estimated the FUV flux $G_0$, in units of the equivalent Habing flux (1.6$\times$10$^{-3}$~erg~cm$^{-2}$~s$^{-1}$), from \citet[][their eq.7]{hollenbach91} as 
\begin{equation}\label{eq:Go}
 G_0 = 3.7\times10^{-3}\tau_{100\mu \rm m}T_d^5,
\end{equation}

\noindent
using the dust opacity at 100~\mum\ ($\tau_{100\mu \rm m}\approx 1.4$) estimated from the dust SED fit of each component, and assuming that the dust temperatures are similar to the actual equilibrium dust temperatures ($T_0$) at the surface of their respective emitting regions.

Assuming most hydrogen is in molecular form, we can also estimate the molecular hydrogen column density from the dust opacity and absorption coefficient  following the formulation by \citet[][their eq.~A.9]{kauffmann08} as
\begin{equation}\label{eq:NH2}
 N({\rm H_2}) = 10^{-4}\times\frac{\tau_{\nu}}{ \mu_{{\rm H_2}} \it m_{{\rm H}} \it k_{\nu}}  ~~[{\rm cm}^{-2}],
\end{equation}

\noindent
where $m_{{\rm H}}$ is the hydrogen atom mass (in kg), and $\mu_{{\rm H_2}}$ is the molecular weight per hydrogen molecule. We use a value of 2.8 for the latter, which is the value needed to compute particle column densities. While the classical value of 2.33, used sometimes in the literature, actually correspond to the mean molecular weight per free particle ($\mu_{\rm p}$), which is used to estimate other quantities, like thermal gas pressure. The factor $10^{-4}$ is used 
to convert the dust absorption coefficient $k_{\nu}$ from units of m$^2$ kg$^{-1}$ (from eq.~\ref{eq:dust-absorption-coefficient}) to cm$^2$ kg$^{-1}$. Combining eq.~(\ref{eq:NH2}) with eq.~(\ref{eq:tau-dust}) and eq.~(\ref{eq:dust-absorption-coefficient}) we obtained $N(\rm{H_2}) = (5.2\pm2.3)\times10^{21}$~cm$^{-2}$.

We can also estimate the visual extinction $A_V$ (mag) from the standard conversion factor 
$N({\rm H_2}) = 9.4\times10^{20} A_V$ from which the atomic hydrogen 
column density can be estimated using the relation $N({\rm H}) = 2.21\times10^{21} A_V$ found 
for the Milky Way \citep{guver09}. We obtained $A_V = 5.5\pm2.5~mag$ and
$N(\rm{H}) = (1.2\pm0.5)\times10^{22}$~cm$^{-2}$. All observed fluxes, dust absorption coefficients and optical depths are summarized in Table~\ref{tab:sed-results}. 

\section{The HF absorption line}\label{sec:appendix-HF}

The formation of Hydrogen fluoride (HF) is dominated by a reaction of F with \hh\ making the HF/H$_2$ 
abundance ratio more reliably constant than \twco/H$_2$, specially for clouds of small extinction $A_v$ \citep{neufeld05}. Therefore HF has been proposed as a potentially sensitive probe of the total column 
density of the diffuse molecular gas \citep[e.g.,][]{neufeld05, monje11}.

 {Because of its very large $A$-coefficient ($A_{10}=2.42\times10^{-2}$ s$^{-1}$), this transition is generally 
observed in absorption \citep[e.g.,][]{neufeld97, neufeld05, phillips10, sonnentrucker10, 
neufeld10, monje11, rangwala11, kamenetzky12, pereira-santaella13}.}
This high $A$-coefficient translates into a simple excitation scenario, 
where most HF molecules are expected to be in the ground $J=0$ state from where they can be excited into the $J=1$ state by absorbing a photon at 1232.5~GHz under ambient 
conditions common to the diffuse and even dense ISM. Only an extremely dense region ($n({\rm H_2}) > 10^9~\3cm$, 
at $\sim$50 K), with a strong radiation field, could excite HF and generate a $J=1\rightarrow0$ feature in 
emission \citep[e.g.,][]{neufeld97, neufeld05, spinoglio12, pereira-santaella13, vdwerf10}.  {For a more extended reference list see \citet[][their Sect. 1]{vdWiel16}.}

\begin{figure}[ht]
\centering
\hspace{-0.6cm}\includegraphics[angle=0,width=0.48\textwidth]{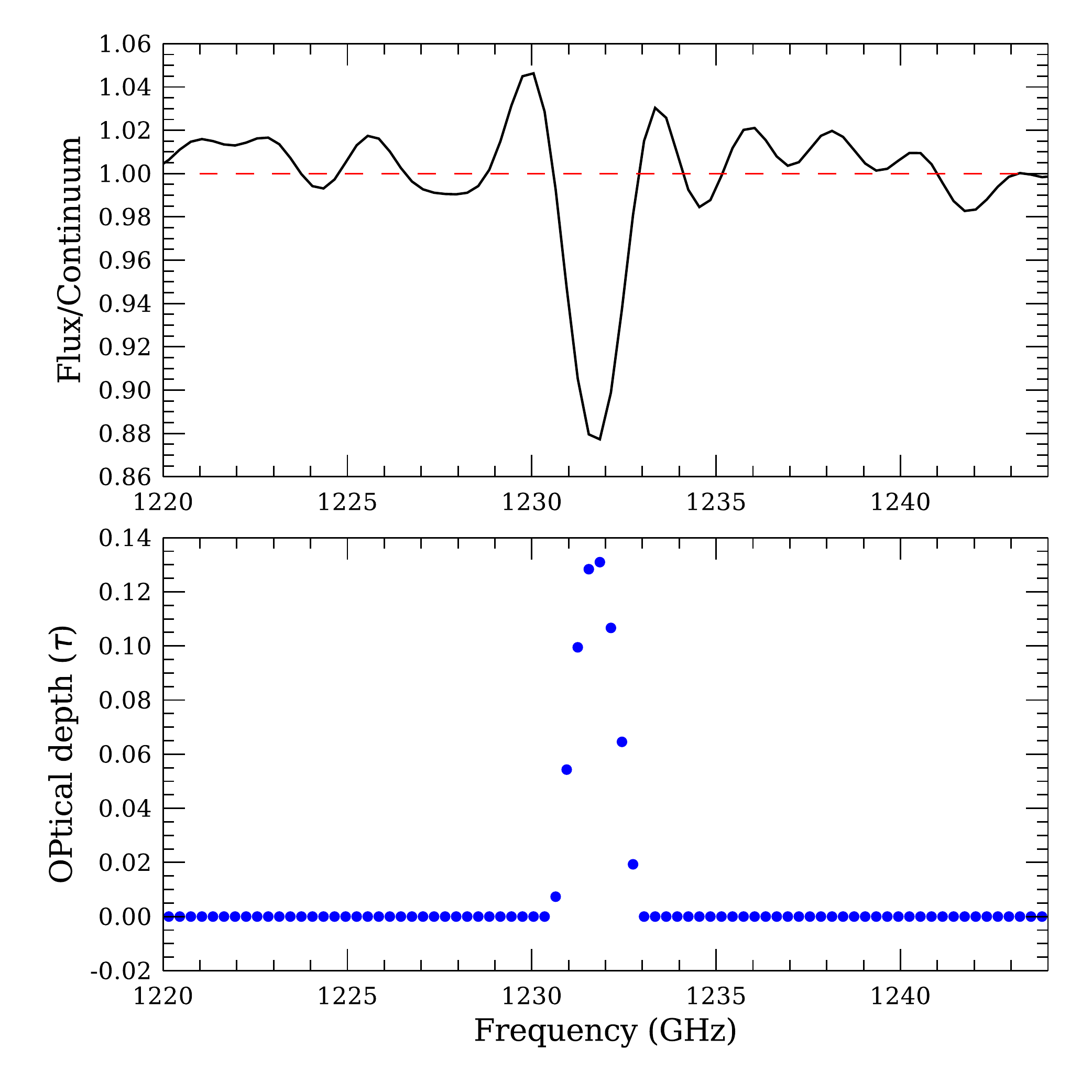}%

\vspace{-0.4cm}

\caption{{\footnotesize \textit{Top panel}: SPIRE spectrum of the $J=1\rightarrow0$ ground-state transition of 
HF toward the nuclear region of NGC~253. This corresponds to the unapodized residual spectrum normalized by the 
continuum, after subtracting the most prominent (\twco, \nii, and \hho) lines from the SSW band. \textit{Bottom 
panel}: Estimated optical depth of HF as a function of frequency.}} 
\label{fig:HF-column}
\end{figure}

From Eq.(3) in \citep{neufeld10}, and assuming all HF molecules are in the ground state, we can estimate the 
total HF column density from the absorption optical depth as:
\begin{equation}\label{eq:HF-column}
 \int \tau d v = \frac{A_{ul} g_u \lambda^3}{8 \pi g_l} N({\rm HF})
\end{equation}

\noindent
where $g_u = 3$ and $g_l = 1$, which yields $\int \tau d\nu = 4.16 \times 10^{-13} N({\rm HF})$ {cm$^2$ km 
s$^{−1}$}. The optical depth of HF can be estimated from a double side band (DSB) receiver as 
$\tau=−ln(2F_l/F_c-1)$, with $F_l/F_c$ the line-to-continuum ratio (Neufeld et al. 2010). In the case of SPIRE 
(a single side band spectrometer), $\tau$ can simply be estimated as $\tau=-ln(F_l/F_c)$ \citep[cf.,][, their sect 4.1; who discuss the caveat of line smearing by a spectrometer that does not resolve the spectral profile]{kamenetzky12, vdWiel16}. Fig.~\ref{fig:HF-column} shows the estimated optical depth of HF (bottom panel) at each frequency element. In order to reduce the uncertainties and noise (ringing effect) introduced by the sinc convolution of the SPIRE FTS, we first fit all the prominent \twco, \nii, and \hho\ lines (including the near by \ph2o\ $2_{20}-2_{11}$ at 1228.8 GHz), and then subtract their combined fluxes from the SSW band to produce the 
residual spectrum (normalized by the continuum) of HF $J=1\rightarrow0$ shown in Fig.~\ref{fig:HF-column} 
(top panel).

Integrating the optical depth (of the normalized absorption feature below unity) we find a 40$''$-beam averaged 
column density $N({\rm HF}) \approx  (1.07 \pm 0.11) \times 10^{14}~\2cm$. The uncertainty for this column was 
estimated as the fraction ($\sim$0.105) between the rms value ($\sim$5.07 Jy), computed for the residual 
spectrum (around the HF line) between 1220 GHz and 1244 GHz, and the peak flux ($\sim$48.23 Jy) of the HF 
absorption feature. This column density is a factor $\sim$2.3 lower than the HF column density derived from the 
velocity resolved HIFI spectrum of HF \citep{monje14}. However, the latter shows blue shifted absorption and 
redshifted emission (i.e., a P-Cygni profile), which are unresolved in the SPIRE spectrum. The P-Cygni profile 
of HF is suggestive of an outflow of molecular gas with a mass of $\sim$10$^7$~\Msun\ and an outflow rate 
$\sim$6.4~\Msun~yr$^{-1}$ \citep{monje14}, which is in agreement with the outflow rate derived from the 
\twco\ $J=1\to0$ high resolution map obtained with ALMA \citep{bolatto13}.

Because of the unresolved line profile, we quote our estimated HF column density as a lower limit. From the $N({\rm HF})/N({\rm H_2})=2.94\times10^{-8}$ abundance ratio observationally determined by \citet[][for the warm component of AFGL 2136 IRS 1, their Table~3]{indriolo13}, which is similar to the value $3.6\times10^{-8}$ predicted by \citealt{neufeld09}, we 
obtain a molecular hydrogen column density of $(3.64\pm0.37)\times$10$^{21}~\2cm$, for the 40$''$ beam.
This hydrogen column density is comparable to the column obtained in Sect.~\ref{sec:continuum} from the dust 
emission at 100~\mum\ (c.f. Table~\ref{tab:sed-results}) .
Besides the unresolved line profile of HF, there are other 
uncertainties to be considered in this calculation. First, the column density we derive is also a lower limit of 
the total column density, since we only observe the HF gas in front of the continuum emission. Second, the 
molecular abundance, and whether or not all HF molecules are truly in the ground state, are arguable assumptions 
since non-equilibrium chemistry could be at play in the environment with enhanced cosmic rays density of the 
nuclear region of NGC~253.

\section{Modeling the CO LSED}\label{sec:model}

\subsection{Note on the CO line widths}\label{sec:line-widths}

From the spectrally resolved HIFI lines of NGC~253, we noticed a variation in the lines' FWHM widths. Even among the \twco\ ladder the FWHM decreases with frequency, i.e., the higher the $J$-transition, the narrower the line width (cf., Table~\ref{tab:hifi-fluxes}).  Although different beam filling factors can cause a variation in the line FWHM, with larger beams covering larger areas, this effect is unlikely to account for the broader line widths observed in the lower-$J$ \twco\ lines (cf., Tables~\ref{tab:hifi-fluxes} and \ref{tab:champ-CO6-5}). Thus, we note that assuming the same FWHM for all the \twco\ lines in any single-component radiative transfer model introduces uncertainties that affect most directly the derived column densities (since the line intensities provided by radiative transfer models are proportional to the column density per assumed line width). From the different line widths observed in the HIFI spectrum of the \twco~$J=9\to8$ and $J=5\to4$, we estimate that such uncertainty should be at least 20\%. Since a broader FWHM would require a larger column density to match the observed flux of a given line, the column densities reported in the following section for the lower-$J$ CO lines should be considered lower limits. We also consider multi-component models that both excitation and linewidth trends suggest are physically more accurate in sec.~\ref{sec:co-lines}.

\subsection{Non-LTE excitation analysis}\label{sec:excitation}

Following previous work in the literature, we used the radiative transfer code 
RADEX\footnote{http://www.sron.rug.nl/$\sim$vdtak/radex/index.shtml} 
\citep{vdtak07} to explore a wide range of possible excitation conditions that can lead to the observed line fluxes of a particular molecule. Those line intensities are sensitive to the kinetic 
temperature ($T_k$), the volume density of the collision partner ($n(\rm H_2)$), and the column density per line width ($N/\Delta V$). For our analysis we use only H$_2$ as collision partner, since it is the most abundant molecule and has the largest contribution to the excitation of the CO lines. The code uses a uniform temperature and density of the collision partner to model an homogeneous sphere.
Therefore, our analysis is not depth dependent. RADEX assumes the LVG (large velocity gradient/expanding sphere) formalism 
for the escape probability calculations. Hence, these models can only reproduce a  
\textit{clump} that represent the \textit{average} physical conditions of the gas from which the CO emission emerges. This is a well fitted model for single dish observations of unresolved emissions 
convolved with the telescope beams.
The physical conditions were modeled using the collisional data available in the 
LAMDA\footnote{http://www.strw.leidenuniv.nl/$\sim$moldata/} database \citep{schoier05}. The collisional rate coefficients for \twco\ and \hho\ are adopted from \citet{yang10} and \citet{daniel11}, respectively.

For the volume density we explored ranges between $10^2~\3cm$ and $10^7~\3cm$, the kinetic temperature varies from 4 K to 300 K, and the column density per line width lies between $10^{10}$ \ndv~and $10^{20}$ \ndv. 
In order to obtain the actual column density, the values reported must be multiplied by the local velocity dispersion (line width) of a single cloud. For comparison, a $\Delta V=23$~\kms\ was derived for the nuclear region of the Active Galactic Nuclei (AGN) driven galaxy NGC~1068 from high resolution maps \citep{schinnerer00}. Since we do not have a good estimate for NGC~253, a conservative value of $\Delta V=10$~\kms\ was adopted.

Since the optical depths obtained from the dust SED fit (Sect.~\ref{sec:continuum}) are not 
negligible (i.e., the dust emission is not optically thin in the whole frequency/wavelength range), 
and considering that the gas and dust must be well mixed in the emitting region, we modified the 
original RADEX code in order to include a more representative background emission $I_{bg}(\nu)$ 
as a diluted blackbody radiation field, in a similar way as done by \citet{poelman05} and 
\citet{pb09}. We considered the first two dust components at 37~K and 70~K (as estimated in Sec.~\ref{sec:continuum}), as well as the contribution from the cosmic microwave background at $T_{{\rm CMB}}$=2.73 K, according to the following equation

\begin{multline}\label{eq:radex-background}
 I_{bg}(\nu) = B_{\nu}(T_{{\rm CMB}}) + \\ 
 \left( 1-e^{-\tau_{\nu}} \right) \left[ B_{\nu}(T_c) + f_w B_{\nu}(T_w) \right],
\end{multline}

\noindent
where $B_{\nu}$ is the Planck function, and the dust optical depth $\tau_{\nu}$ is computed for 
each transition line using eq.~(\ref{eq:tau-dust}), with a fixed dust mass 
$M_{dust}$=3$\times$10$^6$~\Msun\ estimated from the 500~\mum\ photometric flux 
(Sec.~\ref{sec:continuum}). The factor $f_w$ corresponds to the relative contribution of the warm 
component with respect to the cold component, and is defined as the ratio between the corresponding 
area filling factors, $f_w=\Phi_w/\Phi_c=0.02$, (c.f., Table~\ref{tab:sed-fit}). 
The contribution factor $f_w$ is needed in order to mimic the observed 
dust continuum emission in the spherical clump. Otherwise, the warm 
dust component at $T_w=70$~K would dominate the background radiation field in the radiative 
transfer calculations, which would not be realistic. In strict rigour, the second term 
of eq.(\ref{eq:radex-background}) should be multiplied by a geometrical dilution factor $\eta_d$, 
which indicates the fraction of the dust emission actually seen by the molecules. However, we do 
not have a way to constraint this parameter from the convolved (unresolved) emission of the entire 
nuclear region of NGC~253, collected by the single dish of Herschel. Hence, for simplicity we assume $\eta_d=1$, which is equivalent to assume that the dust
and the gas arise from the same volume.

\begin{figure*}[!ht]
 \centering
  \hspace{-0.00cm}\epsfig{file=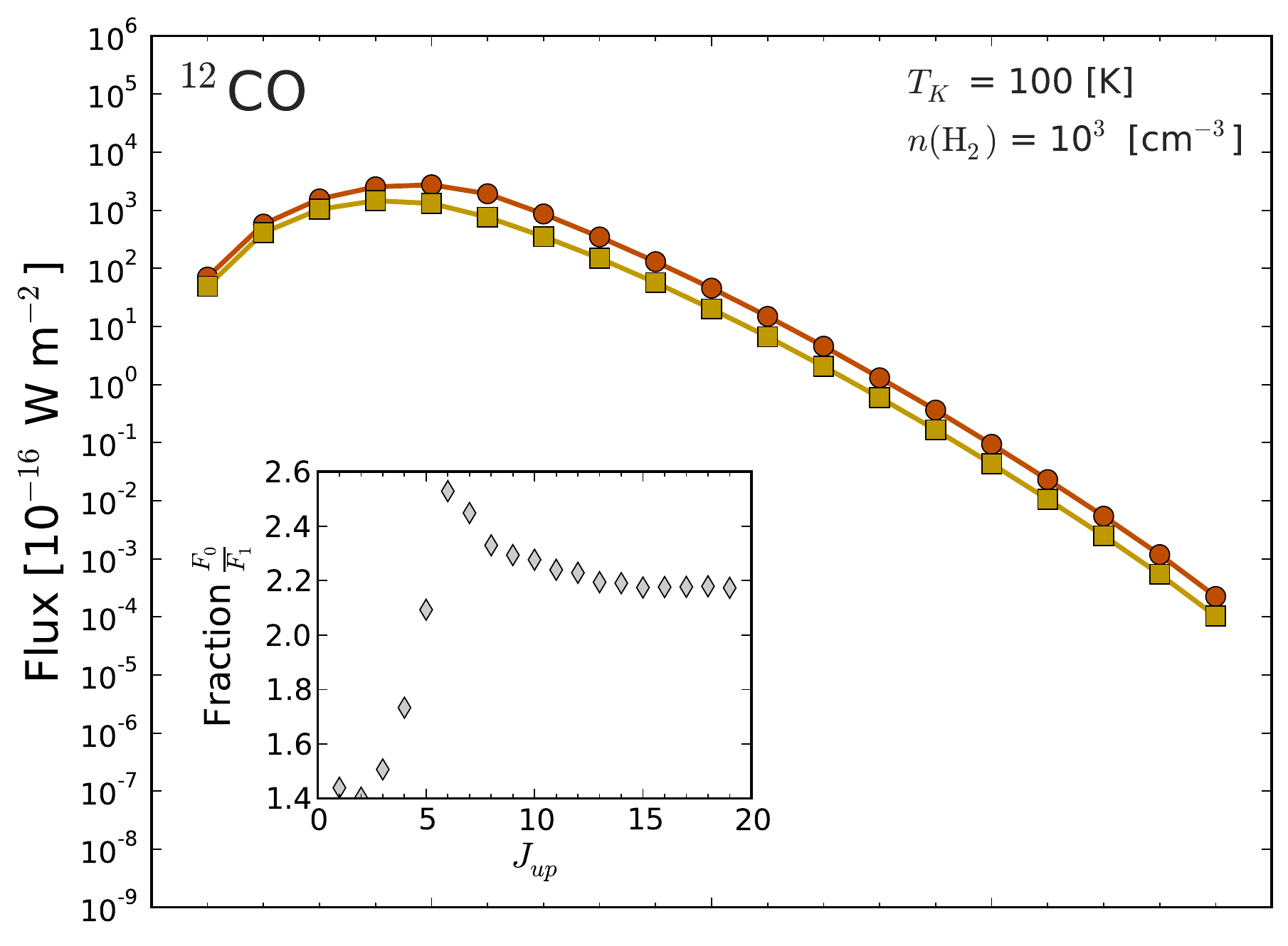,angle=0,width=0.39\linewidth}%
  \hspace{-0.00cm}\epsfig{file=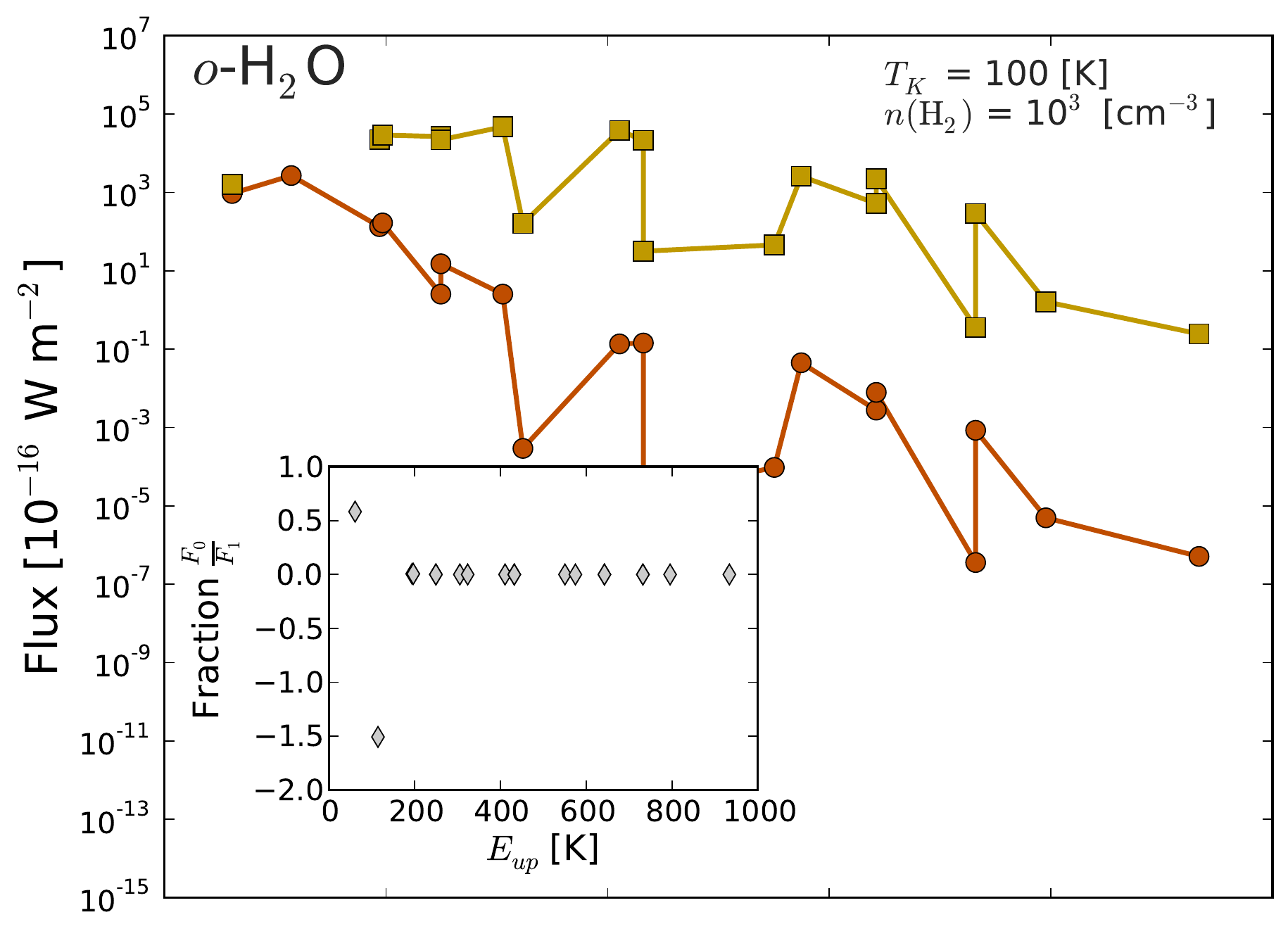,angle=0,width=0.395\linewidth}

  \vspace{-0.0cm}
  \hspace{0.2cm}\epsfig{file=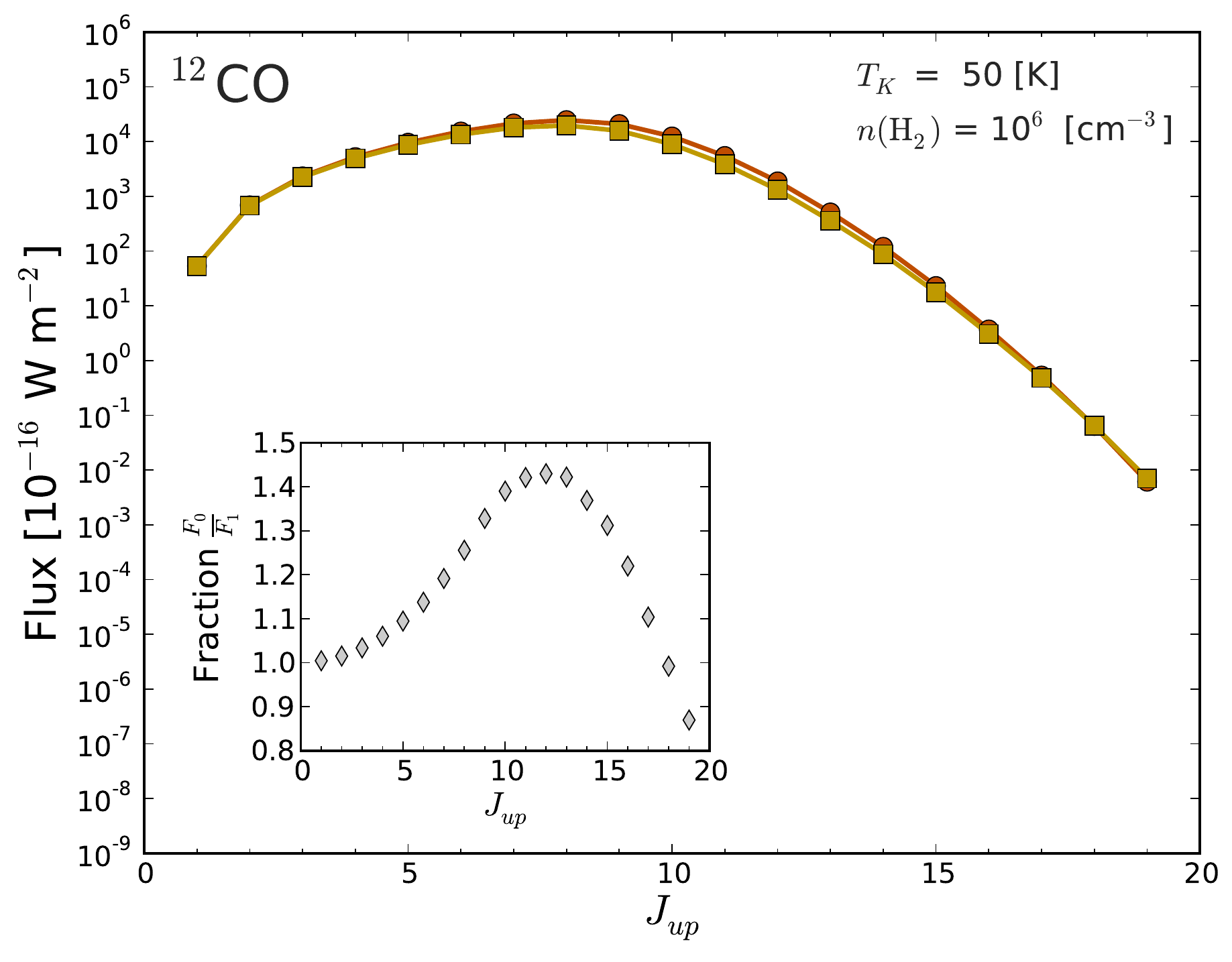,angle=0,width=0.395\linewidth}%
  \hspace{-0.1cm}\epsfig{file=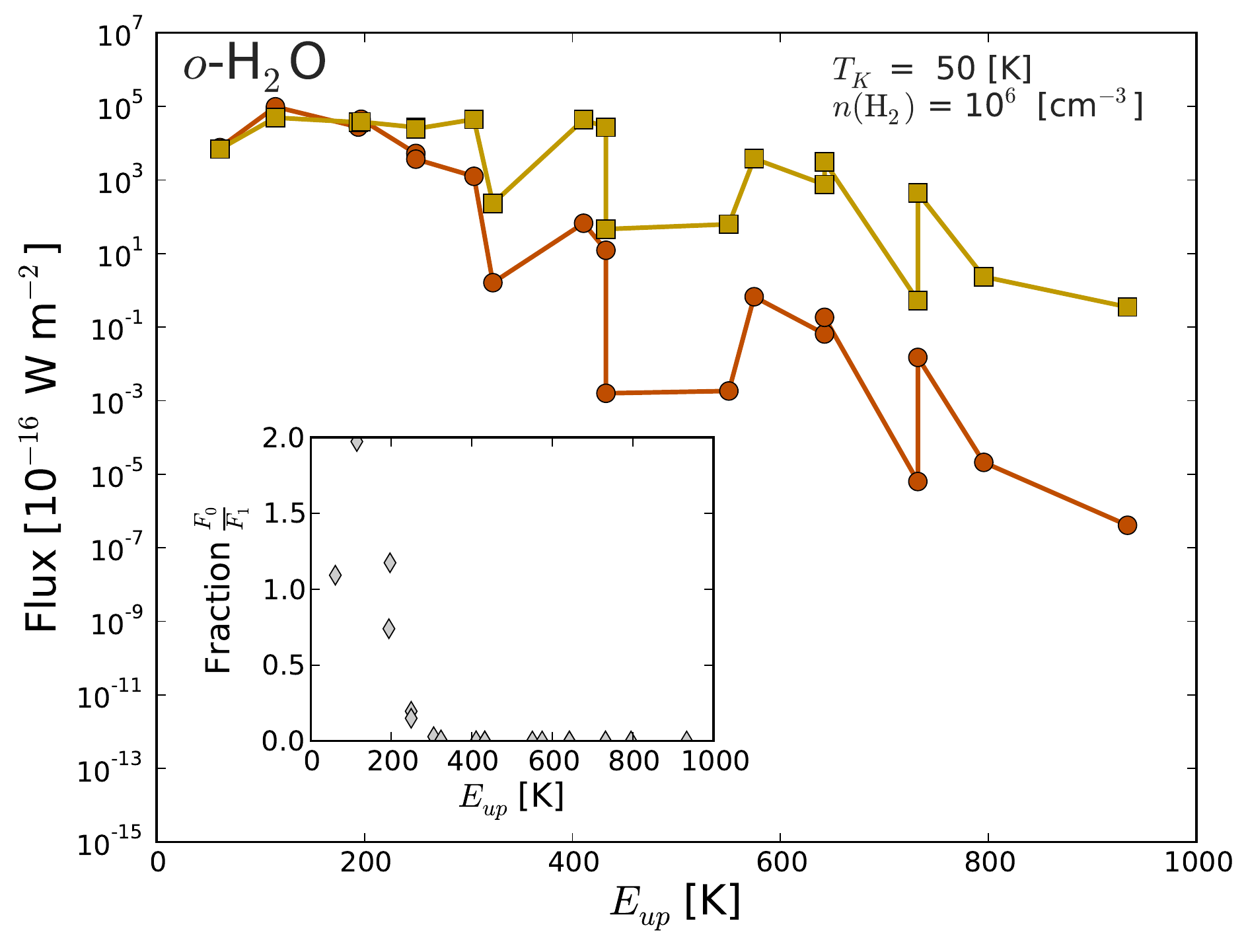,angle=0,width=0.405\linewidth}
 
 \caption{\footnotesize{Fluxes of the \twco\ transitions from $J_{up}=1$ to $J_{up}=19$ (\textit{left panels}) and ortho-H$_2$O transitions from $E_{up}=61.0$ K to $E_{up}=878.2$ K (\textit{right panels}), as estimated with RADEX for different densities $n({\rm H_2})$ and temperatures $T_k$.  The original RADEX fluxes were scaled up assuming a total line width $dV=190$~\kms. The circles correspond to the fluxes using only the cosmic microwave background at 2.73 K as background radiation field for the excitation calculations. The squares show the fluxes obtained when the dust emission is included in the background radiation field, as from eq.(\ref{eq:radex-background}). The inset shows the ratio between the fluxes without dust emission ($F_0$) over the fluxes with dust emission ($F_1$).}}

  \label{fig:dust-background-field}
\end{figure*}

The spectrum in the millimeter regime is usually dominated by the cosmic background black body 
radiation field at 2.73 K, which peaks at 1.871 mm. Therefore, this component of the radiation 
field of eq.(\ref{eq:radex-background}) is generally considered (in the literature) to dominate the 
radiative excitation of the lower-$J$ levels of \textit{heavy molecular rotors}, such as CO, CS, 
HCN, HCO$^+$ and H$_2$CO. Hence, there seems to be a general agreement in the (sub-)millimeter 
astronomy, that knowing the specific background radiation field of a single molecular cloud (or an ensemble of clouds, as in the case of extra galactic astronomy) is not really needed. 

On the other hand, the far- and mid-infrared radiation field (mainly from dust emission, especially 
in circumstellar material or in star-forming regions) is important for molecules with widely spaced 
rotational energy levels (e.g., the \textit{lighter hydrides} OH, H$_2$O, H$_3$O$^+$ and NH$_2$), 
as well as for the higher-$J$ levels of the heavy rotors mentioned before. Since the dust is 
usually at higher temperatures than 2.73 K, its diluted black body radiation field will peak at 
shorter wavelengths (cf. Fig.~\ref{fig:dust-sed-fit}), increasing the radiative excitation of the 
higher-$J$ levels and, hence, leaving \textit{fewer available molecules} to populate the lower-$J$ 
levels. This effect is particularly important for Herschel observations, with which several of the 
higher-$J$ levels in the far- and mid-infrared regime have become available for a number of 
molecules.

The actual effect of a background radiation field (including dust emission) on the redistribution among 
rotational levels, depends on the local ambient conditions of the emitting gas. 
That is because at high densities (or temperatures) the collisions are expected to dominate the 
excitation of the mid- and high-$J$ levels of molecules, such as CO, while at lower densities 
(or temperatures) the radiative excitation, as well as spontaneous decay from higher-$J$ levels, 
are expected to be the dominant component driving the redistribution of the level populations.

To demonstrate this, Fig.~\ref{fig:dust-background-field} shows the fluxes (\Wm) of several 
transitions of the \twco\ and {$ortho-{\rm H_2O}$} molecules, with ($F_1$) and without ($F_0$) 
considering the dust emission in the background radiation field (cf., 
eq.~\ref{eq:radex-background}), for different volume densities and kinetic temperatures. 
The ratio between these 
two fluxes is shown in the inset.
Column densities (per line width) $N/\Delta V=10^{17}$ and $N/\Delta V=10^{16}$ \ndv\ were used for \twco\ and \oh2o, respectively. The original RADEX fluxes were scaled up assuming an \textit{average line width} $\Delta V=190$~\kms\ (from the average FWHM of the \twco\ and \thco~$J=6\to5$ HIFI lines, cf. Table~\ref{tab:hifi-fluxes}) for each transition. 

Although the difference in fluxes of the \twco\ transitions is barely noticed in the logarithmic 
scale, the absolute fluxes obtained \textit{without} using the dust emission in the background 
field are more than 40\% brighter than the fluxes obtained when the dust emission is included in 
the background field, for low densities ($10^3~\3cm$) and moderate temperatures (100 K). On the 
other hand, at high densities ($10^6~\3cm$) and relatively low temperatures (50 K), the fluxes 
$F_0$ of the lower-$J$ levels ($J_{up}<5$) are just a few percent brighter than the fluxes $F_1$.  
The difference in higher-$J$ levels ($J_{up}\geq5$) varies up to $J_{up}=18$, where the relation 
between the two fluxes is inverted. 

In the case of \oh2o\ the relation between the few fluxes $F_0$ and $F_1$ up to the energy level 
$E_{up}\sim200$ K varies depending on the ambient conditions. Above that level, the fluxes obtained 
with the dust emission in the background radiation field are always brighter (for the ambient 
conditions explored) by factors of a few and up to three orders of magnitude.
Since the \hho\ lines observed with SPIRE are not spectrally 
resolved, and knowing (from HIFI spectra) that some of them are blended with other lines,
the excitation analysis and abundance estimates of \hho\ are not addressed here. 
Instead, the analysis and more sophisticated models of the \hho\ lines in NGC~253 (and other galaxies) were presented in a parallel work based on HIFI velocity-resolved spectra by \citep{Liu07}. 
In the next sections we present the excitation analysis of \twco, \thco\ and HCN.

\subsection{Excitation of the CO lines}\label{sec:co-lines}

From the SPIRE and PACS spectra we have \twco\ transitions from $J=4\rightarrow3$ to $J=19\rightarrow18$, 
although the $J=17\rightarrow16$ transition was not detected with PACS because it is found in a very noisy 
spectral range. The lower-$J$ transitions are taken from the values reported for a 43$''$ beam by 
\citet{israel95} and \citet{wall91}, and they were corrected for a 40$''$ beam, assuming the average source size of 16\farcs7 as found from the \twco~$J=6\to5$ map (Sect.~\ref{sec:spire-corrected}).

First we tried to fit the full \twco\ line spectral energy distribution (LSED) using two components. The low-$J$ 
($J\leq7$) lines can be fit with one component, but the second component can fit either the mid-$J$ ($J\leq12$) 
or the high-$J$ ($J\geq13$) transitions, but not both simultaneously. So we need three components to fit the full 
\twco\ LSED simultaneously. The model we use is described by:

\begin{equation}\label{eq:CO-model}
 F_{tot}(\nu) = \Phi_{1}F_1 + \Phi_{2}F_2 + \Phi_{3}F_3
\end{equation}

\noindent
where $\Phi_{i}$ are the beam area filling factors and $F_{i}$ are the estimated fluxes for each component in 
units of \Wm. The estimated fluxes are a function of three parameters: the density of the collision partner 
(usually H$_2$) $n(\rm H_2)$ ($\3cm$), the kinetic temperature of the gas $T_k$ (K), and the column density per 
line width $N/\Delta V$ ($\ndv$) of the molecule in study. These are the input parameters for the modified RADEX 
code that uses the background radiation field as described in Sect.~\ref{sec:excitation}.

In contrast with previous work in the literature, we prefer to fit all the \twco\ LSED simultaneously, so we 
do not have to guess or speculate about up to which transition we should fit first and then subtract the 
modelled fluxes from the remaining higher transitions. Also because the latter method considers the effect of 
the first component on the higher-$J$ lines, but it does not take into account the effect of the second 
component on the lower- and mid-$J$ lines, which we note is not negligible. We also try the excitation 
conditions (temperature, volume and column densities) obtained by \citet{rosenberg14}. In their models, gas 
densities of up to $\sim$3$\times$10$^5~\3cm$ were found for their third component. The beam filling factors 
they reported are larger than one, which we find non-physical for a galaxy with an unresolved source size (see 
discussion in Sect.~\ref{sec:model-constraints}), so we have to  scale our fluxes estimated with RADEX by 
appropriate filling factors. We found 
that the \twco\ $J=14\to13$ and $J=15\to14$ lines observed with PACS are underestimated by factors $\sim$2.5 and 
$\sim$4, respectively, using the excitation conditions from \citet{rosenberg14}, while the higher-$J$ lines are 
underestimated by more than one order of magnitude.

Considering all the transitions, would require twelve parameters to fit the \twco\ LSED alone, so methods like 
the Bayesian likelihood analysis used in the literature \citep[e.g.,][]{ward03, kamenetzky12} 
become impractical due to the large number of combinations of input parameters that need to be 
explored. Instead, we use the simplex method \citep[e.g.,][]{nelder65, kolda03} to minimize 
the error between the observed and estimated fluxes, using sensible initial values and constraints of the 
input parameters as described below. Following \citet{rosenberg14}, we also included all the available \thco\ 
fluxes \citep[from SPIRE and ground based telescopes, e.g.,][]{israel95} to constraint the column density of 
the lower-$J$ lines (up to $J=8\to7$), as well as the HCN fluxes \citep[from ][]{paglione97,knudsen07}, 
to break the dichotomy between density and temperature for the high-$J$ transitions. The RADEX fluxes of the \twco\ and \thco\ lines were corrected by the FWHM (estimated from a Gaussian fit) of $\Delta V=190$ \kms\ (\citealt{wall91}, consistent with the average FWHM of the \twco\ and \thco~$J=6\to5$ HIFI lines, cf. Table~\ref{tab:hifi-fluxes}), and by $\Delta V=120$~\kms\ for the HCN lines \citep[][their Table~2]{paglione97}. All fluxes from ground based telescopes were 
corrected to our 40$''$ beam. 

\subsubsection{Constraints of the Model Parameters}\label{sec:model-constraints}

From the high spatial resolution maps by \citet{sakamoto06, sakamoto11}, and the two-dimensional Gaussian fit of the continuum and \twco\ emission (Sects.~2 and 3) we know that the size of the \twco\ emitting region ($\sim$16\farcs7) is smaller than the beam size (40$''$), so the beam area filling factors $\Phi_{i}$ must be strictly lower than unity, irrespective of the number of clouds or clumps found along the line of sight. Also, high resolution maps \citep{sakamoto11} and SOFIA/GREAT observations of the \twco\ $J=16\rightarrow15$ towards Galactic molecular clouds \citep[e.g.,][]{pb15b} indicate that the size of the CO emitting region decreases with $J$-transition. Therefore, the beam area filling factors of the three components should also decrease. Hence, the following condition  was imposed in the fitting procedure
\begin{equation}\label{eq:phi-constraint}
\Phi_3 \leq \Phi_2 \leq \Phi_1 < 1
\end{equation}

Following \citet{ward03} and \citet{kamenetzky12}, we also restricted the density $n(\rm H_2)$ and column density 
$N(\rm CO)$ to physically plausible values. That is, the total molecular mass  of the emitting region 
($M_{region}$) cannot be larger than the dynamical mass $2.4\times10^9$~\Msun\ of the galaxy \citep{houghton97}, 
and the column lengths cannot be larger than the size of the emitting region. These restrictions eliminate 
models with very large column density and too low volume density. The molecular gas mass contained in the beam is estimated as
\begin{equation}\label{eq:mass-constraint}
M_{mol} = \frac{A_{beam} 1.5 m_{\rm H_2} \sum_{i=1}^{3} \Phi_{i}N_{i} }{X_{\rm max}}
\end{equation}

\noindent
where $A_{beam}$ is the area (in cm$^2$) subtended by the beam size, $\Phi_{i}$ and $N_{i}$ are the beam area filling 
factors and column densities of the three components, and the factor 1.5 multiplying the molecular 
hydrogen mass $m_{\rm H_2}$ accounts for helium and other heavy elements \citep{kamenetzky12}.
Following \citet{ward03}, we assumed a conservative value $X_{\rm max}=5\times10^{-4}$ for the [\twco ]/[\hh ] fractional abundance, since the average value found in starburst galaxies may be even higher than values (e.g., 2.7$\times$10$^{-4}$) measured in warm star-forming molecular clouds like NGC~2024 \citep{lacy94}.

The circumnuclear gas layer extends about 680$\times$255~pc ($\sim$40$''\times$15$''$) at position angle 
58$^{\circ}$ as estimated  {from the CO $J=2\rightarrow1$ map by \citet{sakamoto06}. A smaller extension, however, is expected for the higher excitation gas. From the 2-D Gaussian fit of the \twco\ 
$J=6\rightarrow5$ map we estimate a CO emitting gas extension} of about 350$\times$210~pc 
($\sim$20$''.8\times$12$''$.5), assuming a distance $D$=3.5~Mpc \citep{rekola05}. So we used the 
smallest extent of 210~pc across, to constraint the equivalent length of the \twco\ column density. 
The later can be approximated from the area filling factor $\Phi_{i}$, assuming a circular (Gaussian) homogeneous emitting region of size 210 pc, and a circular homogeneous cloud of size $S_{cloud}\approx N_{i}/(n({\rm H_2}) X_{\rm max})$. In the same way the beam filling factor can be estimated as $(\Omega_{source}/\Omega_{beam})^2$, assuming an homogeneous source size $\Omega_{source}$ and a Gaussian beam size $\Omega_{beam}$, the area filling factor of our models can be estimated as the area of the cloud size over the area of the emitting region. So the cloud size can be constrained using the smallest extension of 210~pc across as upper limit by the following expression

\begin{equation}\label{eq:column-constraint}
\frac{ N_{i} }{ n({\rm H_2}) X_{\rm max} } \leq\ \sqrt{ \Phi_{i} }(210~{\rm pc}).
\end{equation}

From the RADEX documentation, and several practical tests done by us, we know that the cloud excitation 
temperature become too dependent on optical depth at high column densities. So very high optical 
depths can lead to unreliable temperatures due to convergence uncertainties in RADEX. Therefore, we 
excluded column densities that lead to an optical depth $\tau\geq100$ in any of the transition lines. 
For the volume densities and kinetic temperatures explored, we usually met this condition with $log_{10}
(N_{CO}/\Delta V)\gtrsim18.2$~\ndv.

Since \twco\ and \thco\ are supposed to co-exist in the same emitting gas, we used the same volume density and kinetic temperature for \thco\ as obtained in the three components of \twco. For the column density of \thco, we used the $^{12}$C/$^{13}$C isotope ratio of $\sim$40 confirmed by \citet{henkel14}. From the high resolution maps by \citet{sakamoto11} and observations of Galactic molecular clouds \citep[e.g.,][]{pb10, pb12} we know that the \thco\ emission is less extended than that of \twco. Therefore, we restricted the beam area filling factors of \thco\ components to be lower than those of \twco, and they are the only three free parameters used to fit the \thco\ LSED. We found that only the first two components of \twco\ are sufficient to fit the \thco\ LSED, as well as the lower ($J_{\rm up}<9$) transitions of \twco, while the third component contributes significantly for $J_{\rm up}>10$ transitions.

On the other hand, we found that the HCN LSED can be reproduced using the same volume density and temperature of 
the second and third component of \twco, while the HCN column densities are free parameters. Because of the 
comparable critical densities of the mid- and high-$J$ CO lines to those of the low-$J$ HCN lines, and from the 
extension of the HCN $J=4\rightarrow3$ map by \citet{sakamoto11}, we inferred that the beam area filling 
factor of the first HCN component should be $\Phi_1({\rm HCN})\leq \Phi_2(^{13}{\rm CO})$. We set the area 
filling factor of the second HCN component to be equal to $\Phi_3(^{12}{\rm CO})$, given that the extension and 
distribution of the high ($J_{up}>13$) \twco\ transitions is similar to that of the HCN lines, as observed in 
Galactic star-forming regions \citep[e.g., M17~SW,][]{pb15a, pb15b}.

\begin{figure*}[!ht]
 \centering
  \hspace{-0.00cm}\epsfig{file=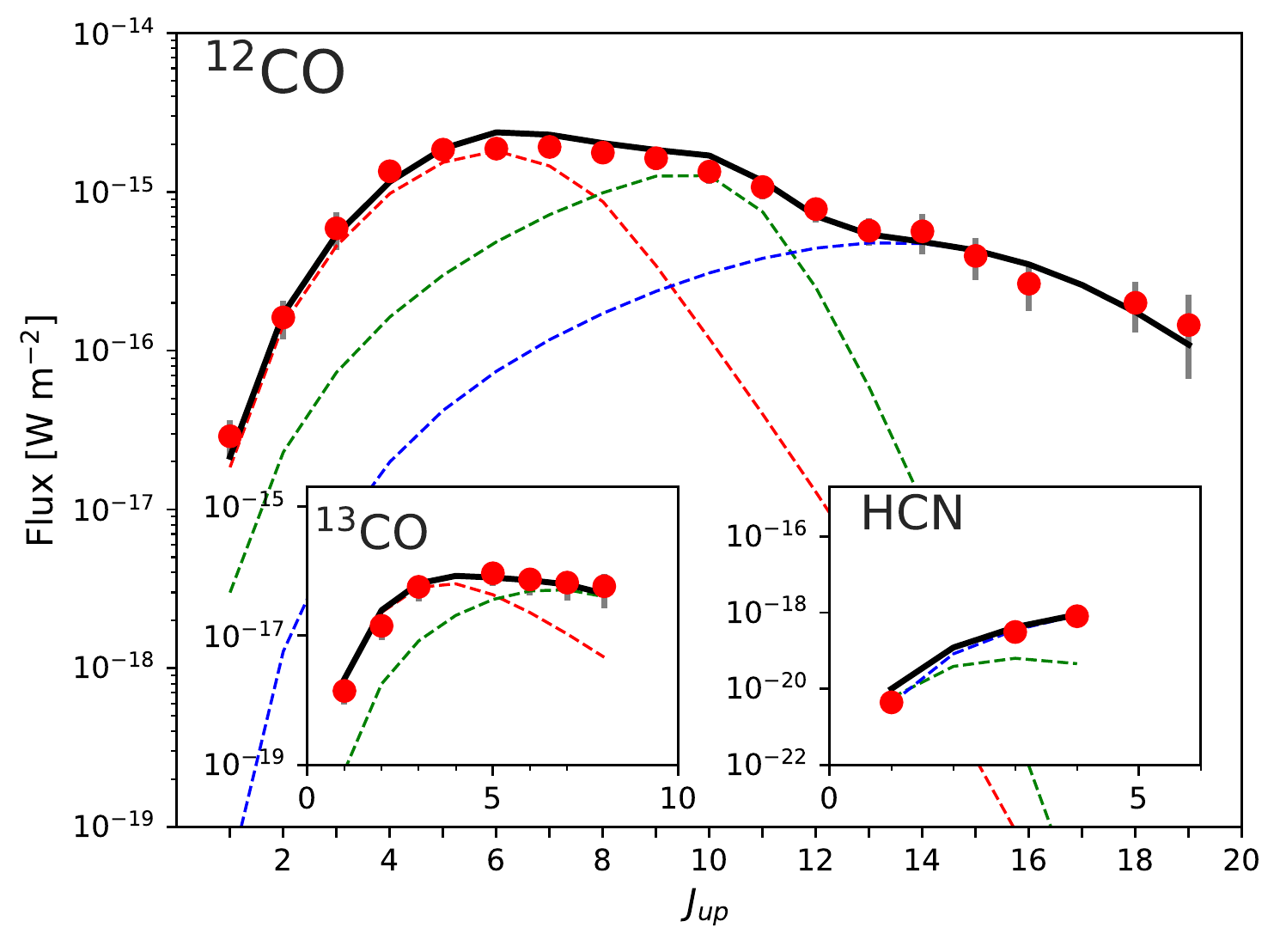,angle=0,width=0.7\linewidth}%

 \caption{\footnotesize{Spectral line energy distribution (SLED) of \twco, including ground-based ($J=1\rightarrow0$ to $J=3\rightarrow2$, from \citet{israel95}, SPIRE ($J=4\rightarrow3$ to $J=13\rightarrow12$) and PACS ($J=14\rightarrow13$ to $J=19\rightarrow18$) observations. The insets shows the SLED of \thco\ and \hcn. The dashed lines (with peaks from left to right) correspond to the 1st, 2nd, and 3rd components. The parameters of the fitted components can be found in Table~\ref{tab:LVG-results}}}

  \label{fig:CO-SED-fit}
\end{figure*}

   \begin{table*}[!ht]
   
   \begin{center}
   
      
     \caption{\footnotesize{LVG Model Results for NGC253.}}\label{tab:LVG-results} 

         \tabcolsep 5.8pt
         \scriptsize
         \begin{tabular}{lccc}
	    \hline\hline
	    \noalign{\smallskip}

	       & \multicolumn{3}{c} { Component Parameters } \\
          \cline{2-4}
          Quantity   &  1$^{st}$ component  &  2$^{nd}$ component  &  3$^{rd}$ component \\
          
	    \noalign{\smallskip}
	    \hline
	    \noalign{\smallskip}
	    
         $N({\rm H_2})~[\2cm]$  &  (8.0$\pm$1.8)$\times$10$^{21}$  &  (1.6$\pm$0.4)$\times$10$^{22}$  &  (3.2$\pm$0.7)$\times$10$^{21}$ \\         
         $S_{cloud}$~[pc]       &   1.6$\pm$0.4 &  (1.5$\pm$0.3)$\times$10$^{-2}$                   & (2.6$\pm$0.6)$\times$10$^{-4}$ \\
         $M_{mol}$~[\Msun]      &  (1.9$\pm$0.4)$\times$10$^{7}$ & (7.6$\pm$1.7)$\times$10$^{6}$ & (6.1$\pm$1.3)$\times$10$^{4}$\\         
                 
        \noalign{\smallskip}
        \hline
        \noalign{\smallskip}
        
	    \multicolumn{4}{c} { \twco } \\
 
        \noalign{\smallskip}
        \hline
        \noalign{\smallskip}        
                 	    
	     $\Phi$  &   (2.8$\pm$0.6)$\times$10$^{-1}$  & (5.5$\pm$1.2)$\times$10$^{-2}$  &  (2.2$\pm$0.5)$\times$10$^{-3}$\\
         $T_{K}$~[K]  &   90$\pm$10  &  50$\pm$6  &  160$\pm$12\\
         $n(\rm H_2)~[\3cm]$  &   (1.6$\pm$0.3)$\times$10$^3$  &  (3.2$\pm$0.8)$\times$10$^5$  &  (3.9$\pm$0.8)$\times$10$^6$ \\
$N(^{12}{\rm CO})~[\2cm]$  &   (4.0$\pm$1.5)$\times$10$^{18}$  &  (7.9$\pm$3.5)$\times$10$^{18}$  &  (1.6$\pm$0.4)$\times$10$^{18}$ \\
         
         
	    \noalign{\smallskip}
	    \hline
	    \noalign{\smallskip}
	    
	    \multicolumn{4}{c} { \thco } \\  
	    
	    \noalign{\smallskip}
	    \hline
	    \noalign{\smallskip}	    
	    
	     $\Phi$  &   (2.5$\pm$0.6)$\times$10$^{-1}$  & (1.2$\pm$0.3)$\times$10$^{-2}$  &  \\
         $T_{K}$~[K]  &   90$\pm$10  &  50$\pm$6  &  \\
         $n(\rm H_2)~[\3cm]$  &   (1.6$\pm$0.3)$\times$10$^3$  &  (3.2$\pm$0.8)$\times$10$^5$  &  \\
         $N(^{13}{\rm CO})~[\2cm]$  &   (1.0$\pm$0.3)$\times$10$^{17}$  &  (2.0$\pm$0.8)$\times$10$^{17}$  &  \\

	    \noalign{\smallskip}
	    \hline
	    \noalign{\smallskip}
	    
	    \multicolumn{4}{c} { HCN } \\  
	    
	    \noalign{\smallskip}
	    \hline
	    \noalign{\smallskip}    
	    
	     $\Phi$  &     & (1.2$\pm$0.3)$\times$10$^{-2}$  &  (2.2$\pm$0.5)$\times$10$^{-3}$\\
         $T_{K}$~[K]  &    &  50$\pm$6  &  160$\pm$12\\
         $n(\rm H_2)~[\3cm]$  &     &  (3.2$\pm$0.8)$\times$10$^5$  &  (3.9$\pm$0.8)$\times$10$^6$ \\
         $N({\rm HCN})~[\2cm]$  &     &  (1.3$\pm$0.5)$\times$10$^{14}$  &  (1.6$\pm$0.6)$\times$10$^{15}$ \\
         
	    \noalign{\smallskip}
	    \hline
	  \end{tabular}
	  
	  \end{center}

\end{table*}

\subsubsection{The LSED Model}

The best model fit for \twco\ is shown in Fig.~\ref{fig:CO-SED-fit}. The insets show the model fit for 
\thco\ and HCN. The resulting parameters of each component are summarized in Table~\ref{tab:LVG-results}.
The third component fitting the higher-$J$ (PACS) lines is a totally new addition surpassing works 
previously reported. It shows that the molecular gas in the central 350$\times$210 pc of NGC~253 is much 
more highly excited than that traced with only the lower- and mid-$J$ lines observed with ground based 
telescopes or even Herschel/SPIRE alone. The HCN fluxes help to constrain well the parameters of the 
third component. We found that temperatures $\gtrsim180$~K do not reproduce the slope described by the 
$J_{up}>11$ \twco\ fluxes, overestimating the $J_{up}>14$. Lower volume densities could compensate for 
the overestimation, but $n({\rm H_2})<10^5~\3cm$ do not reproduce the three available HCN fluxes.

From the $N(\rm CO)$ columns we derive the column 
density of molecular hydrogen for each component. As discussed above, we assumed a \twco\ abundance relative to \hh\ of 5$\times$10$^{-4}$ that lead to values of $N(\rm H_2)$ for the first and third components which are similar (within the uncertainties) to the \hh\ columns estimated from the dust continuum emission (Sect.~\ref{sec:continuum}).
Our assumed [\twco ]/[\hh ] value is a factor 
$\sim$2.3 larger than the relative abundance of 2.2$\times$10$^{-4}$ derived by \citet{harrison99} based on an assumed carbon gas-to-dust ratio and the measured fractions of gaseous carbon-bearing species. On the other hand, our assumed [\twco ]/[\hh ] value is a factor $\sim$6 larger than the value of 8$\times$10$^{-5}$ assumed by \citep{bradford03} for a much smaller 15$''$ beam. If we use the 
latter [\twco ]/[\hh ] value instead, we would get molecular hydrogen column densities that are even larger than the $N(\rm H)$ column density derived from dust continuum emission.


\subsubsection{Gas Mass traced by \twco}

The gas mass of the cloud associated to each component of the model can be estimated using eq.(\ref{eq:mass-constraint}). Using our assumed relative 
abundance value of [\twco ]/[\hh ]$=5\times10^{-4}$, and the adopted local velocity dispersion of 10~\kms, 
we find molecular gas masses of about 2$\times$10$^7$~\Msun, 
8$\times$10$^6$~\Msun\ and 6$\times$10$^4$~\Msun, for the first, second, and third components, respectively (cf., Table~\ref{tab:LVG-results}). 
If we add up the masses of the three components we obtain a total gas mass of $\sim$2.7$\times$10$^7$~\Msun, which is similar (within the uncertainties) to the range of mass (1--5$\times$10$^7$~\Msun) found by \citet{harrison99}, and \citet{bradford03}, based on ground based observations of low- and mid-$J$ \twco\ transitions.
This gas mas, however, is about one order of magnitude lower than the gas mass derived from the 870~\mum\ and 500~\mum\ dust continuum emission of APEX/LABOCA \citep{weiss08} and our Sec.~\ref{sec:continuum}, as well as the gas mass derived by \citet{houghton97} from the \twco~$J=1\to0$ line intensity alone. 

The total mass derived from our \twco\ LSED model is in agreement with the mass found by \citet{bradford03} and the LVG models by \citet{rosenberg14}. As noted by Rosenberg \etal\ this mass values should be considered a lower limit, since CO becomes dissociated in the presence of high radiation fields and, thus, our assumed [\twco ]/[\hh ] abundance ratio may underestimate the actual column of \hh\ gas. In order to match the gas mass obtained from the dust continuum emission at 500~\mum\ (Sect.~\ref{sec:continuum}) a \twco\ to \hh\ relative abundance of 2.2$\times$10$^{-5}$ would be needed, which is a factor $\sim$3.6 smaller than value assumed by \citep{bradford03}.


\subsubsection{The Cloud Sizes and the Relation with Star-forming Regions}

From eq.(\ref{eq:column-constraint}) we obtain a characteristic cloud size (including only the molecular gas) of about 1.6$\pm$0.4~pc for the first \twco\ component.  This is similar to the cloud size of 2~pc found by \citet[][their Sect.4.2]{bradford03}, based on visual extinction arguments and including the atomic gas. The size of this component is comparable to the size of diffuse clouds or individual dark clouds in the Milky Way \citep[e.g., $\zeta$ Ophiuchi;][]{stahler05}. 

The characteristic sizes of the second and third components are much smaller than 1~pc (cf., 
Table~\ref{tab:LVG-results}). The second component has the largest column density, as well as the lowest gas temperature of the three components, and it has a high volume density of $\sim$3$\times$10$^5~\3cm$.
So this component can be associated with starless cores, or dense and relatively cold cores where the star-formation process may be at play. 

From the third component, instead, we derive an equivalent size of about 3$\times$10$^{-4}$~pc 
($\sim$9.3$\times$10$^{9}$~km or $\sim$62~AU). This is just about two orders of magnitude larger than the size of the super giant star Rigel in the Orion constellation, and is about two orders of magnitude smaller than the $\sim$0.5~pc size of the small clouds (SCs) found in some SNRs like IC443, although these SCs are also expected to be clumped and have small filling factors in a 45$''$ and 55$''$ FWHM beams \citep[e.g.,][]{lee12}. This 
estimated size is also much smaller than its estimated Jeans length ($\sim$12~pc) derived from its gas density and temperature (assuming all the gas is molecular), so the objects/clouds this component is associated to are not likely to have a purely gravitational origin.  

\subsubsection{Energetics and Excitation}

The observed CO LSED measures the luminosity of the molecular gas in the central 40$''$ of NCG~253. 
The total cumulative flux of all the available CO lines (cf., Fig~\ref{fig:CO-SED-fit}) is 
1.65$\times$10$^{-14}$~\Wm. This corresponds to $\sim$34\% of the \cii~158~\mum\ and $\sim$42\% of the combined \oi~63~\mum\ and \oi~145~\mum\ intensities (Table~\ref{tab:pacs-emission-fluxes}). 

At a distance of 3.5 Mpc, the molecular (CO) gas flux corresponds to a luminosity of 6.3$\times$10$^6$~\Lsun, giving a luminosity-to-mass ratio of $\sim$0.23~\Lsun/\Msun\ considering the total gas mass contained in our three CO components. 
This ratio is about factor two larger than the ratio found by \citet{bradford03}, considering a distance of 2.5~Mpc instead. 

The total CO flux in the inner 40$''$ region represent a 3.2$\times$10$^{-4}$ fraction of the total IR luminosity of the galaxy observed with IRAS \citep{rice88}. This fraction is considerably large because of the large fraction of the gas mass (represented by the first and second CO component in our model) is highly excited. The $L_{\rm CO}/L{\rm IR}$ observed in NGC~253 is only about a factor two lower than the luminosity ratio found in the luminous infrared galaxy NGC~6240 \citep{meijerink13}, and almost one order of magnitude higher than the ratio found for Mrk~231 \citep{vdwerf10} and Arp~220 \citep{rangwala11}. Although, a large line-to-continuum ratio can be explained by gas compressed and heated by shocks, like in the case of NGC~6240 \citep{meijerink13}, we think this is not the case for NGC~253. That is because the bulk of the \twco\ luminosity is contained in the low- and mid-$J$ lines that have either low density (first component) or low temperature (second component), which do not match a shock or turbulent dominated scenario. Besides, as mentioned by \citet{bradford03}, the near-IR \hh\ emission observed in the nuclear region of NGC~253 is more characteristic of UV fluorescence, rather than the thermal spectrum produced by shock heating, as found in Galactic outflow sources \citep[cf.,][]{engelbracht98}. This was confirmed by high resolution VLT/SINFONI maps of the near-IR \hh\ and \FeII\ emissions, where no strong correlation between \hh\ and \FeII\ (a strong near-IR shock tracer) was found, while a good match between \hh\ and the ISAAC PAH 3.21~\mum\ (a tracer of the fluorescently excited gas, including excitation by both O and B stars) maps was observed \citep{rosenberg13}.  {We believe the above is enough observational evidence to conclude that shocks or turbulence heating is not likely to be the main excitation source of the bulk of the CO line intensities, described by the first and second components in our LSED model. This may seem in contradiction with (mechanically and cosmic ray heated) PDR models previously reported by \citet{rosenberg14} where they concluded that \textit{mechanical heating} is necessary to reproduce the observed CO emission from ground based and SPIRE data alone. However, it is sufficient to recall \citet[][their Sect.~6]{rosenberg14} where they discuss that the relative contribution of mechanical heating is dominant over cosmic ray heating in their models, but they also state that the main source of heating in all the models they tested is actually photoelectric heating. This is then in agreement with our statement above, that shock (or turbulent/mechanical) heating is not the main source of excitation for the low- and mid-$J$ CO emission.} 


On the other hand, the third component in our model describing the \twco\ $J_{up}>13$ lines, do have high 
density and temperature, matching best a shock or turbulent dominated scenario. The emission of the 
high-$J$ \twco\ lines detected with PACS may originate from shocked clumps in SNe remnants, as well as in 
the turbulent dominated clumps along the molecular outflow traced by the \twco~$J=1\to0$ high resolution ALMA observations \citep{bolatto13}.
However, this molecular outflow is not observed in the \twco~$J=2\to1$ (neither its isotopes) nor in the 
\twco~$J=3\to2$ high-resolution observations done with the Submillimeter Array (SMA) by \citep{sakamoto11}.
Although it can be argued that the emission of the $J=2\to1$ and $J=3\to2$ \twco\ transitions may indeed be present in the molecular outflow identified by \citet{bolatto13}, and they may just be below the sensitivity level of SMA (in comparison with ALMA), but it is a fact that the \textit{intensity} of the $J_{up}>13$ lines of \twco\ are actually even fainter than the $J=3\to2$ line, as observed in the PACS fluxes 
(c.f, Fig.~\ref{fig:CO-SED-fit}). That means, the fluxes of the \twco\ $J_{up}>13$ lines observed in our 40$''$ beam must be dominated by the emission arising from the starburst ring. This scenario is well supported if we consider that the high-$J$ CO lines present a remarkable spatial correlation with the \hcn\ and \hcop\ lines, as observed with SOFIA/GREAT in some Galactic molecular clouds \citep[e.g., M17~SW,][]{pb15a}. And the \hcn\ $J=4\to3$ high resolution map by \citet{sakamoto11} does not trace the molecular outflow observed in the \twco~$J=1\to0$. In fact, the actual outflow, originally observed in H$\alpha$, may lack sufficiently dense gas to excite any HCN emission (as well as the high-$J$ CO lines), as shown with the \hcn~$J=1\to0$ OVRO map by \citet[][their Fig.~5]{knudsen07}.

Another plausible scenario could be an internal source of heating  {within the dense gas environment of the starburst ring}. That is, hot 
cores, which have comparable sizes to the one derived above for the third CO component. The detection of H$_2$S 
in NGC~253 by \citet{martin05} can be considered to arise from the massive star forming cores in the 
nuclear starburst. That is because sulfur is largely depleted (by a factor of 100--1000) in the ISM, and the 
major gas-phase formation routes to H$_2$S are mainly endothermic \citep[$\geq$7000 K;][]{pineau93, 
rodgers03}. Therefore, the H$_2$S emission is generally associated with sputtering on dust 
grains due to either intense UV radiation from star-forming regions or shocks generated by young stellar 
objects, as is being observed in the Orion KL outflow \citep{minh90}. Likewise, \citep{garcia-burillo00} reported enhanced abundance of the shock-tracer SiO from high resolution observations, arguing that the 
SiO emission may arise in bipolar outflows powered by young massive stars associated with the nuclear starburst 
and/or due to large-scale shocks induced by the nuclear bar. 
The SiO emission in NGC~253 is located in two regions, between 10$''$ and 20$''$ away from the center and 
opening out in a spiral-like structure, as well as in an inner ring of radius $\sim$4$''$ in the center of 
NGC~253, interpreted as the inner Inner Lindblad Resonance (iILR). The latter SiO emission coincides with the 
high resolution H$_2$S emitting area reported by \citet{minh07}. Thus, like SiO, the H$_2$S emission could 
originate from shock waves. However, the high rotation temperature derived for H$_2$S is considered a signature 
of hot core chemistry, where H$_2$S is released from dust mantles by heating of massive star forming regions 
\citep{rodgers03}. Besides, the detection of the H$_2$S 2$_{2,0}$--2$_{1,1}$ transition, which have an
upper state energy level of 84~K, indicates the presence of hot gas. Therefore, \citet{minh07} favour the 
hot core chemistry scenario for the H$_2$S emission, which is supposed to trace the ongoing star formation 
through hot core activity. A rough estimate indicates that several thousands of Orion KL--like cores may exist 
toward the H$_2$S emitting area in the inner 20$''\times$20$''$ nuclear region of NGC~253 \citep{minh07}.
Therefore, the high-$J$ \twco\ lines may as well be associated to these hot cores.


Nevertheless, as discussed by \citet{rosenberg14}, the effects of cosmic rays (although, perhaps not 
dominant) cannot be ruled out as an external source of heating in hot cores, or in addition to shock generated by 
YSOs, bipolar outflows or mechanical heating. 
The high star-formation activity, along with the relatively high SNe rate in NGC~253, allowed to estimate an 
enhanced cosmic-ray density which, in turn, allowed to predict (and detect) very high energy ($>100$~GeV) gamma 
rays from the nuclear region of NGC~253 \citep[e.g.,][an references there in]{paglione96, domingo-santamaria05, rephaeli10, acero09}. And the gamma-rays are basically the product of 
the enhanced cosmic-ray rates interacting with dense gas \citep[e.g.,][]{paglione96, hewitt09}.
The observed gamma-ray flux in NGC~253 indicates a cosmic ray density that is three orders of magnitude higher 
than that found in the center of the Milky Way and, therefore, it is expected to play a significant role in the 
excitation of the high-$J$ \twco\ lines, as well as in the abundance and line intensities of other species.

\subsection{Molecular Lines as Diagnostic of Enhanced Cosmic Rays}

Because the \hh\ cosmic-ray (CR) dissociation cross sections 
are small ($\sim$3$\times$10$^{-26}~\2cm$), cosmic rays can penetrate deep into molecular cloud cores, 
keeping the gas temperature above that of the cosmic microwave background and enhancing the abundance and 
line intensities of species like \twco, \hcn, \hcop, OH$^+$ and H$_2$O$^+$ \citep[e.g.,][and references therein]{goldsmith78, meijerink06, meijerink11}. 

Enhanced local CR ionization rates in small clumps can explain the production of OH 
molecules behind a C-type shock \citep{wardle99}. The fluxes of the OH lines detected (in emission and absorption) 
with PACS are about one order of magnitude higher than the fluxes of the \hho\ lines 
(Table~\ref{tab:pacs-absorption-fluxes}), which can also be attributed to the enhanced cosmic rays in the nuclear region of 
NGC~253 \citep{meijerink11}.

Likewise, the detection (although only in absorption) of the ionic species OH$^+$ and \hho$^+$ in our 40$''$ beam SPIRE spectrum, are indication of high ionization fractions ($x_e > 10^{−3}$) produced by the enhanced cosmic rays \citep[cf.,][]{meijerink11}.

Other diagnostics, like the HCN/HNC and the HCN/\twco\ line ratios, are expected to be sensitive to mechanical 
heating \citep[e.g.,][]{loenen08, meijerink11}. Both line ratios are expected to increase when 
mechanical heating is important, which is the scenario favoured by \citet{rosenberg14} for the excitation of 
the \twco\ LSED in the nuclear region of NGC~253. However, the interpretation of these ratios is not straight 
forward in environments with high densities and high CR rates (like in the case of NGC~253), where He$^+$ 
effectively destroys both HCN and HNC, and the HCN and \twco\ lines are expected to trace different regions, as 
pointed out by \citet{meijerink11}. We argue, though, that the latter statement holds only for the low- and 
mid-$J$ \twco\ lines. The higher ($J_{up}>10$) transitions, on the other hand, are found to have very similar 
spatial distributions (and thus very similar beam area filling factors) to that of HCN (and \hcop), as observed with SOFIA/GREAT in Galactic molecular clouds \citep[e.g., M17~SW,][]{pb15b}.
Besides, the \twco\ $J=14\rightarrow13$ have almost the same critical density 
($2-3\times$10$^6~\3cm$ for temperatures between 50 K and 200 K) as the HCN $J=1\rightarrow0$ transition. 
Therefore, we still expect the \hcn($1-0$)/\twco($14-13$) line ratio to be a useful diagnostic tool, even in high density and high CR rates environments.

For NGC~253 we obtained a line ratio \hcn($1-0$)/\twco($14-13$)$\sim$2$\times$10$^{-4}$ from the 40$''$ beam 
fluxes (W~m$^{-2}$). The most similar line ratios we find in the model predictions by \citet{meijerink11} are for a CR 
rate of 5$\times$10$^{-14}$~s$^{-1}$ in their models Set 1b (ratio $\sim$7$\times$10$^{-5}$) and Set~1c (ratio 
$\sim$9$\times$10$^{-4}$). These models represent high CR rates scenarios including the effect of mechanical 
heating corresponding to star formation rates of about 140 and 950~\Msun~yr$^{-1}$, respectively, for a Salpeter
IMF, as described in \citet{loenen08}. 
We also compared the \hcn($3-2$)/\twco($14-13$)$\sim$1$\times$10$^{-2}$ observed in NGC~253 with the line 
ratios from \citet{meijerink11}. The only predicted ratio that is comparable to the observed value in 
NGC~253 is $\sim$3$\times$10$^{-2}$ from their model Set~1a, which correspond to the same enhanced CR rate of 
5$\times$10$^{-14}$~s$^{-1}$ but without any mechanical heating effect. 

We note, however, that the HCN $J=3\rightarrow2$ transition is usually optically thicker than the $J=1\rightarrow0$ transition ($\tau_{1\to0}=0.4$ and $\tau_{3\to2}=1.9$ for the second component in our HCN model) and can be affected by self-absorption, as observed in Galactic molecular clouds \citep[e.g., M17~SW,][]{pb15b}.
We also note that the models by \citet{meijerink11} were run using  {a lower total hydrogen density 
(10$^{5.5}~\3cm$, and thus, a lower density of the collision partner \hh)} than what we found for the second and third components of our \twco\ LSED model. So we cannot 
tell how the  {(PDR)} \twco\ and \hcn\ line fluxes depend on even higher densities. Therefore, we 
cannot conclude from the line ratios alone whether mechanical heating or the enhanced CR rates are the dominant source of 
heating and excitation in the \twco\ and \hcn\ lines described by the second and third components of our \twco\ 
LSED model. A more sophisticated analysis including properly constrained PDR/XDR/CR and even shock models, will be deferred to a future paper.

\subsection{Line Ratios as Diagnostic of Ionization, Density and Temperature of the ISM}

Emission-line ratios obtained from pairs of mid- and far-IR lines from the same ionic species have been used for statistical studies including several different type of galaxies \citep[e.g.,][in references there in]{spinoglio15, fernandez-ontiveros16}. Ratios like \nii~205~\mum//\nii~122~\mum\ (hereafter \nii~$_{205/122}$), \siii~$_{33.5/18.7}$, \oiii~$_{88/52}$, and \nev~$_{24.3/14.3}$, have the same ionization potential but different critical densities \citep[cf.,][their Table~1]{fernandez-ontiveros16}, hence they are used as diagnostic for the densities of the ionized gas in the $n_{\rm H}\approx10~\3cm$ to $10^5~\3cm$ range 
\citep[e.g.,][]{rubin94}. The density and temperature of PDRs are usually estimated from the cooling lines of the ionized and neutral gas, \cii~158~\mum, \oi~63,145~\mum, and \ci~370,609~\mum, following the predictions from PDR and XDR models \citep[e.g.][]{tielens85, liseau06, meijerink07}.

   \begin{table}[tp]
            \centering
      \caption{\footnotesize{Line flux (W~m$^{-2}$) ratios from the 40\arcsec\ aperture (SPIRE and PACS) and toward the south-west (SW) position (SOFIA/upGREAT and APEX/CHAMP+).}}
         \label{tab:line-ratios}
         \scriptsize
         \begin{tabular}{lcc}
	    \hline\hline
	    \noalign{\smallskip}
            Line      &   40\arcsec\ central &  15\farcs1 SW     \\
            Ratio    &    region      &   position      \\
	    \noalign{\smallskip}
	    \hline
	    \noalign{\smallskip}

\cii\ / CO(11--10)  &  5.31$\pm$1.12  &  1.23$\pm$0.22$^{~\mathrm{a}}$ \\
\cii\ / CO(6--5)    &  16.29$\pm$3.35  &  5.61$\pm$1.01 \\
CO(11--10) / CO(6--5) &  3.01$\pm$0.55  &  4.45$\pm$0.78$^{~\mathrm{b}}$ \\

	    \noalign{\smallskip}
	    \hline
	  \end{tabular}

\begin{list}{}{}
\scriptsize
\item[$^{\mathrm{a}}$] If a 15\farcs1 beam is considered instead of 22\farcs7 to compute the flux of \twco~$J=11\to10$, this ratio would be a factor $\sim2.25$ larger.

\item[$^{\mathrm{b}}$] If a 15\farcs1 beam is considered for \twco~$J=11\to10$, this ratio would be about 56\% smaller.

\end{list}

\end{table}

The \ci~$_{609/370}$ line ratio is expected to be sensitive to the temperature range 20--100~K in PDRs, while the \oi~$_{145/63}$ line ratio is generally used, in the optically thin limit, as a temperature tracer in the 100--400~K range for the neutral gas \citep{tielens85, kaufman99, liseau06, meijerink07}.
Moreover, the \ci~$_{609/370}$ line ratio is also  {expected to be} sensitive to X-rays since they are able to penetrate deep into the cloud and warm all the neutral carbon, thus lowering the \ci~$_{609/370}$ ratio as the temperature increases \citep{meijerink07,ferland13}.

We found a \ci~$_{609/370}$ ratio of 0.40$\pm$0.06, which is within the median value found by \citep{fernandez-ontiveros16}, consistent with the ratios expected for typical PDR dominated starburst galaxies. 
The inverse ratio (used by some PDR models and authors) \ci~$_{370/609}=2.48\pm0.38$ is expected to be found in gas with total densities between a few times $10^2~\3cm$ and $10^3~\3cm$ for UV fields in the range $10^2$--$10^3~G_0$ (in units of the Habing flux, where $G_0 = 1$ corresponds to 1.6$\times$10$^{-3}$~erg~cm$^{−2}$~s$^{−1}$ , which is the local Galactic interstellar radiation field), but also for lower impinging UV fields of $<100~G_0$, i.e., the remaining of higher UV fields absorbed before reaching higher density gas in the range $10^3$--$10^4~\3cm$ \citep{meijerink07}. 

 {Following the analysis by \citet[][their Fig.~9]{pereira-santaella13}, we found the observed \ci~$_{609/370}$ ratio at either lower density (and higher temperature) or at higher density (and lower temperature) than the first component of our \twco\ SLED model. But the optically thin limit used by Pereira-Santaella et al. underestimate the observed \ci\ line fluxes in our 40\arcsec\ aperture (and estimated filling factors). Larger column densities (i.e., optically thick regime) are needed to also reproduce the observed line fluxes.
On the other hand, using the same excitation conditions as the first component of our \twco\ SLED (Table~\ref{tab:LVG-results}) leads to a  \ci~$_{609/370}$ ratio of 3.5, higher than our observed line ratio. The line ratio and fluxes can be reproduced simultaneously by using the temperature and filling factor of the first component of the \twco\ SLED, but with a lower volume density of $n(\rm H_2)=10^{2.5}~\3cm$ and with $N(\rm C)=5\times10^{18}~\2cm$ (i.e., both $\tau\sim2$). This would agree with the picture of \ci\ emission arising from the C$^+$/C/CO PDR transition layer (as discussed above), where the volume of gas is expected to be more diffuse than the volume of gas where the bulk of the CO emission originates from.}

A \ci~$_{370/609}$ line ratio larger than factor  {three should be expected at densities in the range $10^2$--$10^6~\3cm$, if X-rays would be at play \citep[][their Fig.~3]{meijerink07}}. Thus, we can discard an XDR effect in the \ci\ emission observed with our 40\arcsec\ aperture. Note, however, that this is not evidence to rule out the presence of a strong (nor weak) XDR/AGN in the nuclear region of NGC~253, as suggested in the literature \citep[e.g.,][and references there in]{mohan02, weaver02, fernandez-ontiveros09, muller-sanchez10}. This is because an XDR/AGN would have a strong effect only within  $\lesssim$100~pc radius (or much less if is a weak X-ray source), which will be diluted in our 40\arcsec\ beam (such a spatial scale would have a beam area filling factor of $\lesssim$0.08 in our beam at the distance of NGC~253).

The work by \citealt{spinoglio15} and \citealt{fernandez-ontiveros16} showed that AGNs and starbursts are separated by the \siv~10.5//\siii~18.7 ratio, which is sensitive to the ionization parameter. They concluded, however, that harder radiation fields are not associated with a warmer neutral gas, since they did not find a correlation between the \siv~10.5//\siii~18.7 and the \oi~$_{145/63}$ ratios. An explanation for this is that the \oi~63~\mum\ line can be affected by self-absorption, making its interpretation in extra-galactic environment difficult.
It has been shown that the \oi~63~\mum\ line can be easily absorbed by a relatively small cold layer of hydrogen column of about $N_{\rm H}\sim2\times10^{20}~\2cm$ \citep{liseau06}. This self absorption effect is readily observed under different environments of Galactic molecular clouds \citep[e.g.,][and references there in]{leurini15, gusdorf17, kristensen17}, but it has also been observed in the large scale \oi~63~\mum\ spectra of extra-galactic sources like Arp~220, NGC~4945, NGC~4418, and even in the \oiii~88~\mum\ of IRAS17208-0014  \citep[e.g.,][]{gonzalez-alfonso12, fernandez-ontiveros16}. 
From our 40\arcsec\ beam we found a ratio of \oi~$_{145/63} = 0.12\pm0.03$, which is the same (within uncertainties) as the value 0.13$\pm$0.01 derived from the calibrated data reported by \citealt{fernandez-ontiveros16}. 

The inverse of this ratio (actually used in some theoretical models) is \oi~$_{63/145}=8.06\pm1.72$, which is close to the degeneracy limit (a ratio of ten) between optically thick and optically thin emission, according to the model by \citealt{liseau06}. \citet{fernandez-ontiveros16} noted that a \oi~$_{145/63}$ ratio larger than 0.1 is indicative of optically thick emission according to the model by \citealt{tielens85}. We note that this would be the case only for gas temperatures $<100$~K (based on the same model), which is unlikely to be the case for the warmer gas where the bulk of the \oi\ emission is expected to emerge from. We consider that self-absorption in the \oi~63~\mum\ line is a very plausible reason for the observed ratio, even when a self-absorption feature is not visible in our 40\arcsec\ PACS spectrum, as in the case of the other extra-galactic sources mentioned above. This could be due to a narrow self-absorption feature (or just as broad as the warmer background emission) not resolved in the PACS spectral resolution. This was pointed out for the case of the Galactic source G5.89--0.39 for which SOFIA/GREAT spectrum shows clear absorption features in the \oi~63~\mum\ line \citep{leurini15} while the Herschel PACS observations shows a Gaussian profile (i.e., no hint of absorption) with a spectral resolution of 90~\kms\ \citep[][their Fig.~2]{karska14}. This indicates that spectrally-resolved observations are needed for a better interpretation of the \oi\ lines, which can be obtained with the SOFIA/upGREAT receiver in order to confirm/discard the self-absorption scenario.

As pointed out by \citet{kaufman99} the observed peak line intensities in extra-galactic sources depend on the intensity emitted by each ensemble of clouds collected by the beam, and on the beam area filling factor of the respective emission. Hence, it is important to correct for the different filling factors of the emission from different tracers in order to compare with the line ratios predicted by theoretical models. In the case of the line ratios discussed above, \ci~$_{609/370}$ and \oi~$_{145/63}$, it is assumed that the emission of the lines of the same tracer have the same filling factor (hence no correction is needed). But in the case of the \cii~$_{158}$/\oi~$_{63}$ ratio or between very different CO transitions, one should correct as best as possible for the different area filling factors. That is because the more extended \cii\ emission may fill the beam while the higher-$J$ transitions (as well as the \oi\ emission) is expected to arise from more compact high density and high temperature regions, as shown in maps of the Galactic star-forming region M17~SW \citep[e.g.,][]{pb15a,pb15b}. Then, for instance, one should compute the line intensity ratio between \cii\ and \oi\ as $I({\rm \cii})/I({\rm \oi}) = \left[ F({\rm \cii})/\Omega_{{\rm \cii}} \right]/\left[ F({\rm \oi})/\Omega_{{\rm \oi}} \right]$, with $F$ the observed flux and $\Omega$ the solid angles of the respective emitting regions.
From the 25\arcsec\ resolution maps obtained toward M17~SW \citep{pb12,pb15b} we made some rough estimates of the emitting solid angles for \twco~$J=11\to10$ and $J=6\to5$ relative to \cii\ and CO~$J=6\to5$ as $\Omega_{\rm CO(11-10)}/\Omega_{\rm \cii}\approx0.11$, $\Omega_{\rm CO(6-5)}/\Omega_{\rm \cii}\approx0.59$, and $\Omega_{\rm CO(11-10)}/\Omega_{\rm CO(6-5)}\approx0.19$. This relative beam area filling factors should be considered as rough upper limits since the actual emitting solid angles of these lines can be much smaller in our 40\arcsec\ aperture than in the 3$\times$2~pc$^2$ region map of M17~SW. The line ratios for the 40\arcsec\ aperture central region and for the south-west (SW) position observed with SOFIA/upGREAT are summarized in Table~\ref{tab:line-ratios}. Note that the CO(11--10)/CO(6--5) ratios obtained for the 40\arcsec\ aperture and the SW position are the same (within the uncertainties). However, we estimate that the ratio observed toward the SW position would be about 32\% smaller if the CO(11--10) line would be observed with the same beam of 15\farcs1 as the CO(6--5) line. 
In the model predictions by \citet[][their Fig.~6]{meijerink07} we can use the results for the ratio between CO(10--9) and CO(7--6) since they have similar critical densities and spatial distribution as the CO(11--10) and CO(6--5). Our observed CO(11--10)/CO(6--5) ratio of $\sim0.1$ is expected to be found in PDRs with total densities in the range $10^4$--3$\times$10$^5~\3cm$ and for radiation fields between 10$^2$ and 10$^5$~$G_0$. If X-rays are affecting the excitation of these CO lines, then radiation fields $F_x < 2$~erg~cm$^{-2}$~s$^{-2}$ and densities $>10^6~\3cm$ would be needed to reproduce the observed ratios.
These high densities are to be expected if the higher-$J$ CO lines emerge from the same gas as the HCN~$J=1\to0$ emission \citep{paglione04}.

\section{Summary \& Final Remarks}\label{sec:remarks}

We presented a large set of molecular and atomic lines detected in NGC~253, using the three instruments, SPIRE, PACS and HIFI, on board of the Herschel Space Observatory. About 35 lines were detected and identified in 
the SPIRE spectra (while few lines still remain unidentified), and 30 lines were identified in the four PACS 
spectral ranges. A significant number of lines are still unidentified in the PACS spectra, which will be reported in a follow up work. 

Because NGC~253 is a very rich molecular laboratory 
outside the Milky Way, we were able to detect exotic molecules such CH$^+$ $J=1\to0$ in absorption (with SPIRE), 
and CH$^+$ $J=2\to1$ in emission (with PACS). Other molecules like OH, OH$^+$ and H$^{18}$O were also detected, 
both in emission and in absorption. 

The APEX/SABOCA high resolution map of the dust continuum emission at 350~\mum\ allowed us to estimate an average angular size of 16$''$.7 for the emitting region covered in the 40$''$ equivalent beam size derived for the SPIRE spectra. 
The average source size derived for the continuum emission is in agreement with the source size estimated from the 2-D Gaussian fit of the APEX/CHAMP$^+$ map of the \twco\ $J=6\to5$ line. 

The velocity resolved  spectra of the \cii~158~\mum\ fine-structure line and the high-$J$ CO $J=11\to10$ transition obtained with SOFIA/upGREAT toward a south-west (SW) position in the nuclear region were compared with the corresponding spectra of the CO $J=6\to5$ spectra from APEX/CHAMP$^+$. We found that the corresponding \cii\ obtained from the HIFI map does not match the line shape nor the intensity obtained with SOFIA/upGREAT. We believe that the SOFIA/upGREAT spectrum is correct and that the HIFI spectrum shows excess spectral baseline structure.  The CO(11--10)/CO(6--5) line ratio observed toward the SW position is indicative of high densities, in agreement with the position of the brightest HCN~$J=1\to0$ emission in the nuclear region of NGC~256.

A thorough data reduction and careful combination of the full set of available SPIRE and PACS spectral and 
photometric data, allowed us to merge both data products in order to study the dust continuum SED of NGC~253, as 
seen with Herschel. We found a cold dust component with temperature $\sim$37~K (in agreement with previous 
results quoted in the literature) and a warm dust component at $\sim$70~K, about 20~K higher than previously 
estimated, which is significant when considering a $\pm$10~K error in the temperatures reported. A third component with higher dust temperature of $\sim$188~K was also identified from the continuum flux observed at shorter ($<50$~\mum) wavelengths.
A first order 
estimate of the incident FUV fluxes that heat the three dust components yielded 
values of $G_{0,c}\sim$3.4$\times$10$^5$, $G_{0,w}\sim$8.7$\times$10$^6$ and $G_{0,h}\sim$1.2$\times$10$^9$ (in units of the Habing flux), 
respectively. Total gas mass $\sim$4.5$\times$10$^8$~\Msun\ and column density of $\sim$1.2$\times$10$^{22}$~\Msun\ were estimated from the SPIRE continuum flux observed at 500~\mum.

Combining the SPIRE and PACS data, we obtained \textit{the most extended} \twco\ 
ladder of the 40\farcs\ nuclear region of NGC~253, including the $J=4\to3$ up to $J=13\to12$ transitions from the SPIRE FTS, and the $J=14\to13$ up to $J=19\to18$ transitions from the PACS spectral ranges. A non-LTE excitation analysis showed that at least three 
components are needed in order to fit the \twco\ line spectral energy distribution (LSED). The \thco\ (from 
$J_{up}$=5 to $J_{up}$=8) fluxes detected with SPIRE, as well as ground based observations of the lower-$J$ 
\thco\ and \hcn\ lines, were used to constrain the parameters of the models. 
The total molecular gas mass derived from the three \twco\ components 
is in agreement with the gas mass derived from previous CO observations found in the literature. 
A diffuse and rather warm 
component, with density $\sim$2$\times$10$^3~\3cm$ and temperature $\sim$90~K, was found to fit mostly the 
lower ($J_{up}<7$) \twco\ transitions. The second component correspond to gas with high density 
$\sim$3$\times$10$^5~\3cm$ and a relatively low temperature of $\sim$50~K. Because of their densities and 
temperatures, none of these components can be associated with shocks or mechanical heating. The third component, 
however, that fits mostly the higher-$J$ \twco\ lines detected with PACS, have both high density 
($\sim$4$\times$10$^6~\3cm$) and high temperature ($\sim$160~K) gas, which makes it a better candidate for 
shock/mechanical heating driven gas. We also argue that hot cores are another plausible scenario for 
the excitation of the HCN and PACS \twco\ lines, given the detection and spatial distribution of H$_2$S (likely 
probing hot core chemistry) and \hcn, as observed in high spatial resolution maps. However, the effect and role 
of the enhanced cosmic-rays present in the circumnuclear starburst ring (as derived from SNe rates and gamma-ray 
observations) cannot yet be ruled out, nor accounted for, at this stage. 
The OH lines detected with PACS show fluxes of about one order of magnitude higher than the \hho\ lines, which 
can be a signature of the enhanced cosmic ray ionization rates in the nuclear region of NGC~253. Similarly, the 
detection of the ionic species OH$^+$ and \hho$^+$ are also indicative of high ionization fractions due to the 
enhanced cosmic rays. 

\acknowledgments
SPIRE has been developed by a consortium of insti-
tutes led by Cardiff University (UK) and including: Uni-
versity of Lethbridge (Canada); NAOC (China); CEA, LAM
(France); IFSI, University of Padua (Italy); IAC (Spain);
Stockholm Observatory (Sweden); Imperial College London,
RAL, UCL-MSSL, UKATC, University of Sussex (UK); and
Caltech, JPL, NHSC, University of Colorado (USA). This de-
velopment has been supported by national funding agencies:
CSA (Canada); NAOC (China); CEA, CNES, CNRS (France);
ASI (Italy); MCINN (Spain); SNSB (Sweden); STFC, UKSA
(UK); and NASA (USA).

SOFIA is jointly operated by the Universities Space Research Association, Inc. (USRA), under NASA contract NAS2-97001, and the Deutsches SOFIA Institut (DSI) under DLR contract 50 OK 0901 and 50 OK 1301 to the University of Stuttgart. We thank G.~Sandell, E.~Chambers, and the SOFIA operations crew for their outstanding work during the flight and observing campaigns in Palmdale (USA) and Christchurch (NZ) and for delivering high quality calibrated data.

Dr. J.P. \pb\ (ESO/MPIfR) was sponsored by the Alexander von Humboldt 
Foundation between January 2010 and December 2012, period during which part of this 
work was done. Molecular Databases that have been helpful include the CDMS \citep{mueller05}, 
NASA/JPL \citep{pickett98}, LAMDA \citep{schoier05} and NIST \citep{lovas04}. 

The authors thank the referee for her/his constructive and insightful remarks that helped to improve this work.



\facility{HERSCHEL (SPIRE, PACS \& HIFI), APEX (SABOCA \& CHAMP+), SOFIA (GREAT \& upGREAT)}

\software{HIPE v14,15 \citep{ott10}, KOSMA atmospheric calibrator \citep{guan12}, GILDAS/CLASS \citep{pety05}, RADEX \citep{vdtak07}}



\appendix

\section{SPIRE data reduction}\label{sec:appendix-SPIRE-reduction}

We applied the correction for relative gains for bolometers between the products level 0.5 and 1.0, to obtain a 
better extended map.
In order to extract the integrated fluxes for a given aperture (or beam size), the SPIRE photometry images must 
be converted first from units of Jy/beam to Jy/pixel using the beam areas (in arcsecs$^2$) corresponding to a 
$1''$ beam as assumed in the pipeline, i.e., this process is independent of the spectral index of the source. 
The color corrections instead (needed to extract the fluxes), depend on the spectral index of the source. This 
was estimated as $\alpha=-0.7$ for NGC~253 from high resolution radio observations of SNRs \citep[e.g.,][] {ulvestad97, ulvestad00, tingay04}.
In Table 6.15, Chap.~6.9, {\it Recipe for SPIRE Photometry} (in the HIPE v.15 Help System) there are 
tabulated values for color corrections of the three wavelengths (250 \mum, 350 \mum, and 500 \mum) of the SPIRE 
photometric images, for a number of spectral indexes. We interpolated the tabulated color correction values for alpha=-0.5 and alpha=-1.0 to arrive at one that is applicable for NGC253 at alpha=-0.7. The 40$\arcsec$ aperture integrated fluxes were obtained using the {\it annularSkyAperturePhotometry} task in HIPE. 

\begin{figure*}[htp]
\centering

\hspace{-0.6cm}\includegraphics[angle=0,width=0.48\textwidth]{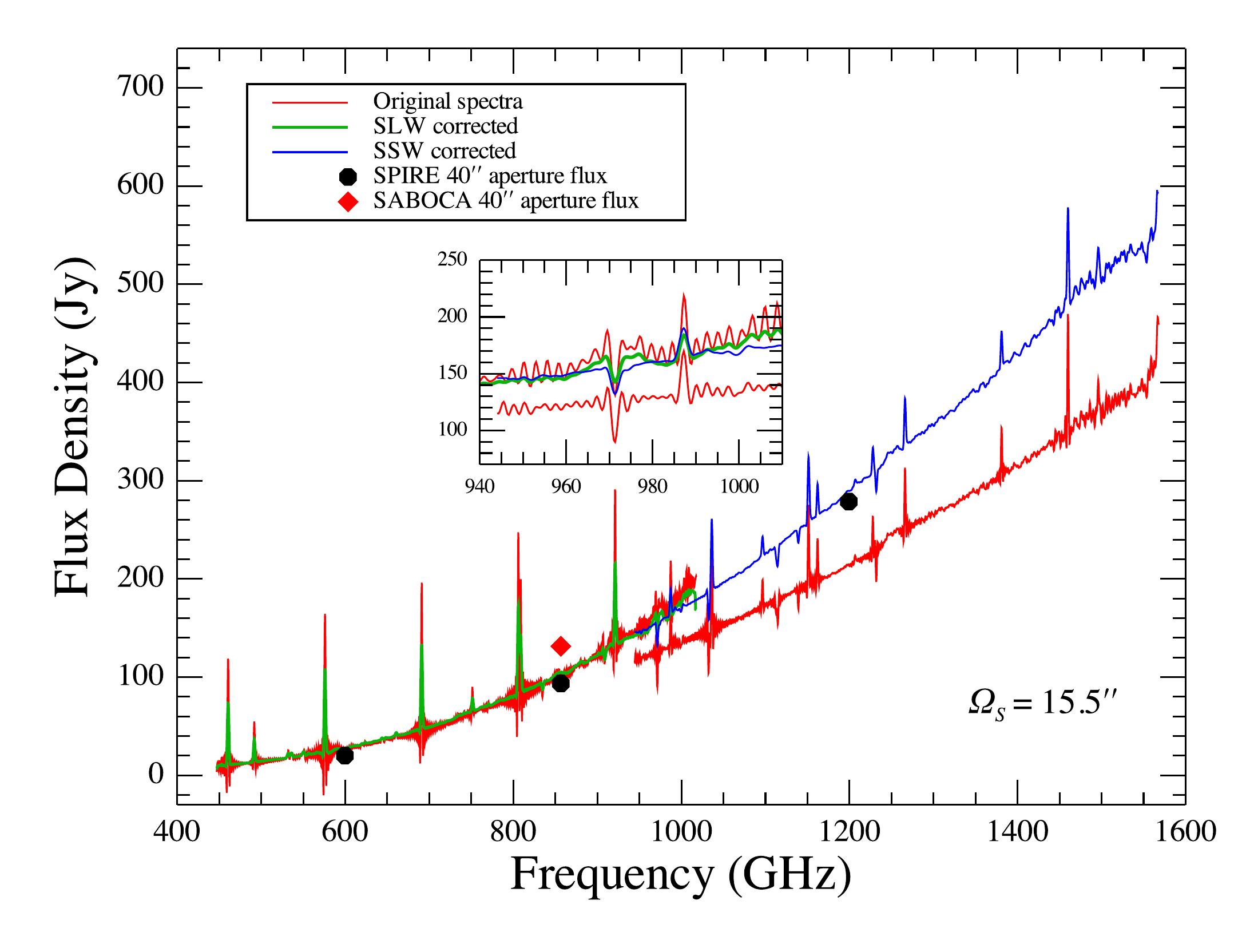}%
\hspace{-0.0cm}\includegraphics[angle=0,width=0.48\textwidth]{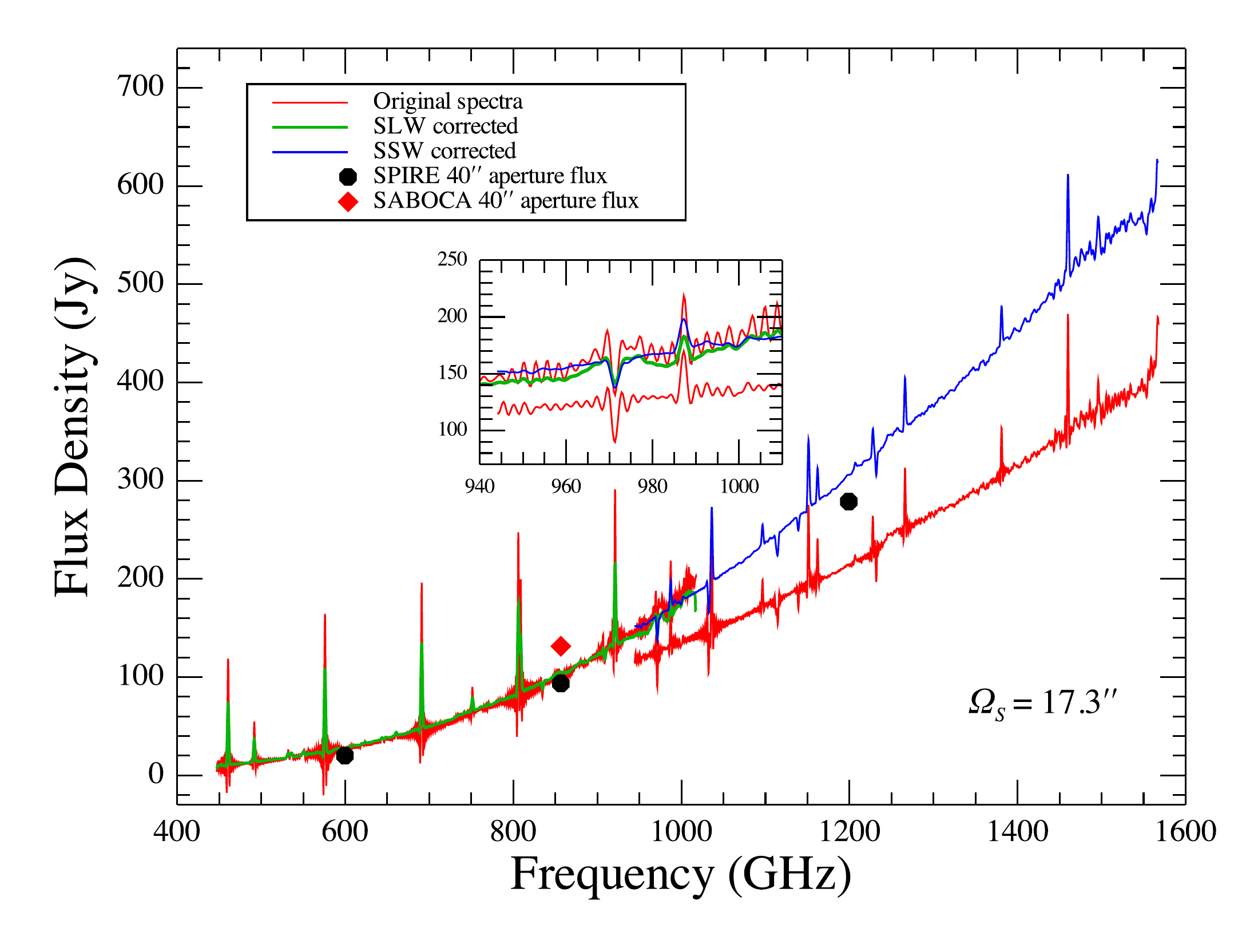}%

\vspace{-0.4cm}

\caption{{\footnotesize SPIRE spectra of NGC253 corrected to a $\sim$40$''$ beam, assuming a source size ($\Theta_S$) with a semi-major axis of FWHM=$16.0''$ (left) and FWHM=$17.3''$ (right) in the {\it semiExtendedCorrector} task. The green and blue lines are the corrected {\it apodized} data from the long- and short-wavelength FTS bands, respectively, while the original {\it unapodized} spectra is shown in red. The dots show the $40''$ aperture integrated fluxes from the SPIRE photometric maps (black) and APEX/SABOCA (red). The inset shows a zoom in to the overlap region between the SLW and SSW bands.}}
\label{fig:corrected-spectrum-comparison}
\end{figure*}

Using a Gaussian shape in the optimization method of the {\it semiExtendedCorrector}, 
and the eccentricity (0.8463) found from the 2-D Gaussian fit of the SABOCA map, we found that a semi-major axis (FWHM) of 15\farcs5 would 
be the optimal fit to match the spectra in the overlap region (between about 960 GHz and 990 GHz).
This is shown in Fig.~\ref{fig:corrected-spectrum-comparison} ({\it left}) with the original and 
corrected spectra using the optimized source size, and a zoom in the overlap region showed in the inset.
On the other hand, using the semi-major axis (FWHM) of 17\farcs3 from the Gaussian fit on the SABOCA map, the corrected spectra 
shows a better match between the OH$^+$ line detected in absorption in both bands 
(right panel in Fig.~\ref{fig:corrected-spectrum-comparison}). 

Note that the change in the line fluxes introduced by the {\it semiExtendedCorrector} tool
are between 4\% and 8\% weaker than without correction in the SLW band, and 22\% --  30\% brighter than without correction in the SSW band. Different source sizes and shapes (or methods) used to correct and match the SPIRE FTS bands will have different effects in the line fluxes.

Since the difference between the SLW continuum level and the photometric flux at 250~\mum\ is the largest ($\sim$27 Jy) of the three wavelengths when using a source size of $17''.3\times9''.2$ 
(while the continuum level at the other two wavelenghts are within the errors), we interpret this 
as an indication that the spectral continuum level at frequencies above 1000~GHz in the SLW band may still be affected by calibration uncertainties. 
Nevertheless, the spectrum of Fig.~\ref{fig:corrected-spectrum-comparison}-\textit{right} is the best calibrated and corrected SPIRE spectrum reported so far for NGC~253, and it will be the spectrum used throughout this work.

\section{PACS data reduction}\label{sec:appendix-PACS-reduction}

\begin{figure*}[htp]
\centering
\hspace{-0.6cm}\includegraphics[angle=0,width=0.48\textwidth]{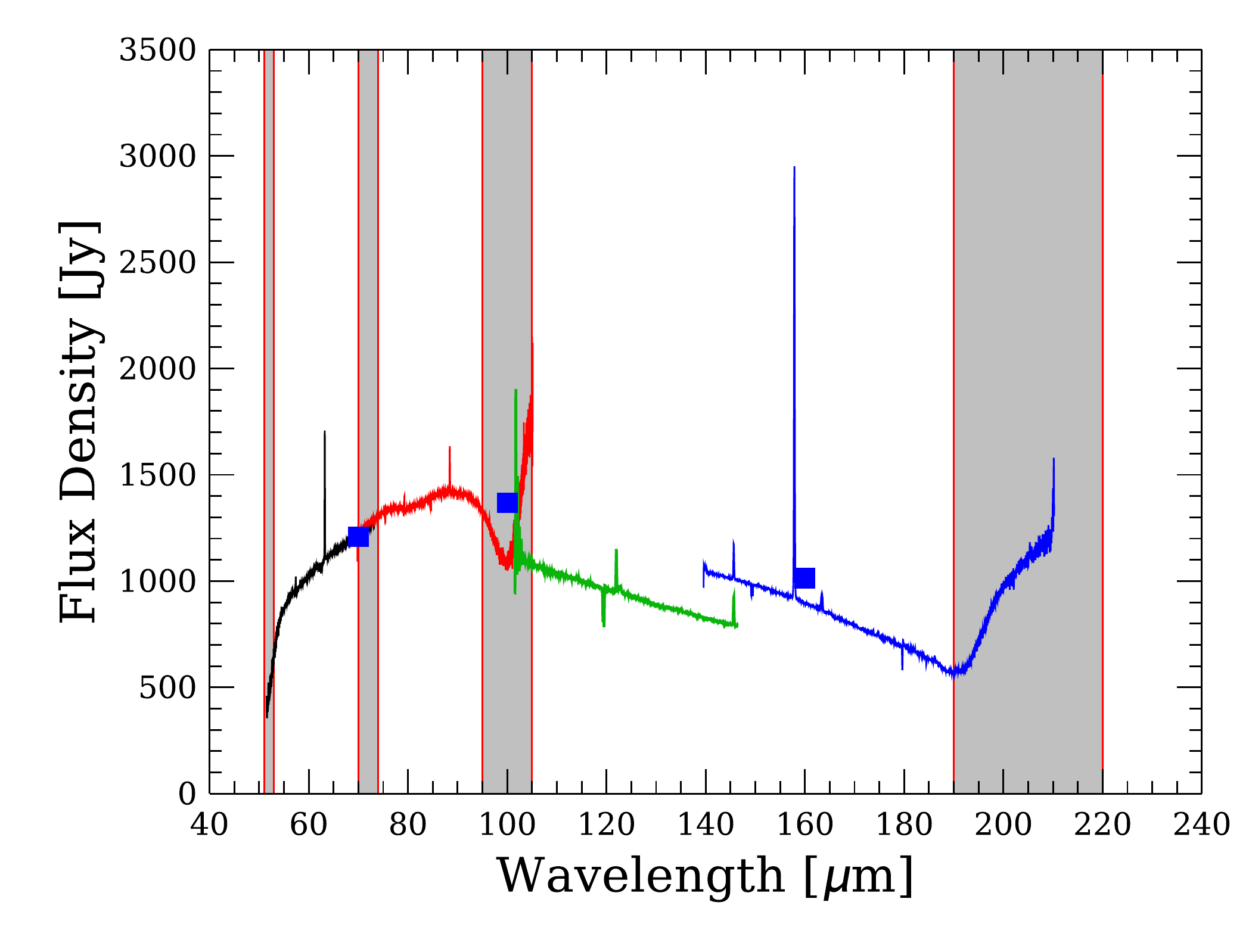}%

\vspace{-0.4cm}

\caption{{\footnotesize PACS spectra of NGC253 extracted from the 5x5 spaxels and corrected 
for point source losses. The spectra obtained with the default pipeline for long range SED 
observations (left panel) shows a poor match with the PACS 40$''$ aperture photometry fluxes 
at 70~\mum, 100~\mum, and 160~\mum\ (squares), and between the wave ranges.  
Using the background normalization (right panel), and excluding the spectral leakage regions (vertical filled areas) from the \textit{specFlatFieldRange} procedure, results in a better match with the photometry fluxes and between the wavelength ranges.}}
\label{fig:apendix:pacs-5x5-spectra}
\end{figure*}

The PACS spectral data was  re-processed using the default pipeline for \textit{Chopped Large Range Scan SED} observations including the telescope background normalization. The leakage regions of the spectra was excluded in the \textit{specFlatFieldRange} procedure of the pipeline, and a polynomial of order three was used instead of the 
default order five (used in HIPE versions $<$ v.14) or wavelet (in the latest versions of HIPE), in order to improve the normalization of the population of spectra used to 
obtain a single spectrum for each spaxel.

For consistency with the 40$''$ \textit{beam corrected} SPIRE spectra (Sect.~\ref{sec:spire-corrected}), 
the PACS spectral ranges were obtained as the total cumulative spectra from all the 
spaxels, corrected by the 3$\times$3 \textit{point source correction factor} included in the PACS SPG 
v14.2.0 calibration tree. Using the 5$\times$5 \textit{point source correction factor} overestimate the continuum flux as compared with the corresponding photometric fluxes at 70~\mum, 100~\mum\ and 160~\mum.

We compared the continuum level of the PACS spectra
with the corresponding 40$''$ aperture fluxes of the PACS photometry maps (from the HSA, obs. IDs 
1342221744 and 1342221745) at 70~\mum, 100~\mum, and 160~\mum. The aperture fluxes were obtained using the \textit{annularSkyAperturePhotometry} task, with inner and outer radius of 750$''$ and 800$''$, respectively. The PACS flux uncertainties include the errors estimated with the annular sky aperture and the 7\% absolute point-source flux calibration for scan maps \citep{balog14}.

The 5$\times$5 corrected PACS spectra, obtained with the modified \textit{Chopped Large Range Scan SED} 
pipeline, are shown in Fig.~\ref{fig:apendix:pacs-5x5-spectra} (\textit{left}). 
The continuum level of some wavelength ranges do 
not match the PACS photometry fluxes, specially the wave range between 100~\mum\ and 150~\mum. 
This mismatch is due mainly to the fact that NGC~253 is a bright source, the default pipeline used in earlier versions of HIPE were not optimized for bright sources. The mismatch is corrected when using the same modified 
pipeline mentioned above, but using the \textit{background normalization} procedure. This is now included in the standard SPG of HIPE v.15. 

The final 5$\times$5 corrected, and background normalized, PACS spectra used in this work are shown in Fig.~\ref{fig:spire-pacs-corrected-spectrum} (\textit{right}). The section of the spectrum affected by spectral leakage are shown by grey filled bands, and they were not used in our analysis.

\section{HIFI single pointing spectra}\label{sec:appendix-HIFI-single}

\begin{figure*}[htp]
\centering
\hspace{-0.0cm}\includegraphics[angle=0,width=0.33\textwidth]{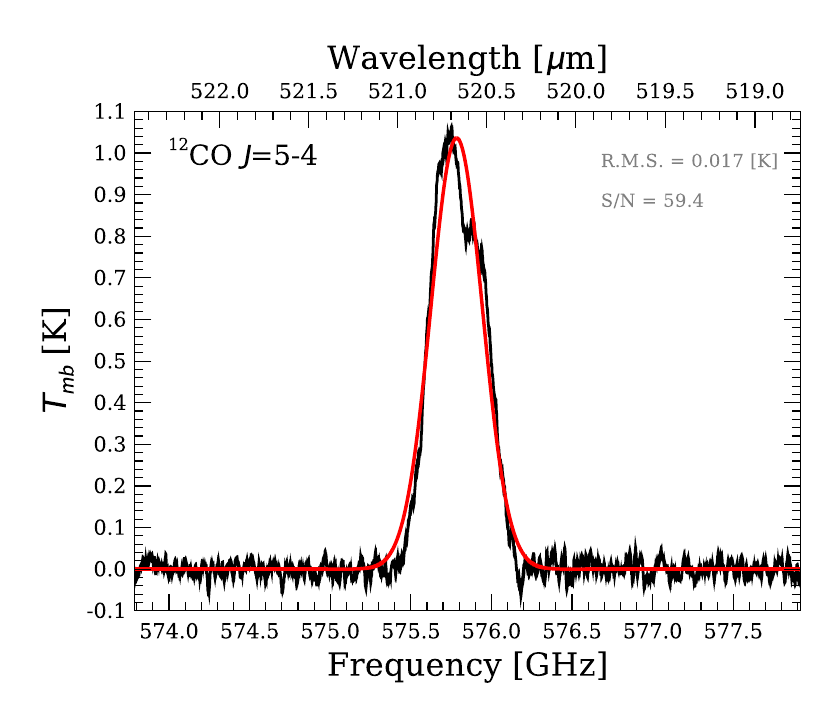}%
\hspace{-0.0cm}\includegraphics[angle=0,width=0.33\textwidth]{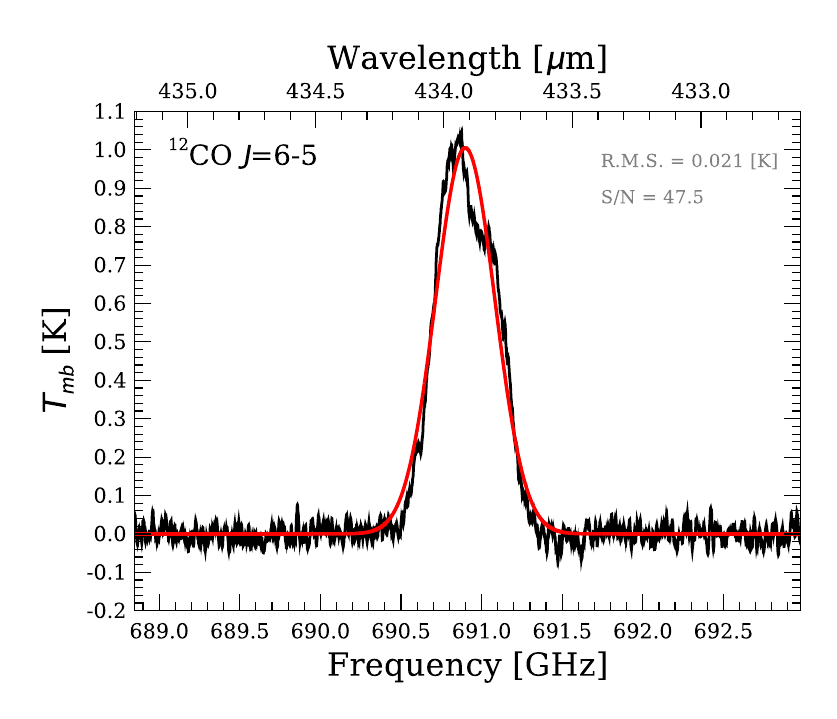}%
\hspace{-0.0cm}\includegraphics[angle=0,width=0.33\textwidth]{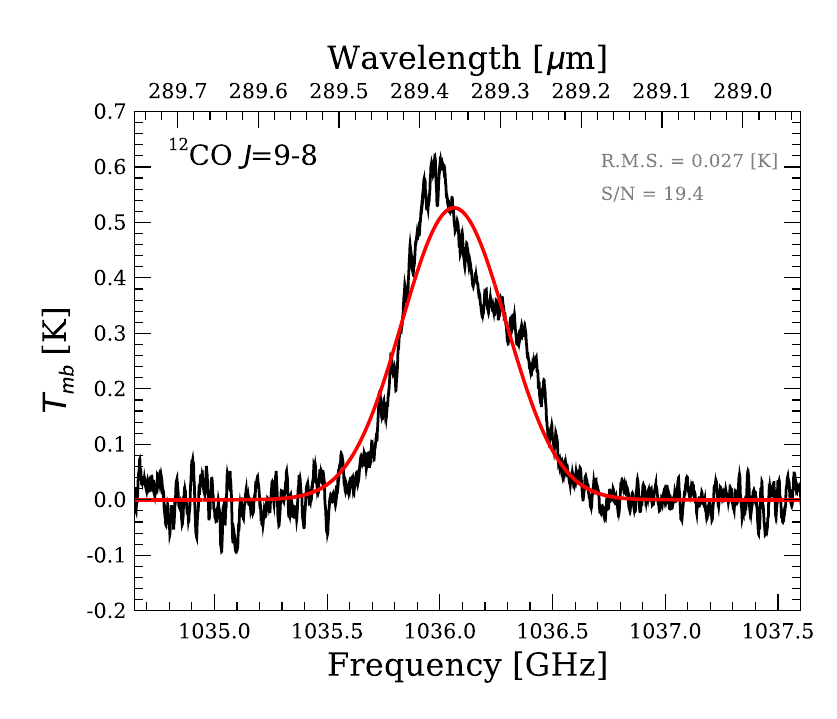}%

\hspace{-0.0cm}\includegraphics[angle=0,width=0.33\textwidth]{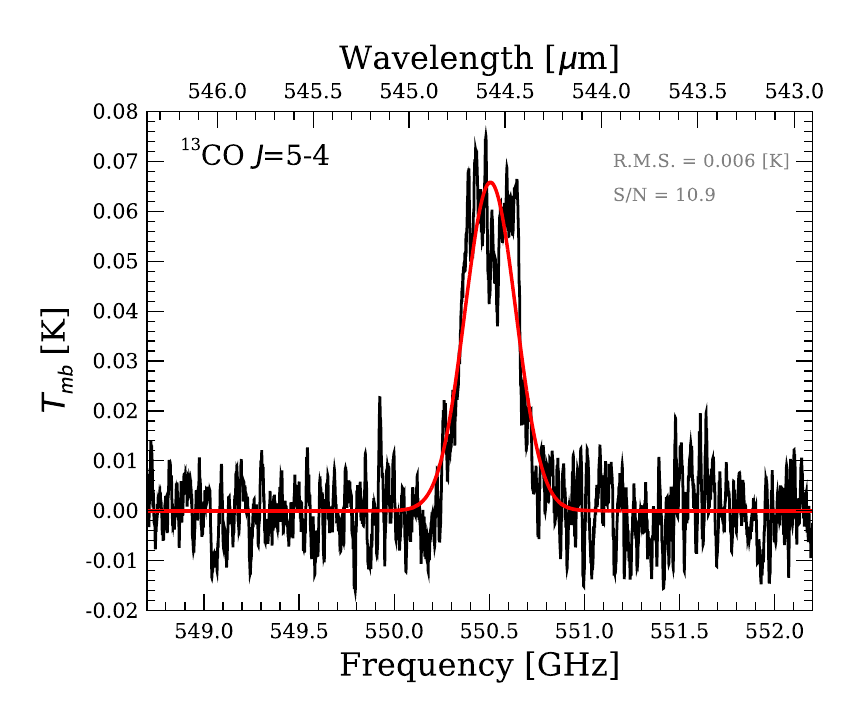}%
\hspace{-0.0cm}\includegraphics[angle=0,width=0.33\textwidth]{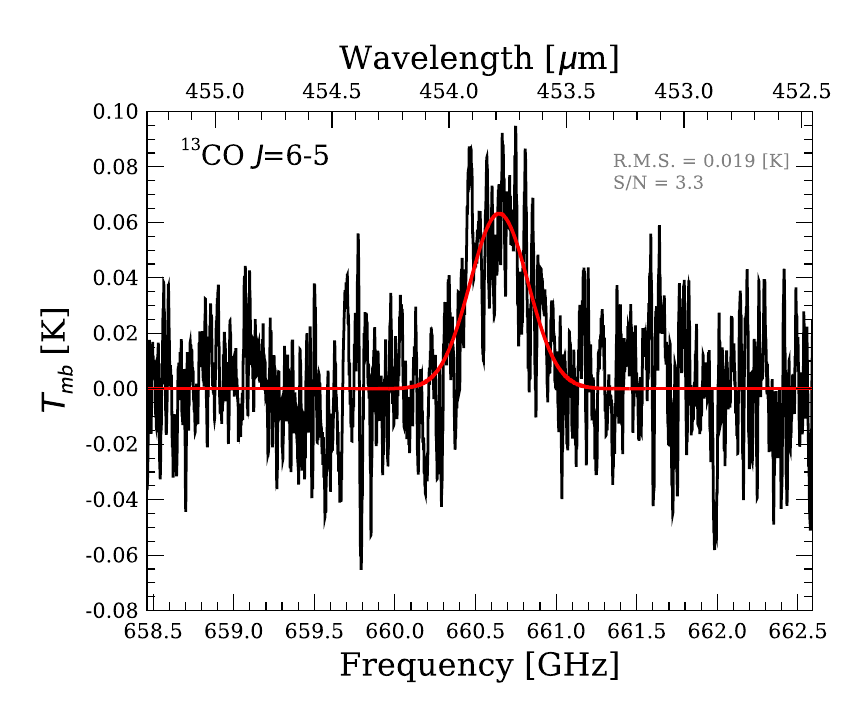}%
\hspace{-0.0cm}\includegraphics[angle=0,width=0.33\textwidth]{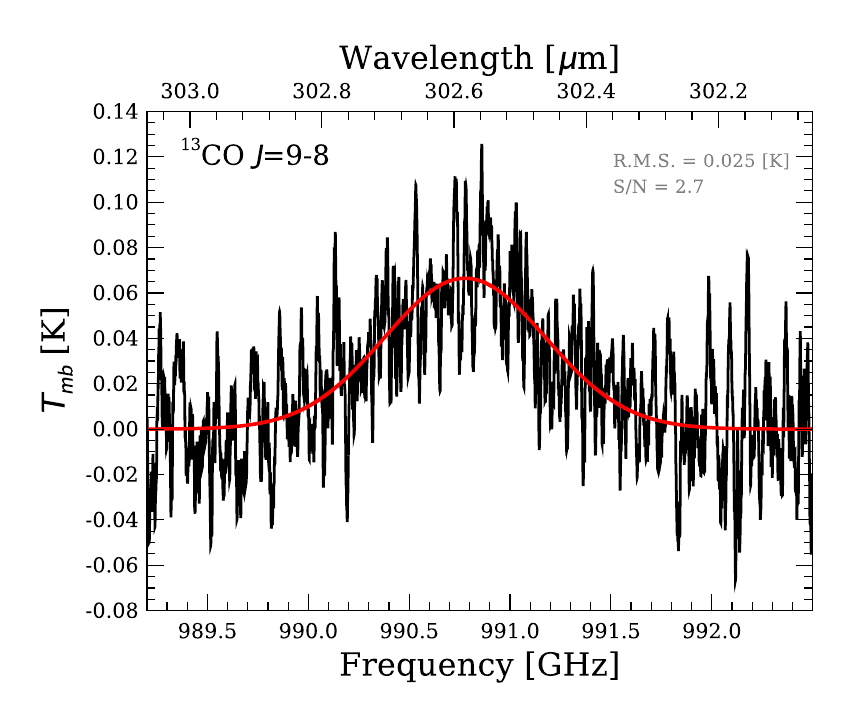}%

\hspace{-0.0cm}\includegraphics[angle=0,width=0.33\textwidth]{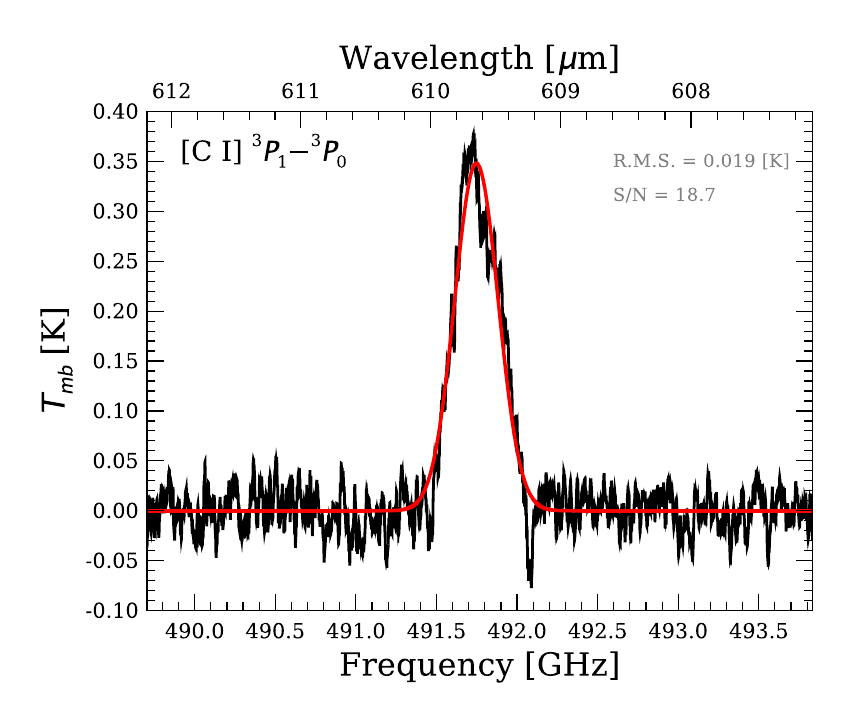}%
\hspace{-0.0cm}\includegraphics[angle=0,width=0.33\textwidth]{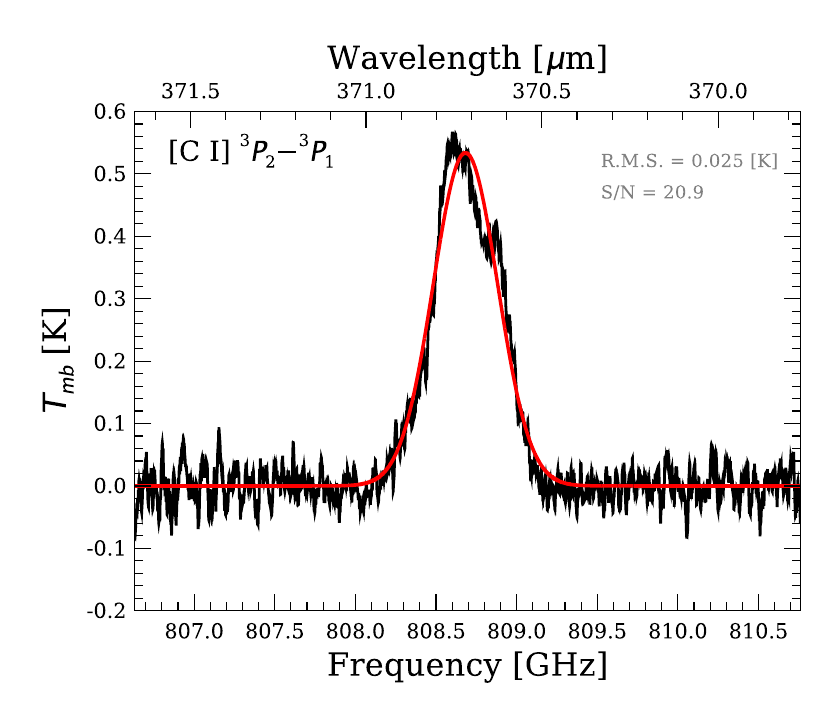}%
\hspace{-0.0cm}\includegraphics[angle=0,width=0.33\textwidth]{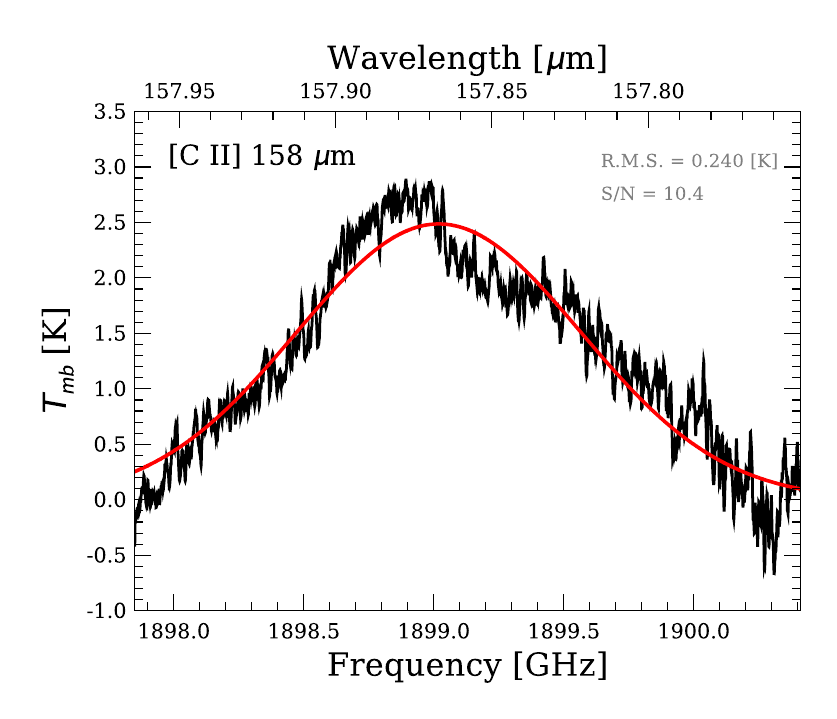}%

\vspace{-0.4cm}

\caption{{\footnotesize HIFI single pointing spectra of several lines at their respective beam sizes. The peak of the \thco~$J=9\to8$} does not correspond to the systemic velocity of NGC~253 and the fitted FWHM is broader than that of the \twco~$J=9\to8$; thus it is most likely a standing wave.}
\label{fig:hifi-single-point-spectra}
\end{figure*}

\newpage




\bibliographystyle{mn2e}
\setlength{\bibsep}{-2.1pt}
\bibliography{NGC253}

\end{document}